\documentclass[reprint,aps,prb,twocolumn,superscriptaddress]{revtex4-1}
\usepackage{}
\usepackage{chngcntr}
\usepackage{amssymb}
\usepackage{upgreek}
\usepackage{graphicx}
\usepackage{mathrsfs}
\usepackage{textcomp}
\usepackage{mathcomp}
\usepackage[utf8]{inputenc}
\usepackage{color}
\usepackage{siunitx}
\usepackage{url}
\usepackage{multirow}
\usepackage{placeins}
\usepackage{gensymb}
\usepackage[utf8]{inputenc}
\usepackage{float}
\usepackage[paperwidth=210mm,paperheight=297mm,centering,hmargin=2cm,vmargin=2cm]{geometry}
\usepackage{lipsum}
\usepackage{circuitikz}
\usepackage{fancyhdr}
\usepackage{mathtools}
\usepackage{bm}
\usepackage[normalem]{ulem}
\usepackage{upgreek}
\usepackage{comment}
\DeclareSIUnit\angstrom{\text {Å}}

\begin{document}

\title{Effect of Ni substitution on the fragile magnetic system ${\text{La}_{5}\text{Co}_{2}\text {Ge}_{3}}$}

\author{Atreyee Das}
\affiliation{Ames National Laboratory, U.S. DOE, Iowa State 
University, Ames, Iowa 50011, USA}
\affiliation{Department of Physics and Astronomy, Iowa State University, Ames, Iowa 50011, USA}

\author{Tyler J. Slade}
\affiliation{Ames National Laboratory, U.S. DOE, Iowa State 
University, Ames, Iowa 50011, USA}

\author{Rustem Khasanov}
\affiliation{Laboratory for Muon Spin Spectroscopy, Paul Scherrer Institut, CH-5232 Villigen, Switzerland}

\author{Sergey L. Bud'ko}
\affiliation{Ames National Laboratory, U.S. DOE, Iowa State 
University, Ames, Iowa 50011, USA}
\affiliation{Department of Physics and Astronomy, Iowa State University, Ames, Iowa 50011, USA}

\author{Paul C. Canfield}
\affiliation{Ames National Laboratory, U.S. DOE, Iowa State 
University, Ames, Iowa 50011, USA}
\affiliation{Department of Physics and Astronomy, Iowa State University, Ames, Iowa 50011, USA}

\begin{abstract}
    $\text{La}_{5}\text{Co}_{2}\text{Ge}_{3}$ is an itinerant ferromagnet with a Curie temperature, $T_C$, of $\sim$ 3.8 K and a remarkably small saturated moment of  0.1 $\mu_{B}/\text{Co}$. Here we present the growth and characterization of single crystals of the ${\text{La}_{5}\text{(Co}_{1-x}\text {Ni}_{x})_2\text {Ge}_{3}}$ series for 0.00 $\leq x \leq$ 0.186. We measured powder X-ray diffraction, composition as well as anisotropic temperature dependent resistivity, temperature and field dependent magnetization along with  heat capacity on these single crystals. We also measured muon-spin rotation/relaxation ($\mu \text{SR}$) for some Ni substitutions ($x$ = 0.027, 0.036, 0.074) to study the evolution of internal field with Ni substitution. Using the measured data we infer a low temperature, transition temperature-composition phase diagram for ${\text{La}_{5}\text{(Co}_{1-x}\text {Ni}_{x})_2\text {Ge}_{3}}$. We find that $T_{C}$ is suppressed for low dopings, $x \leq 0.014 $; whereas for $0.036 \leq {x} \leq 0.186 $, the samples are antiferromagnetic with a Neel temperature, $T_{N}$, that goes through a weak and shallow maximum ($T_N \sim$ 3.4 K for $ x \sim$ 0.07) and then gradually decreases to 2.4 K by $x$ = 0.186. For intermediate Ni substitutions, $0.016 \leq {x} \leq 0.027 $, two transition temperatures are inferred with $T_N > T_C$. Whereas the $T-x$ phase diagram for ${\text{La}_{5}\text{(Co}_{1-x}\text {Ni}_{x})_2\text {Ge}_{3}}$ and the $T-p$ phase diagram determined for the parent $\text{La}_{5}\text{Co}_{2}\text{Ge}_{3}$ under hydrostatic pressure are grossly similar, changing from a low doping or low pressure ferromagnetic (FM) ground state to a high doped or pressure antiferromagnetic (AFM) state, perturbation by Ni substitution enabled us to identify an intermediate doping regime where both FM and AFM transitions occur.
\end{abstract}

\maketitle

\section{Introduction}
\label{sec:Introduction}

Itinerant, metallic, magnetic systems with low transition temperatures have attracted much attention in recent years. The reason can be attributed to the fact that many of these materials allow for suppressing their transition temperatures by application of pressure, changing the chemical composition, or by magnetic field \cite{Canfield2016PreservedMagnetism}. Suppressing phase transitions to zero temperature is often studied as several exotic physical phenomenon such as unconventional superconductivity, non Fermi liquid, etc. are found in proximity of the quantum critical point (QCP) for a second-order transition or a quantum phase transition (QPT) in the case of a first order transition. \cite{Canfield2016PreservedMagnetism, Brando2016MetallicFerromagnets, Pfleiderer2001Non-Fermi-liquidFerromagnets, Huy2007SuperconductivityUCoGe,  Dagotto1994, Paglione2010High-temperatureMaterials, Pfleiderer2001, Gegenwart2008QuantumMetals, Levy2007AcuteURhGe, Uemura2007PhaseSr1xCaxRuO3, Pfleiderer2004PartialMnSi, Ubaid-Kassis2010Quantum/math, Pfleiderer2009SuperconductingCompounds, Saxena2000SuperconductivityUGe2, Aoki2001CoexistenceURhGe, Cheng2015PressureMnP, Ran2019NearlySuperconductivity}. 
Intensive studies have shown that although antiferromagnetic (AFM) transitions in many metallic systems can be continuously suppressed to zero temperature by the non-thermal tuning parameters discussed above \cite{Canfield2016PreservedMagnetism, Shibauchi2014APnictides, Gegenwart2008QuantumMetals, Lohneysen1994Non-Fermi-liquidInstabilityb, Friedemann2009DetachingInYbRh2Si2, Schmiedeshoff2011MultipleYbAgGe, Tokiwa2013QuantumYbAgGe, Mun2013Magnetic-field-tunedYbPtBi, Nandi2010AnomalousCrystals},  the situation becomes quite different for ferromagnetic (FM) transitions. The current theoretical models suggest that stoichiometric systems with minimum disorder generally avoid a FM QCP at zero field, and either the transition becomes a first order through a tricritical point, or a long wavelength AFM phase appears. \cite{Belitz1999FirstFerromagnets, Chubukov2004InstabilityFerromagnets, Conduit2009InhomogeneousFerromagnetism,  Karahasanovic2012QuantumPoints, Pedder2013ResummationPoints,  Belitz2017QuantumMagnets, Brando2016MetallicFerromagnets,Rech2006QuantumPoint, Wysokinski2019MechanismMagnets}. The evolution of a first order FM transition to a quantum phase transition (QPT) has been experimentally verified in several systems, \cite{Brando2016MetallicFerromagnets, Huxley2000MetamagneticUGe2, Pfleiderer2002PressurePresented, Uhlarz2004QuantumZrZn2, Niklowitz2005Spin-fluctuation-dominatedPressure,Kaluarachchi2017TricriticalPressure, Gati2021FormationLaCrGe3} and the transition from ferromagnetic to a modulated magnetic phase has been experimentally observed as well \cite{Brando2016MetallicFerromagnets, Kotegawa2019Indicationsub4/sub, Niklowitz2019Ultrasmall/math, Jesche2017AvoidedCeFePO} . An exception to this scenario was reported in a recent theoretical work, where it was proposed that a FM QCP is achievable even in clean systems with non-centrosymmetric symmetry and sufficiently strong spin-orbit interactions \cite{Kirkpatrick2020FerromagneticSystems}. Although hydrostatic pressure, or in some AFM systems, applied magnetic field, are the preferred choices for suppression of transition temperature, the reason being that they do not introduce any additional disorder, chemical substitution is broadly accepted as a viable tuning parameter to access QCP/QPT and, more generally, tune the ground state in magnetic systems. The complex and intriguing physics associated with FM-QPT and AFM QCP described above motivates the search for new metallic ferromagnetic systems and the tuning of their magnetic properties. \par

$\text{La}_{5}\text{Co}_{2}\text{Ge}_{3}$ is a recently discovered itinerant ferromagnetic compound\cite{Saunders2020ExceedinglyLa5Co2Ge3}. It belongs to the family of $\text{R}_{5}\text{Co}_{2}\text{Ge}_{3}$, (R = La-Sm), which crystallizes in a monoclinic structure (space group C2/m) \cite{Lin2017PolarClusters}. Transport, magnetization, and heat capacity measurements along with muon-spin rotation/relaxation ($\mu$SR) studies at ambient pressure reveal that $\text{La}_{5}\text{Co}_{2}\text{Ge}_{3}$ undergoes a ferromagnetic transition at 3.8 K, and has a very low saturated moment of $\sim$ 0.1 $\mu_{B}/\text{Co}$ \cite{Saunders2020ExceedinglyLa5Co2Ge3}. With an effective moment of $\sim$ 1.0 $\mu_{B}/\text{Co}$, it has a Rhodes-Wohlfarth ratio \cite{Rhodes1963TheAlloys} of 4.9, higher than several well established itinerant ferromagnetic systems.\cite{Santiago2017ItinerantMetals, Saunders2020ExceedinglyLa5Co2Ge3}. All of this makes $\text{La}_{5}\text{Co}_{2}\text{Ge}_{3}$ a very promising candidate for tuning or modifying its magnetic transition. \par

 Initial tuning of the magnetic transition temperature of $\text{La}_{5}\text{Co}_{2}\text{Ge}_{3}$ was done via the application of hydrostatic pressure. The temperature-pressure (T-p) phase diagram up to 5.12 GPa was determined from resistivity, magnetization, and heat capacity measurements \cite{Xiang2021AvoidedLa5Co2Ge3}. It was revealed that up to $\sim$ 1.7 GPa, the system remains ferromagnetic and its $T_{C}$ is suppressed to 3 K. Upon further increase of pressure, $\text{La}_{5}\text{Co}_{2}\text{Ge}_{3}$ enters a different low temperature ground state where the transition temperature depends non-monotonically on the applied pressure. This new state is predominantly antiferromagnetic in nature which suggests that $\text{La}_{5}\text{Co}_{2}\text{Ge}_{3}$ may be another example of a metallic ferromagnetic system which avoids quantum criticality by the stabilization of a new non-ferromagnetic phase. 
\par
Motivated by the desire to further study the evolution of magnetic order in $\text{La}_{5}\text{Co}_{2}\text{Ge}_{3}$, in this work, we report the synthesis and physical properties of ${\text{La}_{5}\text{(Co}_{1-x}\text{Ni}_{x})_2\text {Ge}_{3}}$ with ${0.00 \leq x_{EDS} \leq 0.186}$, where $x_{EDS}$ is the substitution value determined from EDS measurements. Substitution by Ni changes the band filling and clearly introduces a different type of perturbation than pressure into the $\text{La}_{5}\text{Co}_{2}\text{Ge}_{3}$ system. 

\par

Single crystals of ${\text{La}_{5}\text{(Co}_{1-x}\text{Ni}_{x})_2\text {Ge}_{3}}$ were grown and their temperature and field dependent magnetization, temperature dependent resistance and heat capacity were thoroughly investigated. Our studies reveal that for the low substitution levels, between ${0.00 \leq x_{EDS} \leq 0.014}$, the system remains ferromagnetic and the transition temperature is suppressed to 3.2 K. For higher substitution levels, ${0.036 \leq x_{EDS} \leq 0.186}$, the magnetization measurements indicate the appearance of an AFM state. We also see the emergence of an upturn in the resistance that can be associated with the opening of a super-zone gap at temperatures similar to that of the AFM transition, which is found to first weakly increase to a shallow maximum $\text{T}_N \sim$ 3.5 K near $ x \sim$ 0.07 and then slowly decrease to T = 2.4 K for $x$ = 0.186. The intermediate Ni substitutions,  $ 0.016 \leq {x} \leq 0.027 $, show both anti- and ferromagnetic transitions with $\text{T}_N > \text{T}_C$. Muon-spin rotation/relaxation ($\mu \text{SR}$) measurements were done on ${x_{EDS} = 0.027,~ 0.036 ~\text{and}~ 0.074}$ samples to compare the evolution of the internal field due to Ni substitution and compliment the results obtained from the above transport and thermodynamic studies. \par

\section{Experimental Details}
\label{sec:Experimental Details}

\begin{figure}[h!]
    \centering
    \includegraphics[width=\linewidth]{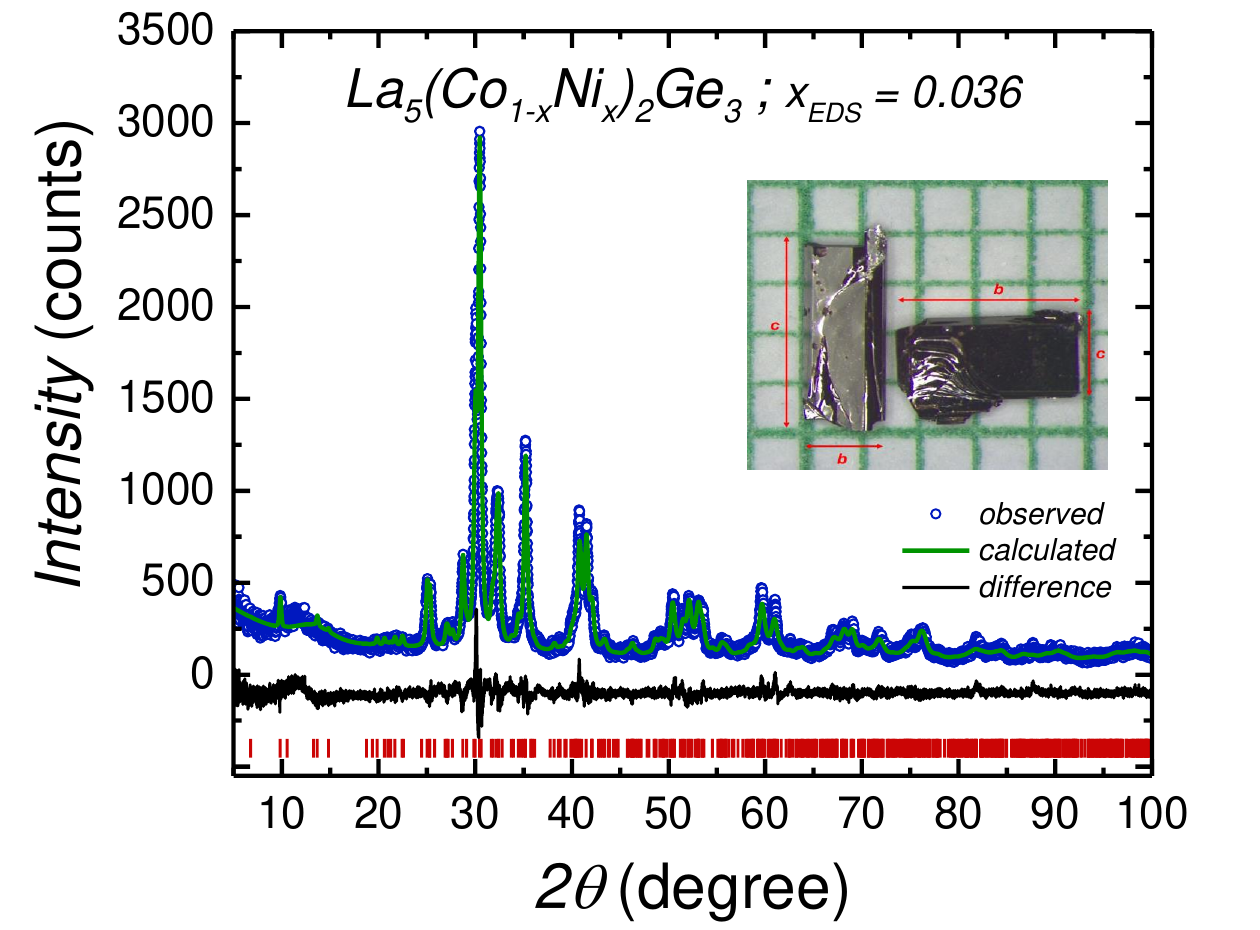}
    \caption{\footnotesize{(Color online) Powder X-Ray diffraction data for ${\text{La}_{5}\text{(Co}_{1-x}\text {Ni}_{x})_2\text {Ge}_{3}}$ for $x_{EDS} = 0.036 $. The data for other $x$-values are qualitatively the same. The vertical red lines represent the expected peak positions for the C2/m ${\text{La}_{5}\text{Co}_{2}\text {Ge}_{3}}$ structural model. (Inset: Typical crystals of ${\text{La}_{5}\text{(Co}_{1-x}\text {Ni}_{x})_2\text {Ge}_{3}}$ for $x_{EDS}$ = 0.036 on a mm grid. The crystallographic $b-$ and $c-$ directions are indicated by red arrows next to the crystals. Crystals of the same batch were used for obtaining the powder pattern shown here.) }}
    \label{fig:PXRD}
\end{figure}

Single  crystals of ${\text{La}_{5}\text{(Co}_{1-x}\text {Ni}_{x})_2\text {Ge}_{3}}$; with ${x_{nominal}}$ = 0.0, 0.01, 0.0175, 0.02, 0.0225, 0.025, 0.03, 0.04, 0.06, 0.08, 0.10, 0.12 and 0.15 were synthesized using a self flux solution growth method \cite{Canfield2001High-temperatureQuasicrystals,Canfield2020NewPhysics,Canfield1992GrowthFluxes} in a manner similar to the growth of pure $\text{La}_{5}\text{Co}_{2}\text{Ge}_{3}$ \cite{Saunders2020ExceedinglyLa5Co2Ge3}. Small pieces of Lanthanum (Materials Preparation Center - Ames National Laboratory $99.99\%$), Cobalt (American Elements $99.95\%$), Nickel (Alfa Aesar $99.98\%$) and Germanium (Alfa Aesar $99.999\%$) with a stating composition of ${\text{La}_{45}\text{(Co}_{1-x}\text {Ni}_{x})_{45}\text {Ge}_{10}}$, were weighed and then placed in a Tantalum crucible \cite{Canfield2001High-temperatureQuasicrystals, Canfield2020NewPhysics}, and sealed in a fused silica ampoule under a partial argon atmosphere. The ampoule was heated in a box-furnace to $1180\tccentigrade$ over 5 hours, held at the temperature for 10 hours, quickly cooled down to $900\tccentigrade$ followed by a slow cool to $800\tccentigrade$ over 80 hours. After dwelling at $800\tccentigrade$ for a few hours, the excess solution was then decanted using a centrifuge \cite{Canfield1992GrowthFluxes, Canfield2020NewPhysics, Canfield2001High-temperatureQuasicrystals}. The typical dimensions of the crystals are ${5~\text{mm} \times 3~\text{mm}}$ with an average thickness of ${0.5 ~\text{mm}}$ and are of the same morphology as that of undoped compound. (See the inset to Fig \ref{fig:PXRD} below for a picture of a representative crystal with the $b$- and $c$-axes identified.) Higher Ni substitution levels did not yield well formed, or usable, single crystals and hence we stopped attempts of substitution at ${x_{nominal}}$ = 0.15. \par

The Ni substitution levels $x_{EDS}$ of the ${\text{La}_{5}\text{(Co}_{1-x}\text {Ni}_{x})_2\text {Ge}_{3}}$ crystals were determined by Energy Dispersive Spectroscopy (EDS) quantitative chemical analysis using a Thermo Noran Microanalytical system Model C10001 EDS attachment to a JEOL JSM-5910LV scanning-electron microscope (SEM). An acceleration voltage of 21 kV, working distance of 10mm and take off angle of 35${\degree}$ were used for measuring all standards and crystals with unknown composition. Undoped, single crystalline, $\text{La}_{5}\text{Co}_{2}\text{Ge}_{3}$ was used as a standard for \text{La, Co and Ge} whereas single crystalline $\text{La}\text{Ni}_{2}\text{Ge}_{2}$ was the standard for \text{Ni} quantification. The composition of the plate like crystals was measured at 3-4 different spots on the crystal's face, revealing good homogeneity of each crystal. The spectra were fitted and the average compositions as well as the error bars were obtained using NIST-DTSA II Lorentz 2020-06-26 software, accounting for both inhomogeneity and goodness of fit of each spectra \cite{Newbury2014RigorousDTSA-II}. \par

\begin{figure}[h!]
    \centering
    \includegraphics[width=\linewidth]{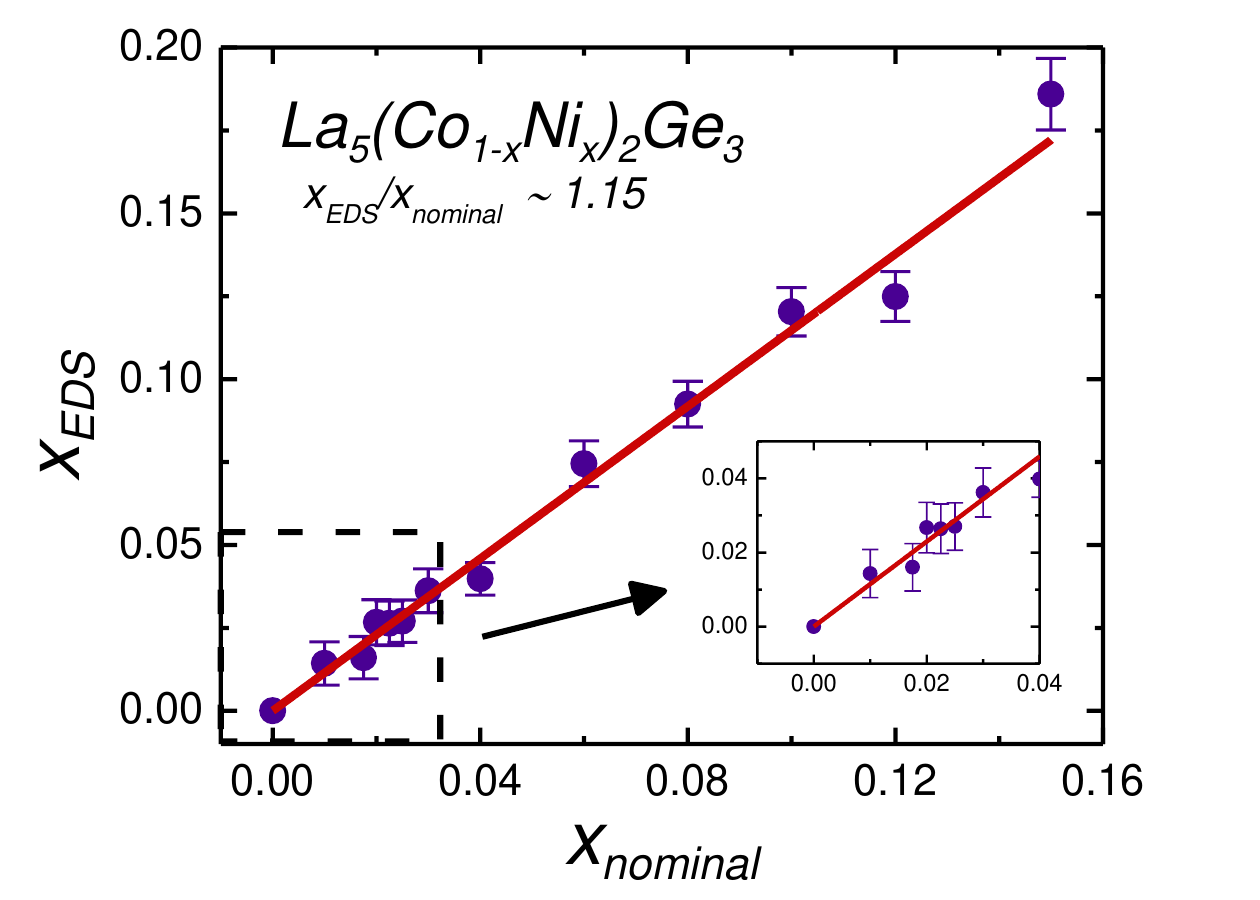}
    \caption{\footnotesize {Measured Ni substitution level, ($x_{EDS}$) values, for $\text{La}_{5}\text{(Co}_{1-x}\text {Ni}_{x})_2\text {Ge}_{3}$ crystals versus $x_{nominal}$ values used in growth solution. The red line is the linear fit across the data points with the intercept fixed to (0,0). (Inset : The lower dopings, nominal ${0.01 \leq x \leq 0.04}$ of the growths done, are zoomed in for clarity).}}
    \label{fig:EDS}
\end{figure} 

The crystal structure was inferred using a Rigaku Miniflex-II powder diffractometer using Cu ${K_{{\alpha}}}$ radiation $(\lambda = 1.5406~\si{\angstrom} )$. The crystals were ground to a fine powder and sieved through a $90~{\mu}m$ mesh  U.S.A. Standard Testing Sieve to reduce preferential orientation. The sieved powder was then mounted and measured on a single crystal Si, zero-background sample holder. The patterns were refined and the lattice parameters were determined using GSAS II software. \cite{Toby2013Package}. \par

Temperature and field dependent magnetization measurements were done in a Quantum Design Magnetic Property Measurement System (QD-MPMS) SQUID magnetometer with samples mounted on a diamagnetic Poly-Chloro-Tri-Fluoro-Ethylene (PCTFE) disk, which snugly fits inside a straw, using a small amount of superglue. The signal from the disk was measured beforehand and subtracted from the combined sample-disk magnetization so as to obtain the value of the moment due to the sample. $M(H)$ measurements were done in magnetic fields from 0 kOe to 50 kOe at a temperature of 1.8 K. Zero field cooled (ZFC) $M(T)$ measurements were done by first lowering the temperature to 1.8 K, and then applying a magnetic field of 1 kOe, after which measurements were done up to 300 K. In addition, for a few substitutions (x = 0.027, 0.036 and 0.074), a 5 quadrant $M(H)$ was done at 1.8 K to a maximum field of 10 kOe as well a zero field cooled (ZFC) and field cooled (FC) $M(T)$ upto 10 K, at a low field of 20 Oe. But before, both $M(H)$ and $M(T)$ data collection, the sample was centered at a temperature of 20 K in a field of 1 kOe, followed by a demagnetization procedure, to reduce remanent magnetic field in the magnetometer. For both these measurements the magnetic field was applied perpendicular to the face of the crystal, $H||a^*$, and in the plane of the plate-like crystal, $H||b~ \text{and} ~H||c$ as done for the undoped compound. \cite{Saunders2020ExceedinglyLa5Co2Ge3}. \par

Temperature dependent resistance measurements were done in the standard four-probe geometry in a Quantum Design Physical Property Measurement System (QD-PPMS) using a 3 mA excitation with a frequency of 17 Hz. Plate like samples of each $x$ of ${\text{La}_{5}\text{(Co}_{1-x}\text{Ni}_{x})_2\text {Ge}_{3}}$ were cut perpendicular to each other with a wire saw so as to apply current either along the $b$ or $c$-axis of the crystal. Electrical contacts were made using 25$\mu m$ diameter platinum wires attached to the bar shaped samples using Epotek-H20E epoxy. $R(T)$ was measured in the range of $1.8~ \text{K} \leq T \leq 300~\text{K}$ on cooling under zero magnetic field. Due to irregularities in the sample shape, there are uncertainties in the measurements of the sample's dimensions. For this reason, $R(T)$ data will be presented as normalized resistance either by its value at 300 K or at 10 K, for clarity in low temperature data. \par

Temperature dependent heat capacity measurements were performed in a QD-PPMS. A plate-like sample was mounted on the micro-calorimeter platform using a small mount of Apeizon N-grease and measured in the range of $1.9~\text{K} \leq T \leq 25~\text{K}$ under zero magnetic field. The addenda (contribution from the grease and the sample platform) was measured separately and subtracted from the total data to obtain the contribution only due to the sample using the PPMS software. For $x_{ EDS}$ = 0.036, $C_{p}(T)$ was measured down to 0.5 K using a 3-He insert to the PPMS.\par

\begin{figure}[h!]
    \centering
    \includegraphics[width=\linewidth]{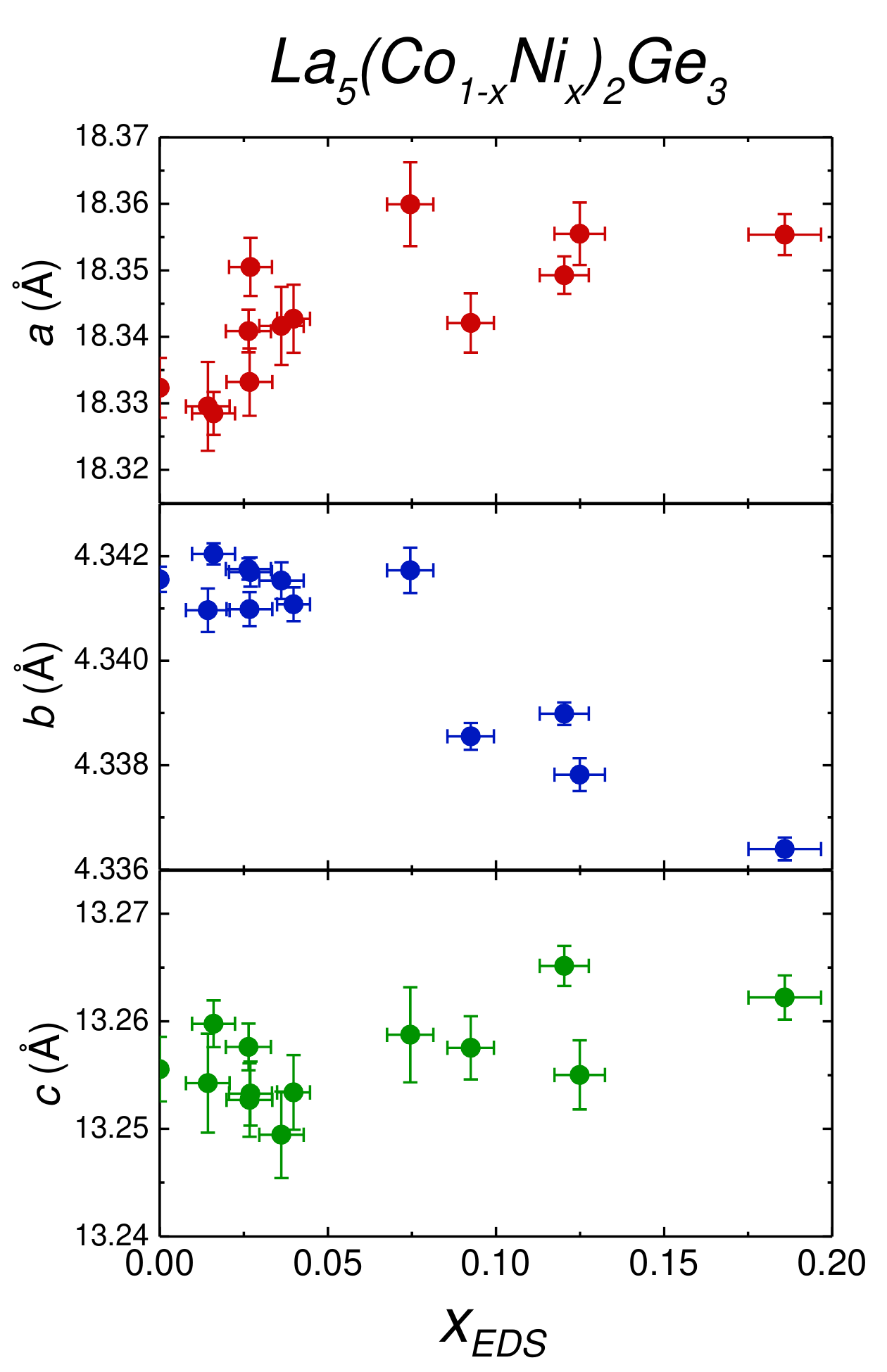}
    \caption{\footnotesize(Color online) Change in the monoclinic lattice parameters a,b,c of the unit cell with change in $x_{EDS}$ in $\text{La}_{5}\text{(Co}_{1-x}\text {Ni}_{x})_2\text {Ge}_{3}$ obtained from the refinement of the powder x-ray diffraction patterns.}
    \label{fig:lattice parameters}
\end{figure}

\begin{figure*}[htbp]
    \centering
    \includegraphics[width=\textwidth]{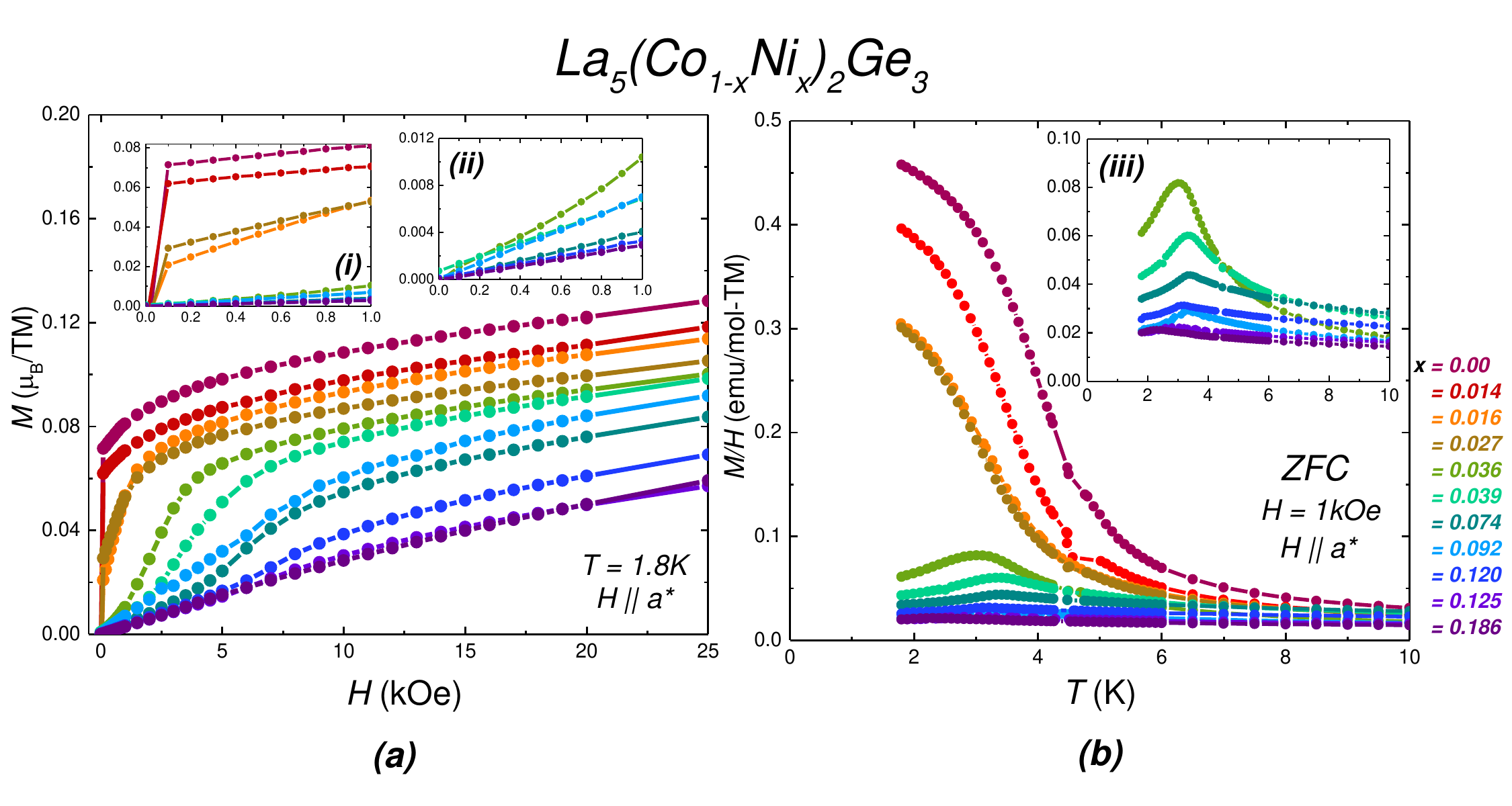}
    \caption{\footnotesize (Color online) (a) The results of the field dependent magnetization at T = 1.8 K for all Ni substituted samples for H$||a^*$ plotted for $0 \leq H \leq 25~\text{kOe}$. The samples were cooled to 1.8 K in the absence of an external field (ZFC) and $M(H)$ measurements were then done in field increasing from zero. (Inset: (i) $M(H)$ data up to 1 kOe for all the dopings. (ii) Zoomed in $M(H)$ data for the AFM samples ($x_{EDS} \geq $ 0.036)). (b) ZFC $M(T)$ measured at 1 kOe plotted for T $\leq$ 10 K. The feature around $\sim$ 4.2 K is an artifact associated with the QD MPMS classic system used for data collection. (Inset: (iii) $M(T)$ data for the AFM Ni substituted $\text{La}_{5}\text{(Co}_{1-x}\text {Ni}_{x})_2\text {Ge}_{3}$ ($x_{EDS} \geq$ 0.036).}
    \label{fig:mag}
\end{figure*}

$\mu \text{SR}$ experiments were done at the $\pi$M3 beam line using the GPS (General Purpose Surface) spectrometer at the Paul Scherrer Institut, Switzerland \cite{Amato2017TheBeam}. The samples were glued on a oxygen free copper (OFC) mount using GE varnish with care that all the single crystals are oriented along the same direction. The crystals were secured with kapton tape covering a 8x8 mm square grid. Both zero-field (ZF) and weak transverse-field (wTF) measurements were done at temperatures ranging from $\simeq$ 1.5 K to $\simeq$ 8 K. 100 ${\%}$ spin polarized muons were implanted into the sample along the $a$* direction (perpendicular to the plate-like face of the crystals). For wTF an external field of 30 Oe was applied along the $b-c$ plane of the crystals. The experiments were performed in a spin-rotated mode, suitable for single crystal samples, allowing us to probe independently the time evolution of the perpendicular and parallel components of the muon-spin polarization $(P(t))$ and determine the direction of magnetic moments in the samples.

\section{Results}
\label{sec:Results}

\subsection*{Composition and Lattice parameters}
\label{sec:Crystallography}

The Ni substitution levels for the different crystals, determined by the EDS measurements $(x_{EDS})$ are plotted as a function of the nominal Ni levels $(x_{nominal})$, used for the growth are shown in Fig \ref{fig:EDS}. As $x_{nominal}$ is increased, $x_{EDS}$ increases in a monotonic manner. The data in Fig \ref{fig:EDS} can be fit well with a straight line, going through the origin, with a slope of $1.15 \pm 0.03$. The inset to Fig \ref{fig:EDS} shows the data for $0.00 \leq \text{x}_{nominal} \leq 0.04 $. It should be pointed out that the values of $x_{EDS}$ for  $0.020 \leq \text{x}_{nominal} \leq 0.025 $  are essentially the same. The detailed measurement results of all these Ni substitution levels will be shown in the Appendix. We will show only the results for one of the substitutions, $\text{x}_{nominal}$ = 0.0225 ($\text{x}_{EDS}$ = 0.027) in the main text for clarity. From this point onwards, in this paper, $x$ and $x_{EDS}$ refer to the $x$ values determined by EDS measurements in $\text{La}_{5}\text{(Co}_{1-x}\text {Ni}_{x})_2\text {Ge}_{3}$  when referring to Ni substitutions. In the rare case when we need to refer to the nominal Ni concentration in the growth melt we will explicitly use the $x_{nominal}$ notation.  \par

Figure \ref{fig:PXRD} shows a representative powder x-ray pattern for ${\text{La}_{5}\text{(Co}_{0.964}\text {Ni}_{0.036})_2\text {Ge}_{3}}$. Similar x-ray patterns for crystals of each Ni substitution have been refined and the lattice parameters $a$, $b$, $c$, and $\beta$ are obtained after refinement of the powder diffraction data using GSAS II\cite{Toby2013Package}. The peaks are matched with the expected peaks of $\text{Pr}_{5}\text{Co}_{2}\text{Ge}_{3}$ monoclinic structure with space group C2/m \cite{Lin2017PolarClusters}. Figure \ref{fig:lattice parameters} shows the change in the lattice parameters $a$, $b$ and $c$ of $\text{La}_{5}\text{(Co}_{1-x}\text {Ni}_{x})_2\text {Ge}_{3}$ versus $x$. There is resolvable decrease of the $b$-lattice parameter, perhaps a modest increase in the $a$-lattice parameter but essentially no resolvable change in the $c$-lattice parameter. $\beta$, the angle between $a$ and $c$, remains almost constant as $x$ increases. Given that the difference in ionic radii between Co and Ni is small and the substitution level is less than 0.20, it is not surprising that there is little resolvable change in the lattice parameters. The exact values of the lattice parameters with change in $x$ values are shown in the Table \ref{tab:lattice} in the Appendix.  \par

\subsection*{Magnetization}
\label{subsec:Magnetization}

\begin{figure}[H]
    \centering
    \includegraphics[width=\linewidth]{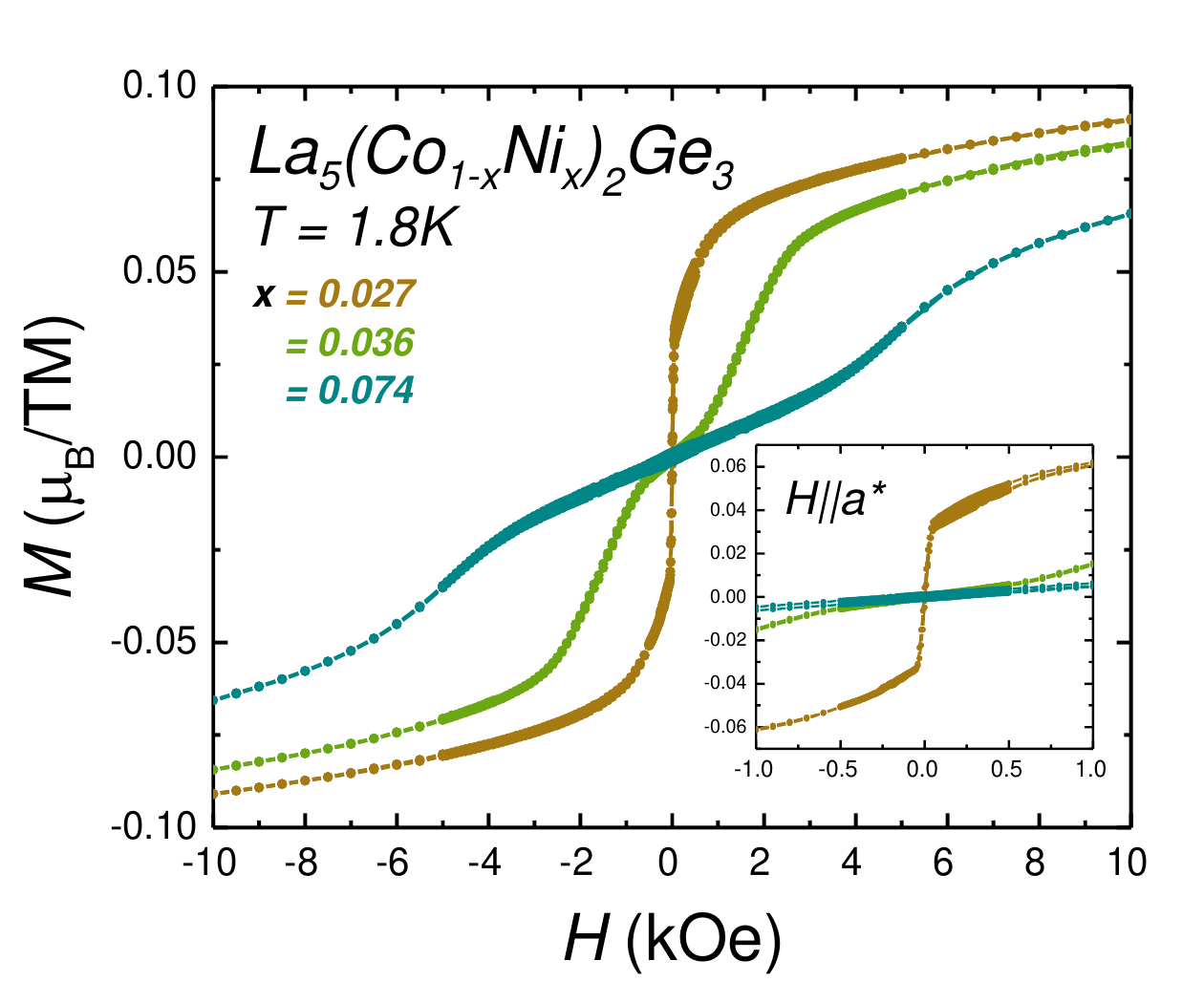}
    \caption{\footnotesize{The 5 quadrant $M(H)$ taken at T = 1.8 K for $x$ = 0.027, 0.036 and 0.074 for $\text{La}_{5}\text{(Co}_{1-x}\text {Ni}_{x})_2\text {Ge}_{3}$ for H$||$a*. (Inset: The $M(H)$ data is zoomed in to show the low field feature).}}
    \label{fig:5 segment}
\end{figure}

Anisotropic field and temperature dependent magnetization measurements were performed on single crystalline samples for each Ni substitution. As was the case for the pure $\text{La}_{5}\text{Co}_{2}\text {Ge}_{3}$ ($x$ = 0.00), for finite $x$, $a*$ is the easy magnetic axis, $b$ is the hard axis, and $c$ is intermediate \cite{Saunders2020ExceedinglyLa5Co2Ge3}. To compare the evolution of magnetism with $x$ thoroughly, we will present a summary of the $M(H)$ and the $M(T)$ data for the field along the $a$* direction in the main text for clarity. For each Ni substitution, the anisotropic $M(H)$ and $M(T)$ along all the three directions is shown separately in the Appendix (Figs \ref{fig:All_0Ni} - \ref{fig:All_15Ni}).  The T = 1.8 K, $M(H)$ data in Fig \ref{fig:mag}(a) suggests that there is a sudden change in the low field response of $\text{La}_{5}\text{(Co}_{1-x}\text {Ni}_{x})_2\text {Ge}_{3}$, from clear ferromagnetic (FM) behavior for $x \leq$ 0.027 to antiferromagnetic (AFM)-like $M(H)$ for $x \geq$ 0.036. This is further evident as we look at the 5 quadrant $M(H)$ taken at T = 1.8 K for $x$ = 0.027, 0.036 and 0.074 in Fig \ref{fig:5 segment}. A finite spontaneous moment is present for $x$ = 0.027 only. We do not observe any significant hysteresis associated with domain pining for any of the Ni substitutions. \par

\begin{figure}[h]
     \centering
     \includegraphics[width=\linewidth]{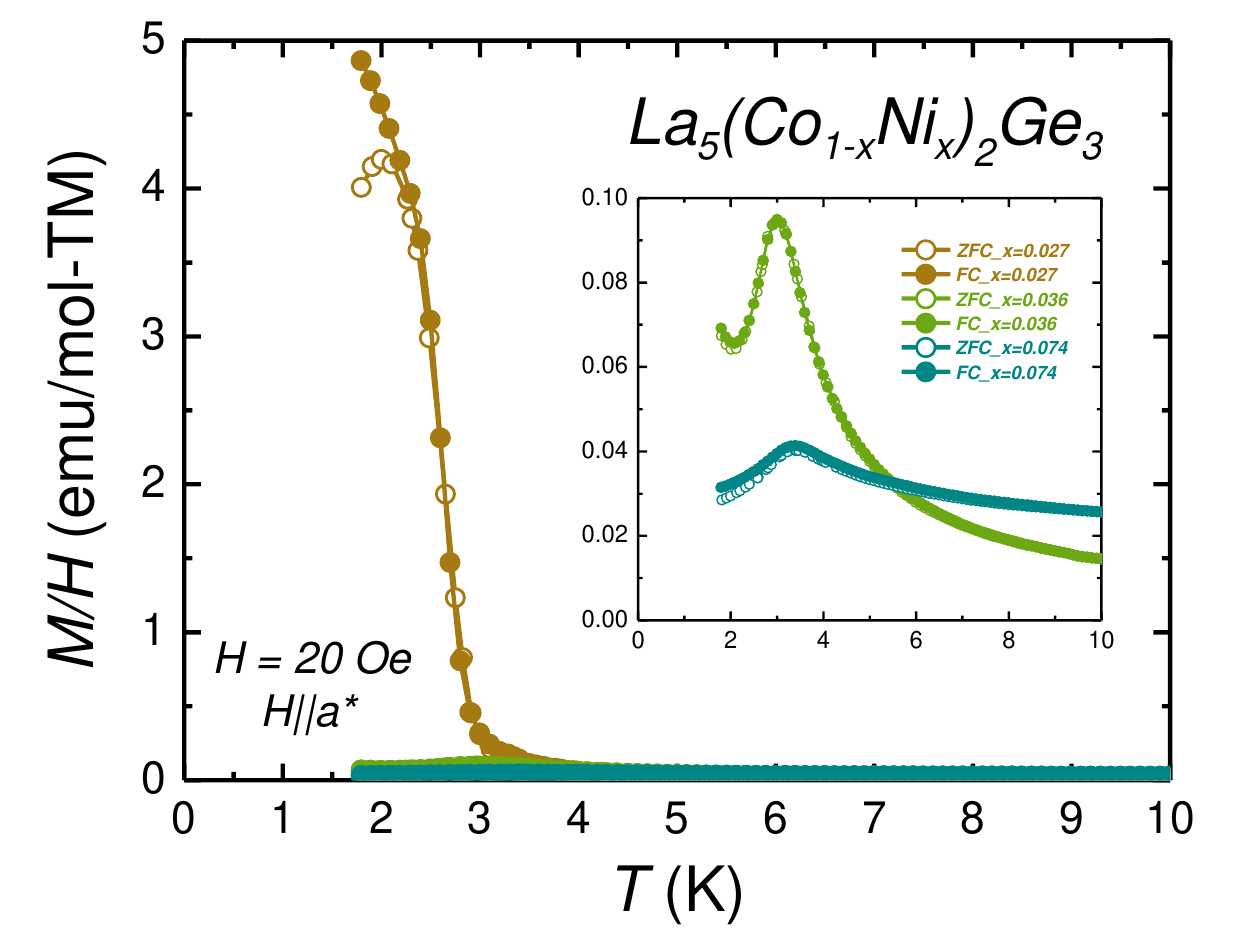}
     \caption{\footnotesize {Temperature dependent magnetization for $x$ = 0.027, 0.036 and 0.074 in $\text{La}_{5}\text{(Co}_{1-x}\text {Ni}_{x})_2\text {Ge}_{3}$ taken in a low field of H = 20 Oe for H$||$a* collected in ZFC-FC mode. (Inset: The same data for the AFM Ni substituted ($x$ = 0.036 and 0.074) samples.}}
     \label{fig:zfc-fc}
 \end{figure}

The H = 1 kOe, ZFC $M(T)$ data shown in Fig \ref{fig:mag}(b) reveal a change in behavior from FM to AFM as $x$ increases. For $x \geq$ 0.036 there is, upon cooling, an increase in $M(T)$ followed by a local maximum and then a decreasing $M(T)$ that is consistent with a antiferromagnetic transition. Low field $M(T)$ was measured at H = 20 Oe for $x$ = 0.027, 0.036 and 0.074 (Fig \ref{fig:zfc-fc}) for H$||a$*. Even though there is a slight irreversibility for $x$ = 0.036 and 0.074, the ZFC-FC splitting is much clearer below the ordering temperature for $x$ = 0.027 again suggesting a transformation of the magnetic ground state. Thus, all of the above evidences in the $M(H,T)$ data suggest that there is an evolution from a low temperature ferromagnetic state for $x$ = 0.00 to an antiferromagnetic state for $x \gtrsim$ 0.04.

\par

Although the high temperature paramagnetic regime was fit to a Curie-Weiss like behavior in the temperature range of $20 ~\text{K} \leq T \leq 100~\text{K}$ for the parent compound \cite{Saunders2020ExceedinglyLa5Co2Ge3}, the doped compounds do not follow a Curie-Weiss like behavior with a temperature independent contribution, $\chi~=~\frac{C}{T-\Theta} + \chi_{0} $ for the above temperature range at a field of 1 kOe, and hence, we did not fit our data and study the change of $\mu_{eff}$ and $\Theta$ with $x$.

\begin{figure*}[t]
    \centering
    \includegraphics[width=0.9\textwidth]{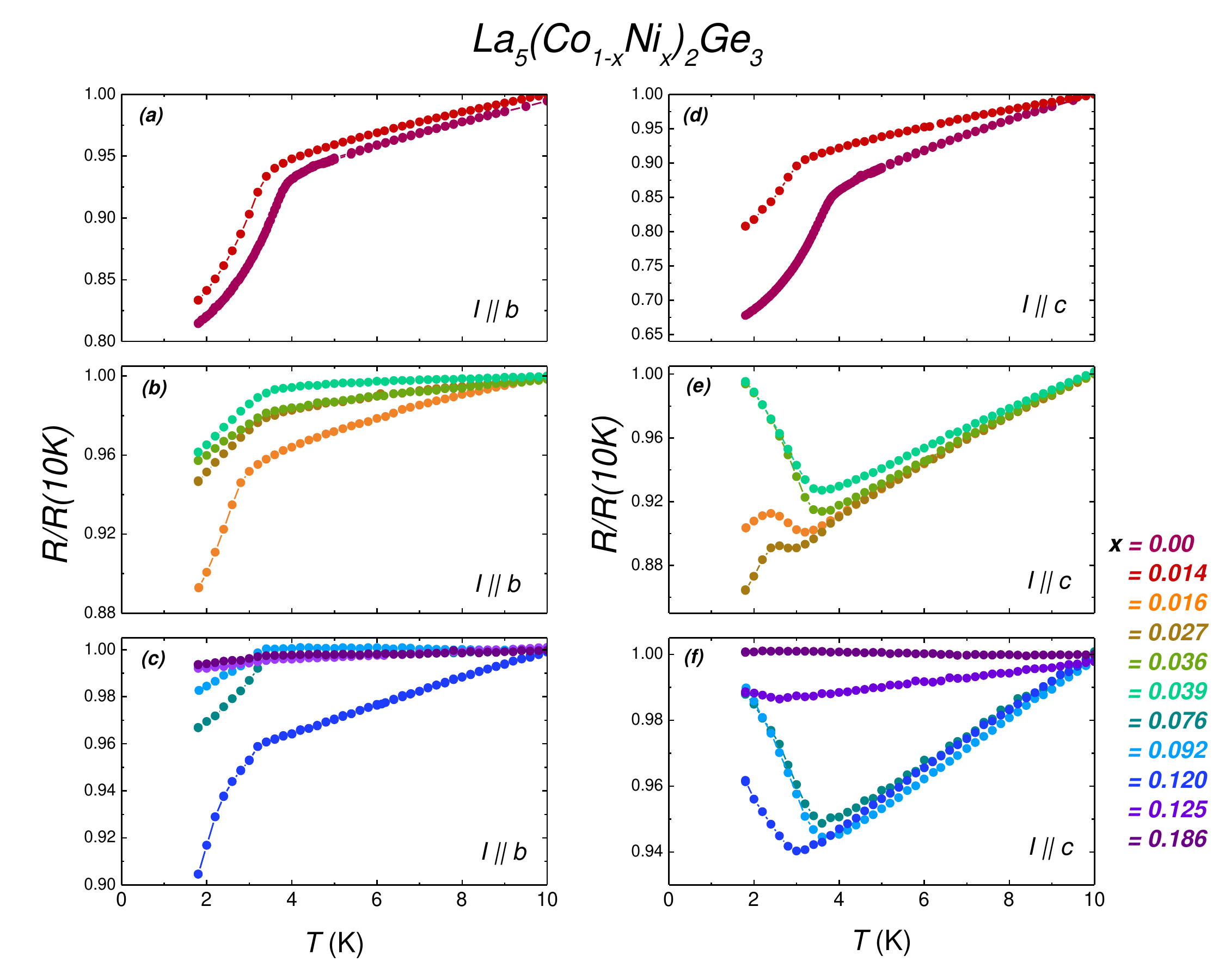}
    \caption{\footnotesize {(Color online) The anisotropic, low temperature resistance data normalized at 10 K for  $\text{La}_{5}\text{(Co}_{1-x}\text {Ni}_{x})_2\text {Ge}_{3}$ single crystals. The left hand column, (a-c) presents I$||b$ data and the right hand column (d-f) presents I$||c$ data. For each current direction the data are divided into three groupings based on the Ni concentrations. Panels (a) and (d) are for $x$ = 0.0 and 0.014. (b) and (e) for $x$ = 0.016, 0.027, 0.036 and 0.039. (c) and (f) for the remaining x upto 0.186 of $\text{La}_{5}\text{(Co}_{1-x}\text {Ni}_{x})_2\text {Ge}_{3}$.}}.
    \label{fig:RT}
\end{figure*}

\par

\subsection*{Resistance}
\label{subsec:Resistance}

The results of the anisotropic, zero-field, in-plane resistance measurements normalized to T = 10 K, plotted for the temperature range $1.8~\text{K} \leq T \leq 10~\text{K}$ are shown in Fig \ref{fig:RT} (Normalized resistance data (${R(T)}/{R(300~\text{K})}$) is shown over the full $1.8~\text{K} - 300~\text{K}$ temperature range in the Appendix in Figs \ref{fig:All_0Ni} - \ref{fig:All_15Ni}). For all the measured samples, $\text{La}_{5}\text{(Co}_{1-x}\text {Ni}_{x})_2\text {Ge}_{3}$ shows metallic behavior for current applied both along the -$b$ and the -$c$ axis for temperatures above the magnetic ordering. The residual resistance ratio $(RRR = \rho(300~\text{K})/ \rho(1.8~\text{K}))$ ranges between 2 - 4, the value being $\sim$ 4 for $x$ = 0.00   \cite{Saunders2020ExceedinglyLa5Co2Ge3} and decreases as $x$ increases to 0.186. In all the 6 panels in Fig \ref{fig:RT} we see a change in the behavior in resistance below $\sim$ 4 K as seen at ambient and high pressure in the parent $\text{La}_{5}\text{Co}_{2}\text{Ge}_{3}$ compound \cite{Saunders2020ExceedinglyLa5Co2Ge3,Xiang2021AvoidedLa5Co2Ge3}. For $x >$ 0.014 a very clear, qualitative anisotropy develops in the magnetically ordered state, becoming even clearer for $x \geq$ 0.036. \par

In Figs \ref{fig:RT} (a) and (d), for $x$ = 0.00 and 0.014, we see a sharp drop in the resistance that is associated with a loss of spin-disorder scattering below the FM transition at around 3.8 K for current applied both along $b$ and $c$ axis. The situation drastically changes as we increase the value of 0.036 $\leq x \leq$ 0.186. When the current is along the $b$-axis (Fig \ref{fig:RT}(b) and (c)), we still see a drop in resistance indicating loss of spin disorder scattering, but when the current is along the $c$-axis for 0.036 $\leq x \leq$ 0.186, (Fig \ref{fig:RT} (e) and (f)), there is an increase in resistance below the transition temperature. This anisotropic behavior of loss of spin disorder scattering along the $b -$ axis and increase of resistance along the $c -$ axis persists as we increase the amount of $x$ up to 0.125 and finally for the highest doping we studied, $x$ = 0.186, we can no longer resolve a resistive feature, associated with a transition particularly along the $c$-axis (See Fig \ref{fig:RT}(f) and Fig \ref{fig:All_15Ni}). For the intermediate Ni dopings, the situation however, is more interesting. For, $x$ = 0.016 and 0.027, when the current is applied along the $c$-axis, there is first an increase in resistance, followed by a decrease upon further cooling (See Fig \ref{fig:RT}(e)).\par 

The persistent loss of spin-disorder scattering for I$||b$ coupled with the evolution of an increasing resistance just below the transition temperature for I$||c$ strongly suggest that the magnetic transition changes from ferromagnetic (FM) to antiferromagnetic (AFM). The increase of resistance at a particular temperature is suggestive of the formation of super zone-gap due to the nesting of the Fermi surface.\cite{Monceau2012ElectronicOverview, Budko2000RotationalState, Elliott1963TheoryMetals}. This in turn reveals that as we increase the doping from 0.036 and beyond, the system enters a low-temperature (T $\sim$ 1.8 K) state which is different from the FM state and this new state is most likely characterized by an AFM component. $\text{La}_{5}\text{(Co}_{1-x}\text {Ni}_{x})_2\text {Ge}_{3}$ with $x$ between 0.016 and 0.027 show both features, suggesting competing interactions characteristic to both FM and AFM behavior. The transformation from a FM ground state to an AFM ground state and a distinct directional anisotropy in the temperature dependent resistance measurements is similar to that of $\text{La}_{5}\text{Co}_{2}\text{Ge}_{3}$ under pressure for both $b$ and the $c$ directions.\cite{Xiang2021AvoidedLa5Co2Ge3} \par

\subsection*{Heat Capacity}
\label{subsec:Heat Capacity}

To further investigate the low temperature phase transition(s) in $\text{La}_{5}\text{(Co}_{1-x}\text {Ni}_{x})_2\text {Ge}_{3}$ with change of $x$, zero-field, temperature dependent heat capacity was measured on representative $x$ valued samples. We specifically chose $x$ = 0.00, 0.014, 0.016, 0.027, 0.036, 0.039 and 0.074 as these lie on the either side of the transition boundary between FM and AFM ground state. \par

\begin{figure}[h]
    \centering
    \includegraphics[width=\linewidth]{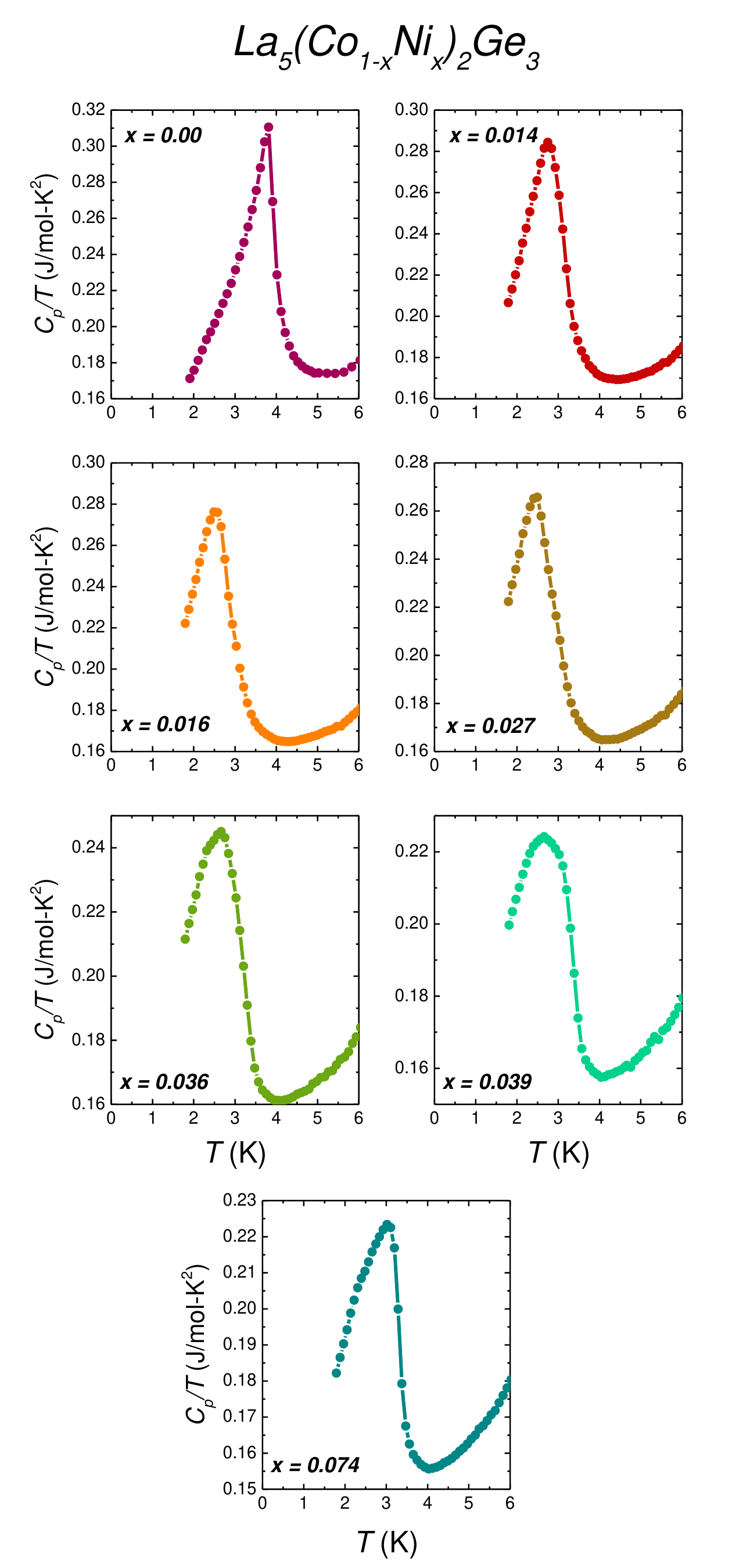}
    \caption{\footnotesize{(Color online) Temperature dependent specific heat data for $x$ = 0.00, 0.014, 0.016, 0.027, 0.036, 0.039 and 0.074 in $\text{La}_{5}\text{(Co}_{1-x}\text {Ni}_{x})_2\text {Ge}_{3}$ shown in different panels for clarity.}}
    \label{fig:Cp_all}
\end{figure}

 The results of the specific heat measurements are shown in Fig \ref{fig:Cp_all}, where we see a $\lambda - $ like feature for all the samples measured, which changes subtly but clearly as we change $x$. The anomaly is consistent with the second order nature of the transition. For $x$ = 0.00, the anomaly associated with the ferromagnetic transition is sharp and seen at $\sim$ 3.9 K. \cite{Saunders2020ExceedinglyLa5Co2Ge3} For higher $x$ the anomaly is less steep and broader. 
 
\par 

The evolution of the transition temperatures can be seen more clearly in the $\frac{d(C_p(T)}{dT}$ data plotted in Fig \ref{fig:heat capacity}. We can see a single transition for $x$ = 0.00, 0.014 and 0.016 which splits into two transitions for $x$ = 0.027 and again becomes a single transition for the high dopings ($x$ = 0.074). \par

\par

\subsection*{Phase Diagram}
\label{subsec:Phase Diagram}

\begin{figure}[ht]
    \centering
    \includegraphics[width=\linewidth]{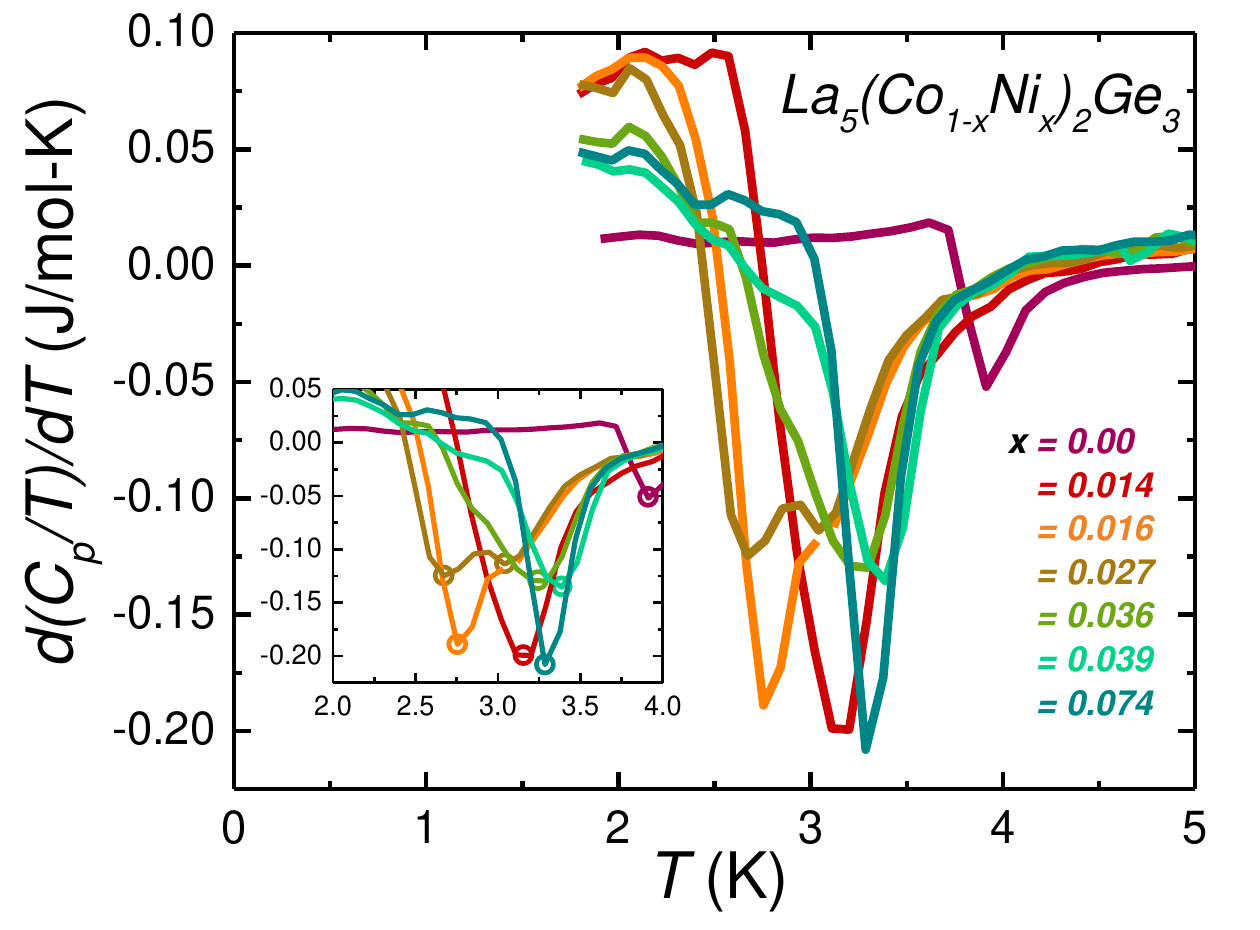}
    \caption{\footnotesize{(Color Online) $\frac{d(C_{p}/T)}{dT}$ versus T for x = 0.014, 0.016, 0.027, 0.036, 0.039 and 0.074. The minima of the curves are taken as a measure of the transition temperature ($T_{mag}$). (Inset: The data has been zoomed in clarity where the $T_{mag}$ has been marked with circles. The two minima for x = 0.027 are evident which gives rise to the two transition temperature for this Ni-substitution.}}
    \label{fig:heat capacity}
\end{figure}

\begin{figure*}
    \centering
    \includegraphics[width=\textwidth]{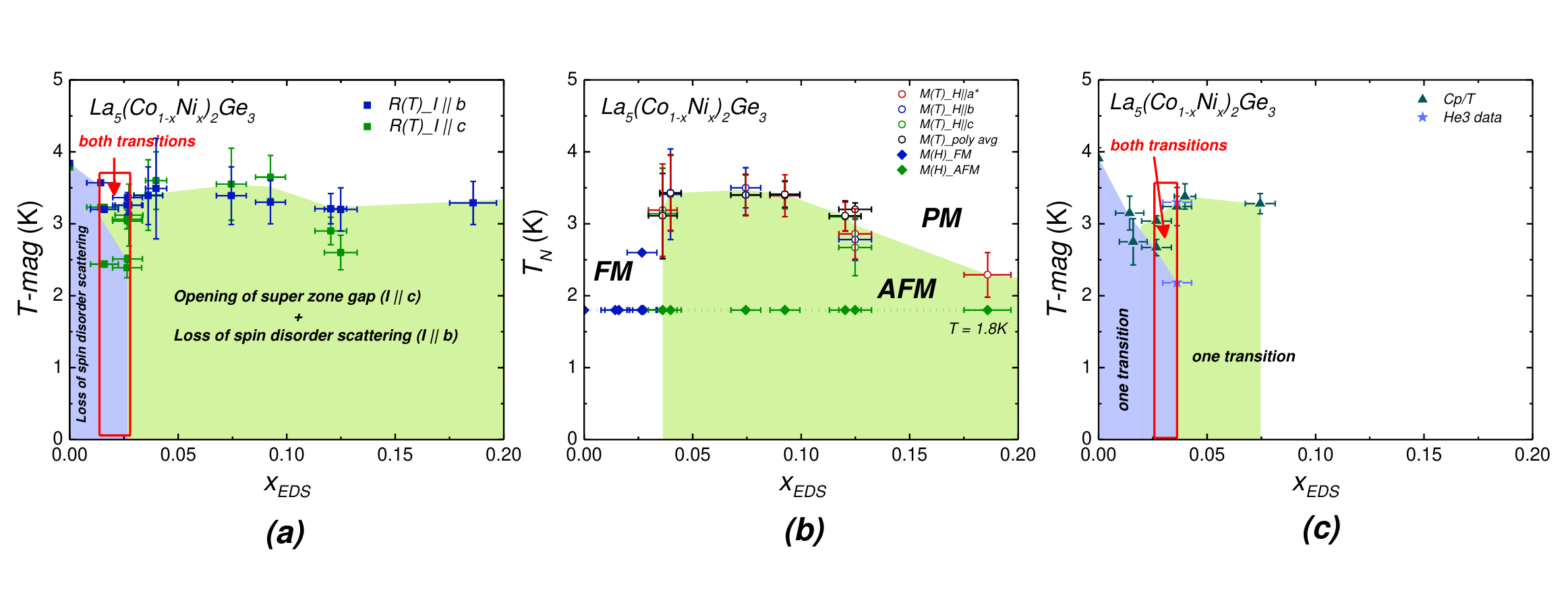}
    \caption{\footnotesize {The T-x phase diagram of $\text{La}_{5}\text{(Co}_{1-x}\text {Ni}_{x})_2\text {Ge}_{3}$ is plotted with the transition temperature obtained from different measurements. (a) $T_{mag}$ is plotted based on the results of temperature dependent resistance data, (b) $T_N$ is plotted based on the results of temperature dependent magnetization data. The magnetic ground state based from $M(H)$ at 1.8 K and the FM transition temperature obtained from the Arrott plots for $x$ = 0.027 are also included. (c) $T_{mag}$ is plotted based on the results of heat capacity measurements. The details on determination of the transition temperatures and the  different regions marked in the phase diagram are discussed in detail in the text.}}
    \label{fig:phase diagram}
\end{figure*}

From the resistance, magnetization and heat capacity measurements, the temperature-composition (T-x) phase diagram of $\text{La}_{5}\text{(Co}_{1-x}\text {Ni}_{x})_2\text {Ge}_{3}$ can be constructed. The phase diagrams obtained from each of these measurements will be presented separately first, for clarity, and then combined together to summarize the effect of Ni substitution on this system.
\par

 In the T-x phase diagram obtained from the $R(T)$ measurements (Fig \ref{fig:phase diagram}(a)), for the low Ni dopings, (${0.00 \leq x \leq 0.014}$) which have a loss of spin disorder scattering in the data for measurements both along I$||b$ and I$||c$ directions (Fig \ref{fig:RT})(a) and (d)) and manifest a FM ground state, (Fig \ref{fig:mag}) the transition temperature, $T_{mag}$, is determined from the intersection of the two dashed lines as shown in Fig \ref{fig:R(T)_low x_criteria} in the Appendix \cite{Saunders2020ExceedinglyLa5Co2Ge3, Xiang2021AvoidedLa5Co2Ge3}. For the intermediate dopings, (0.016 $\leq$ x $\leq$ 0.027), where features characteristic to both loss of spin disorder as well as opening of a superzone gap are observed for measurement along I$||c$  (Fig \ref{fig:RT} (b) and (e)), the $T_{mag}$ is determined following the same protocol as above for both directions. The transition temperature ($\text{T}_{mag}$) associated with the antiferromagnetic Ni substitutions which manifest an opening of a superzone gap are determined from the $\frac{d\rho}{dT}$ data \cite{Fisher1968ResistivePoints} and is shown in Fig \ref{fig:criteria_3Ni new} - Fig\ref{fig:criteria_15Ni}.  Based on all this, in the phase diagram (Fig \ref{fig:phase diagram}(a)) the low Ni dopings having a loss of spin disorder along both $b$ and $c$ direction are shown in blue and the high dopings which show opening of a superzone gap along I$||$c and loss of spin disorder along I$||$b, in green. The intermediate substitutions which show both features along I$||$c are highlighted using a red box. \par

\begin{figure}[h]
    \centering
    \includegraphics[width=\linewidth]{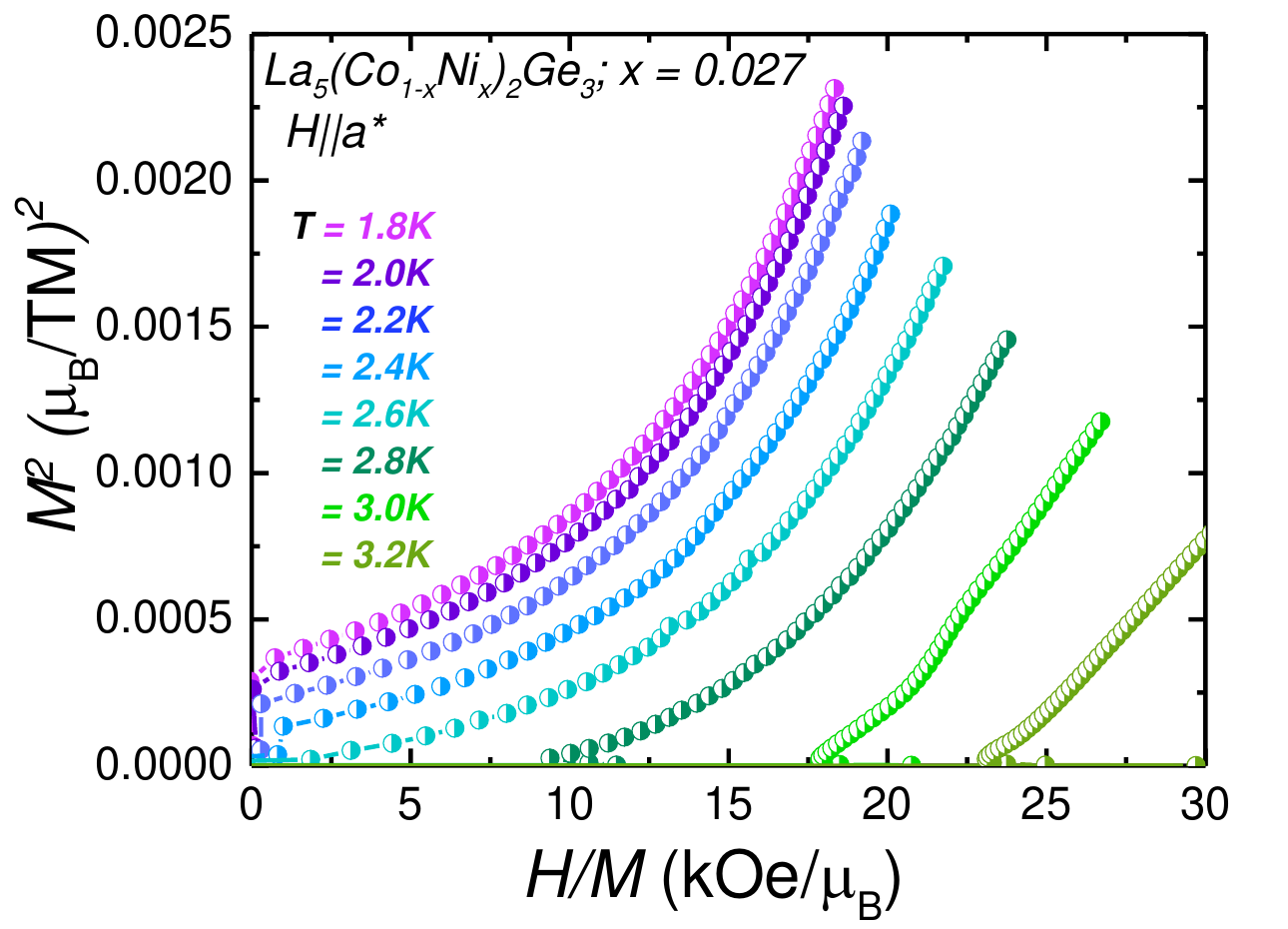}
    \caption{\footnotesize{(Color Online) $M^2$ as a function of H/M (Arrott Plot) obtained by measuring field-dependent magnetization at different, constant temperatures for x = 0.027 in $\text{La}_{5}\text{(Co}_{1-x}\text {Ni}_{x})_2\text {Ge}_{3}$. The data used to make the Arrott Plot is shown in Fig \ref{fig:MH_2Ni} in the Appendix.}}
    \label{fig:arrott}
\end{figure}

In order to understand the nature of the two transitions for the intermediate dopings (0.016 $\leq x \leq$ 0.027) in $\text{La}_{5}\text{(Co}_{1-x}\text {Ni}_{x})_2\text {Ge}_{3}$, field dependent magnetization was measured along H$||$a* at different constant temperatures for $x$ = 0.027. The results of these measurements are shown in Fig \ref{fig:MH_2Ni} and explained in Appendix. Based on the results, Arrott plots, shown in Fig \ref{fig:arrott}, were constructed to determine the ferromagnetic transition temperature ($T_{C}$). ${M}^{2}$ vs $H/M$ plotted for $T<T_{C}$ have a positive ${M}^{2}$ intercept, and have a negative intercept ${M}^{2}$ for $T>T_{C}$, and the curve at $T \sim T_{C}$ passes through the origin \cite{Arrott1957CriterionIsotherms}. From our data (Fig \ref{fig:arrott}), the FM $T_{C}$ is $\sim$ 2.6 K. This temperature is consistent with the lower transition temperature obtained from the $R(T)$ data. From all the above discussed reasons, we can say that the lower temperature transition of the intermediate dopings (0.016 $\leq x \leq$ 0.027) in $\text{La}_{5}\text{(Co}_{1-x}\text {Ni}_{x})_2\text {Ge}_{3}$ is ferromagnetic in nature. \par

The T-x phase diagram obtained from the magnetization measurements is shown in Fig \ref{fig:phase diagram}(b), there is a clear demarcation of the AFM region, shown in green for $x$ = 0.036 and higher Ni substitutions. The field dependent magnetization at T = 1.8 K identifies the low temperature state of this system, which is also shown. The blue points denotes a FM ground state or the state with FM component, and the green data points are for the AFM ground state. The transition temperature, $T_{N}$, for the AFM Ni-substituted $\text{La}_{5}\text{(Co}_{1-x}\text {Ni}_{x})_2\text {Ge}_{3}$ is obtained from the $M(T)$ data by using $\frac{d(\chi T)}{dT}$ as a proxy \cite{Fisher1962RelationAntiferromagnet}, following the same procedure as used to determine $\text{T}_{mag}$ from the $R(T)$ data for the AFM samples by taking the point of maximum slope. Figures \ref{fig:criteria_3Ni new}(b) - \ref{fig:criteria_15Ni}(b) in Appendix, show the $\frac{d(\chi T)}{dT}$ data and the criteria for $T_{N}$ for each direction of the applied field as well as the polycrystalline average for the other AFM Ni substitutions. The FM transition temperature for $x$ = 0.027, obtained from the Arrott plots is also shown. For the lower x values, with a FM ground state, (${0.00 \leq x \leq 0.027}$), the analysis of the $M(T)$ data does not provide quantitative insight into the value of the transition temperature and hence there are no data points in the phase diagram for the FM Ni substituted $\text{La}_{5}\text{(Co}_{1-x}\text {Ni}_{x})_2\text {Ge}_{3}$. \par

 The phase diagram, Fig \ref{fig:phase diagram}(c), is obtained from the heat capacity measurements. The transition temperature $T_{mag}$ is obtained from the $\frac{d(C_{p}/ T)}{dT}$ data. The point of maximum slope for $C_p(T)$ and the minima of $\frac{d(C_{p}/ T)}{dT}$ are similar, within the error bar, and hence for simplicity the minima point is taken as $T_{mag}$. Figure \ref{fig:heat capacity} shows the $\frac{d(C_{p}/ T)}{dT}$ data where the $T_{mag}$ is marked with circles with each $x$. The two minima for $x$ = 0.027, observed at T = 2.7 K and 3.0 K are similar to the two transitions observed in the resistivity data for the same doping when measured along I$||$c direction, indicating the presence of two transitions. Heat capacity for the x = 0.036 sample was also measured down to T = 0.5 K using a He-3 inset, details are discussed in the Appendix. The two transitions are observed at 3.3 K and $\sim$ 2 K (Fig \ref{fig:He3-cp}) and has been marked in the phase diagram as well. \par

From all the above discussed data, it is clear that the transition temperatures obtained from magnetization, resistance and heat capacity are similar and consequently the phase diagrams (Fig \ref{fig:phase diagram}) are qualitatively and quantitatively similar.

\subsection*{Muon Spin Rotation/Relaxation ($\mu$SR)}
\label{subsec:Musr}

\subsubsection*{Zero-field (ZF)}

The data and the phase diagrams discussed above show that the magnetic ordering in $\text{La}_{5}\text{(Co}_{1-x}\text {Ni}_{x})_2\text {Ge}_{3}$ evolves from a FM state to an AFM state with increasing x, with an intermediate range of x where both transitions are observed. The magnetic phases suggested by our phase diagrams can be further explored by conducting both weak transverse field (wTF) and zero field (ZF) $\mu$SR experiments for $x$ = 0.027, 0.036 and 0.074 single crystals by studying the time evolution of the muon asymmetry \cite{Reotier1997MuonMaterials}. The results of the ZF measurements will be discussed here whereas the results of the wTF data will be shown in the Appendix. For completeness, in this work, we will compare the results of our ZF measurements with the $\mu$SR data on the parent $\text{La}_{5}\text{Co}_2\text {Ge}_{3}$ samples \cite{Saunders2020ExceedinglyLa5Co2Ge3}. In both the measurements, muons were implanted into the sample along the $a$* direction (perpendicular to the plane of the plates). 

\par

\begin{figure}[h]
    \centering
    \includegraphics[width=\linewidth]{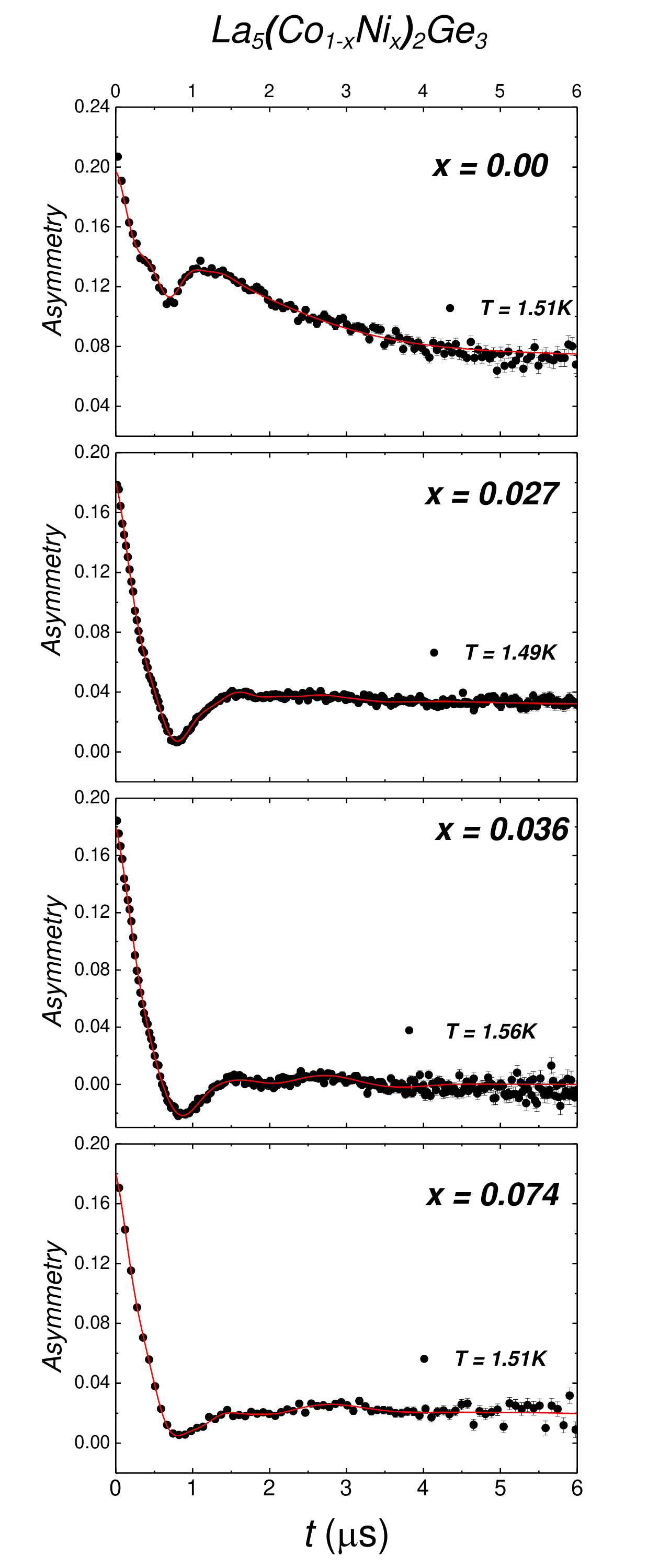}
    \caption{\footnotesize{ZF-$\mu$SR time dependent asymmetry for $x$ = 0.00, 0.027, 0.036 and 0.074 in $\text{La}_{5}\text{(Co}_{1-x}\text {Ni}_{x})_2\text {Ge}_{3}$ below the transition temperature. The $x$ = 0.00 data has been taken from Ref.\cite{Saunders2020ExceedinglyLa5Co2Ge3}. The solid lines are fit to Eq.\ref{eq:zf} (see text for details).}}
    \label{fig:zf all}
\end{figure}

\begin{figure}[H]
    \centering
    \includegraphics[width=\linewidth]{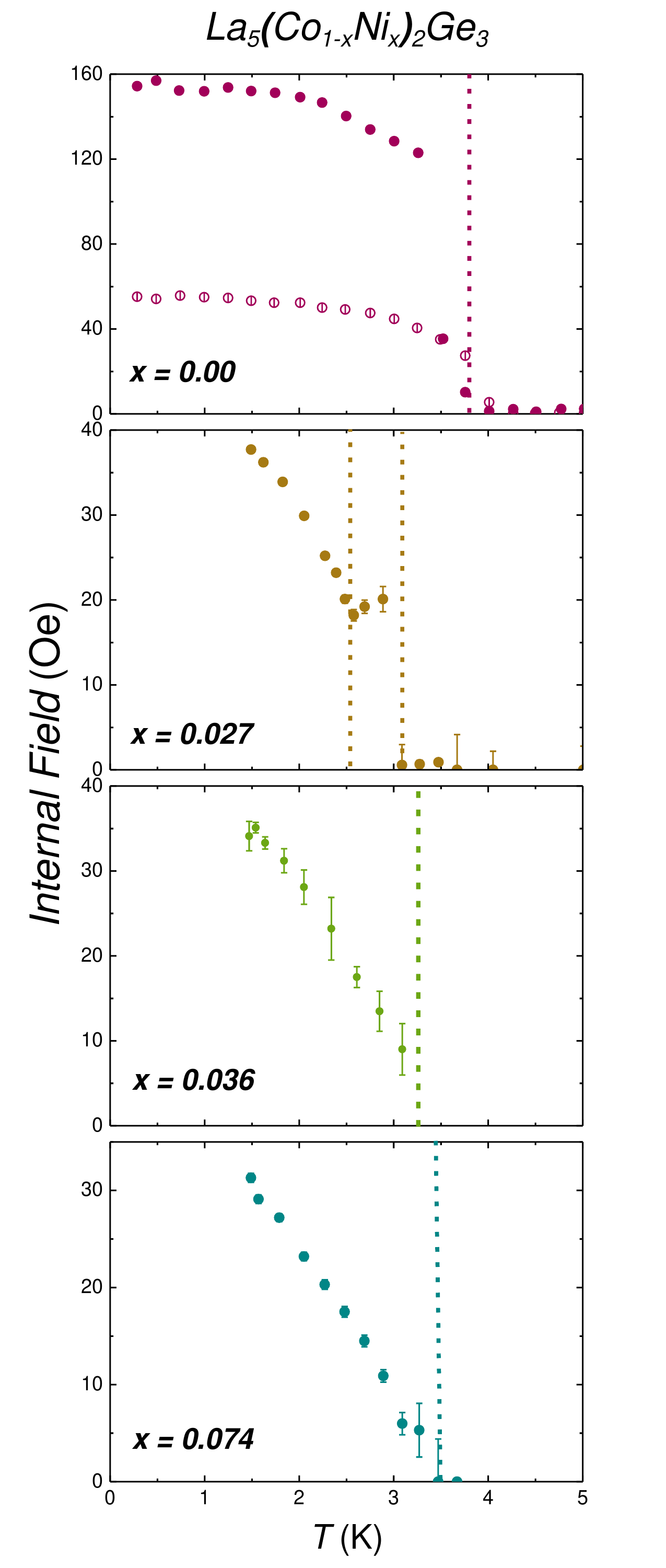}
    \caption{\footnotesize{The internal field ($B_{int}$) obtained by fitting Eq. \ref{eq:zf} to the ZF-$\mu$SR time spectra for $x$ = 0.00, 0.027, 0.036 and 0,074 in $\text{La}_{5}\text{(Co}_{1-x}\text {Ni}_{x})_2\text {Ge}_{3}$. The vertical dashed lines in panel for $x$ = 0.027, 0.036 and 0.074 are the average of the transition temperatures obtained from the transport, magnetization and heat capacity measurements discussed earlier. The data for $x$ = 0.00 is from Ref.\cite{Saunders2020ExceedinglyLa5Co2Ge3}. The two curves for $x$ = 0.00 is due to the assumption of two muon stopping sites ($\sim 20\%$ at the higher internal field site and $\sim 80\%$ at a lower field site).}}
    \label{fig:int field}
\end{figure}

In ZF-$\mu$SR experiments, the muon spins precess about the internal field at the muon stopping site and oscillations are observed only when the initial polarization is perpendicular to the field and is absent for a parallel internal field \cite{Reotier1997MuonMaterials}. As muons were implanted along the $a$* direction of the crystals, we will label the component of the polarization perpendicular [parallel] to $a$* as P$\perp$a* [P$||$a*] respectively. Figure \ref{fig:wTF zf 2Ni} in the Appendix shows the time evolution of muon asymmetry for $x$ = 0.027 both perpendicular and parallel to $a$*. As oscillations are observed in both P$\perp$a* and P$||$a* sets of data, it can be concluded that the internal field along the muon stopping sites has components both parallel and perpendicular to the crystallographic $a$ direction of the $\text{La}_{5}\text{(Co}_{1-x}\text {Ni}_{x})_2\text {Ge}_{3}$ crystals. \par

Figure \ref{fig:zf all} show the ZF-$\mu$SR time spectra for $x$ = 0.00, 0.027, 0.036 and 0.074 in $\text{La}_{5}\text{(Co}_{1-x}\text {Ni}_{x})_2\text {Ge}_{3}$ in the ordered state (T $\sim$ 1.5 K). The data for $x$ = 0.00 has been published in our previous work on this system \cite{Saunders2020ExceedinglyLa5Co2Ge3}. The muon asymmetry for the parent compound is visually different from the Ni substituted $\text{La}_{5}\text{(Co}_{1-x}\text {Ni}_{x})_2\text {Ge}_{3}$ samples with the oscillations becoming much broader for all the substituted ones. 

\par

The ZF asymmetry can be analyzed by taking contributions from oscillating as well as non-oscillating components and the data can be fit using,

\begin{equation}
    \begin{split}
       A_{ZF}(t) & = A(0)[(1-f) e^{-\lambda_{L}t} \\
                 &+f\sum\limits_{i=0}^{n} e^{-\lambda_{T, i}} \cos(\gamma_{\mu} B_{int,i}t)]
    \label{eq:zf} 
    \end{split}
\end{equation}

where, $A(0)$ is the initial asymmetry, $f$ is the oscillating fraction, $\lambda_{L}$ is the longitudinal relaxation rate and $\gamma_{\mu}/2 \pi$ = 135.5 MHz/T is the gyromagnetic ratio of the $\mu$SR signal. $B_{int,i}$ and $\lambda_{T,i}$ are respectively the internal field and the transverse relaxation rate of the $i^{th}$ component at the muon stopping site in the sample. The ZF data set for the parent ($x$ = 0.00) compound was analyzed by assuming two inequivalent muon stopping sites ($\sim 20\%$ at the higher internal field site and $\sim 80\%$ at a lower field site) \cite{Saunders2020ExceedinglyLa5Co2Ge3}. However, for the Ni substituted $\text{La}_{5}\text{(Co}_{1-x}\text {Ni}_{x})_2\text {Ge}_{3}$ samples the data are fit to Eq.\ref{eq:zf} by assuming one stopping site. When fit using two sites, the obtained parameters are very close to each other and hence for simplicity we use only a single site. 

\par

The value of the internal field at the muon stopping position is obtained by fitting Eq. \ref{eq:zf} to the ZF spectra. The internal field is determined by the surrounding magnetic moments and is proportional to the value of ordered moments for the different Ni substituted  $\text{La}_{5}\text{(Co}_{1-x}\text {Ni}_{x})_2\text {Ge}_{3}$ samples studied. Consequently, the temperature dependence of $B_{int}$ reflects the evolution of the magnetic order parameter for each $x$ as shown in Fig \ref{fig:int field}. \par

For $x$ = 0.027, the internal field decreases with increasing temperature up to T = 2.54 K after which we see a change in the behavior that continues until T = 3.1 K, above which the internal field abruptly drops to zero, indicating the onset of the paramagnetic phase. The temperatures at which we see abrupt changes in the internal field are consistent with the two transition temperatures (2.66 K and 3.05 K) observed in the $R(T)$ and ${C}_{p}(T)$ measurements. Thus, the $\mu_{SR}$ data clearly suggests that there is 1) a low T state with the highest internal field , 2) a high T PM state with no internal field and 3) an intermediate temperature magnetic state with a finite internal field having a different temperature dependence. Combining this with the previous data magnetization and transport data, it can be interpreted that the lowest T state below ($\sim$ 2.66 K) is FM with emergence of spontaneous moment in $M(H)$ isotherms (Fig \ref{fig:MH_2Ni} in the Appendix) whereas, the intermediate T region in the $\mu_{SR}$ data (2.66 K $\leq x \leq$ 3.05 K) is AFM. For the other two Ni substituted ($x$ = 0.036 and 0.074) measured samples, we see a monotonic decrease of the internal field with increasing temperature up to the transition. The vertical dashed lines in each of the panel in Fig \ref{fig:int field} are magnetic transition temperatures obtained from the previously discussed transport, magnetization and heat capacity measurements. The T dependence of the internal field for $x$ = 0.00 is also included for comparison which has been published in our previous work \cite{Saunders2020ExceedinglyLa5Co2Ge3} where the two internal fields, due to two muon stopping sites ($\sim 20\%$ at the higher internal field site and $\sim 80\%$ at a lower field site) decreases monotonically up to 3.8 K at the onset of the paramagnetic phase. \par

\section{Conclusion and Discussion}
\label{sec:Conclusion}

\begin{figure}[h]
    \centering
    \includegraphics[width=\linewidth]{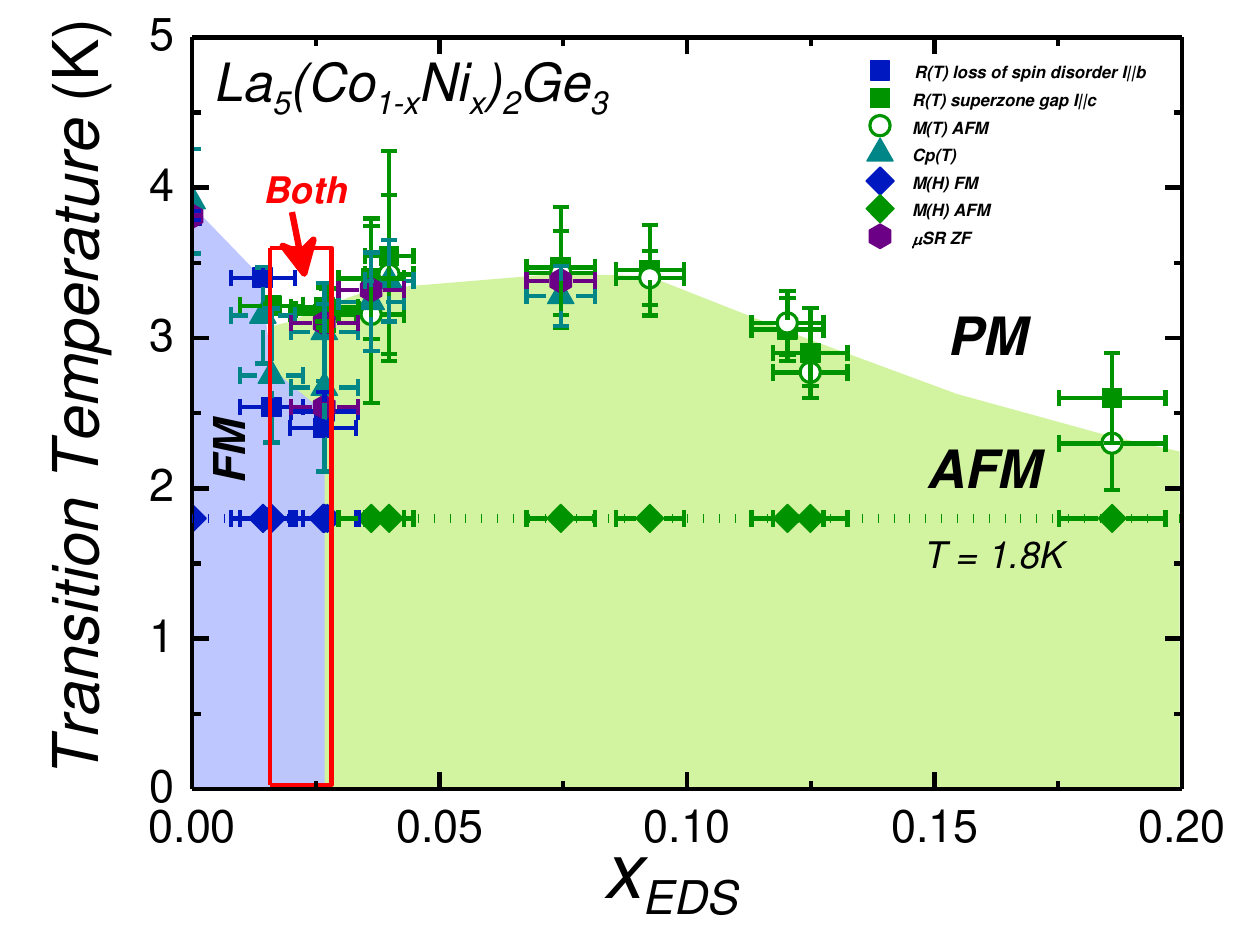}
    \caption{\footnotesize (Color online) The T-x phase diagram of $\text{La}_{5}\text{(Co}_{1-x}\text {Ni}_{x})_2\text {Ge}_{3}$ obtained from anisotropic resistance and magnetization as well as specific heat and $\mu$SR data. The three distinct regions observed are the low $x$ ferromagnetic (FM) marked by blue, the high $x$ antiferromagnetic (AFM) marked by green and a high temperature paramagnetic (PM) region. The intermediate $x$ $(0.016 \leq x  \leq 0.027)$ have two transitions and exhibit properties characteristic to both FM and AFM behavior.}
    \label{fig:T-x}
\end{figure}

Based on all the three separate phase diagrams (Fig \ref{fig:phase diagram}) and the ZF $\mu$SR data analysis, a composite T-x phase diagram is constructed to summarize the behavior of $\text{La}_{5}\text{Co}_{2}\text{Ge}_{3}$ as Co is substituted with Ni (Fig \ref{fig:T-x}). The phase diagram has three regions namely, the high temperature paramagnetic (PM) region $(T \gtrsim 4K)$ shown in white, the low $x$ ferromagnetic (FM) region $(x \leq 0.016)$ in blue and the high $x$ $(0.036 \leq x  \leq 0.186)$ antiferromagnetic (AFM) region marked in the green shaded area. Although we do not have enough low temperature data (T $<$ 1.8 K), we anticipate that the FM state may well exist below 1.8 K for $x \gtrsim$ 0.05, but we do not know its precise fate. For the intermediate $x$ $(0.016 \leq x  \leq 0.027)$ we observe features with two transitions and properties characteristic to both FM and AFM behavior, which is highlighted in a red box. The FM transition temperature decreases with $x$, whereas the AFM transition temperature increases very slightly to 3.4K for $x \sim 0.04$ and then decreases to roughly 2.5K for $x \sim 0.19$. \par

$\text{La}_{5}\text{Co}_{2}\text{Ge}_{3}$, an itinerant ferromagnet with a small saturated moment of $0.1~\mu_{B}/\text{Co}$, a low transition temperature of 3.8 K and a Rhodes-Wohlfarth ratio of 4.9 was presumed to be a promising candidate to suppress the magnetic order to low temperatures \cite{Saunders2020ExceedinglyLa5Co2Ge3}. When perturbed with hydrostatic pressure it was observed that the system remains FM up to a pressure of 1.5 GPa and the $\text{T}_{C}$ is weakly suppressed to $\sim$ 3.0 K. After this the magnetic state transforms to a new, non FM-state where the transition temperature depends non monotonically on the applied pressure, \cite{Xiang2021AvoidedLa5Co2Ge3} indicating that $\text{La}_{5}\text{Co}_{2}\text{Ge}_{3}$ is another FM system that avoids the quantum critical point through the emergence of a new, spatially modulated phase. Substitution of Co with Ni introduces a different type of perturbation by introducing additional electrons; however, like what is observed under applied pressure, we also find quantum criticality is again avoided and that Ni substitution transforms the magnetic state from FM to AFM. The T-x phase diagram of $\text{La}_{5}\text{(Co}_{1-x}\text {Ni}_{x})_2\text {Ge}_{3}$ reported here is qualitatively similar to the T-p phase diagram of $\text{La}_{5}\text{Co}_{2}\text {Ge}_{3}$ under hydrostatic pressure \cite{Xiang2021AvoidedLa5Co2Ge3}.

\par.

\par

The zero field (ZF) $\mu$SR measurements support the temperature dependent magnetization, resistance and specific heat measurements. Signature of two transitions are detected for $x$ = 0.027 in the temperature evolution of internal field. The transition temperatures obtained for $x$ = 0.036 and 0.074 agree our other measurement results. The ZF time spectra differs between the parent and the Ni substituted samples. Based on the values of the obtained internal field, we do not infer a large change in the size of the magnetic moment with Ni doping. \par

To summarize, single crystals of $\text{La}_{5}\text{(Co}_{1-x}\text {Ni}_{x})_2\text {Ge}_{3}$ were synthesized with $x$ varying between 0.00 and 0.186. Powder X-Ray diffraction and EDS measurements confirmed the phase and an estimate of the Ni going in the system respectively. Magnetization, resistance and heat capacity measurements were performed at ambient pressure and a T-x phase diagram of the system was constructed. For the low dopings, between ${0.00 \leq x \leq 0.014}$, the system remains ferromagnetic and the transition temperature is suppressed to 3.2 K. A change in the ground state occurs at ${x = 0.036}$ when the magnetization measurements indicate the appearance of an AFM state at around 3.3 K. The single AFM transition is found to first weakly increase to a shallow maximum $\text{T}_N \sim$ 3.4 K near x $\sim$ 0.07 and then slowly decrease to T = 2.4 K for x = 0.186. The intermediate Ni substitutions,  $ 0.016 \leq x \leq 0.027 $, show both transitions with $\text{T}_N > \text{T}_C$. ZF $\mu$SR measurements and wTF measurements (in the Appendix) on $x$ = 0.027, 0.036 and 0.074 give a measure of transition temperature which agrees with the other results. Given that relatively a small substitution level of Ni is needed to drive the magnetic ordering to the AFM state ($x \sim$ 0.03), such lightly doped samples are in themselves promising materials to study under pressure to see if an AFM QCP can be more readily reached. \par

\section*{ACKNOWLEDGEMENTS}
\label{sec:ACKNOWLEDGEMENTS}

We would like to thank E. Gati and D.H. Ryan for useful discussions regarding this system and $\mu$SR analysis. This work was done at Ames National Laboratory and supported by the U.S. Department of Energy, Office of Science, Basic Energy Sciences, Materials Sciences and Engineering Division. Ames National Laboratory is operated for the U.S. Department of Energy by Iowa State University under Contract No. DE-AC02-07CH11358.

\newpage

\appendix
\counterwithin{figure}{section}

\renewcommand{\thefigure}{A\arabic{figure}}

\setcounter{figure}{0}

\section*{APPENDIX}
\label{sec:Appendix}

\begin{figure}[H]
    \centering
    \includegraphics[width=\linewidth]{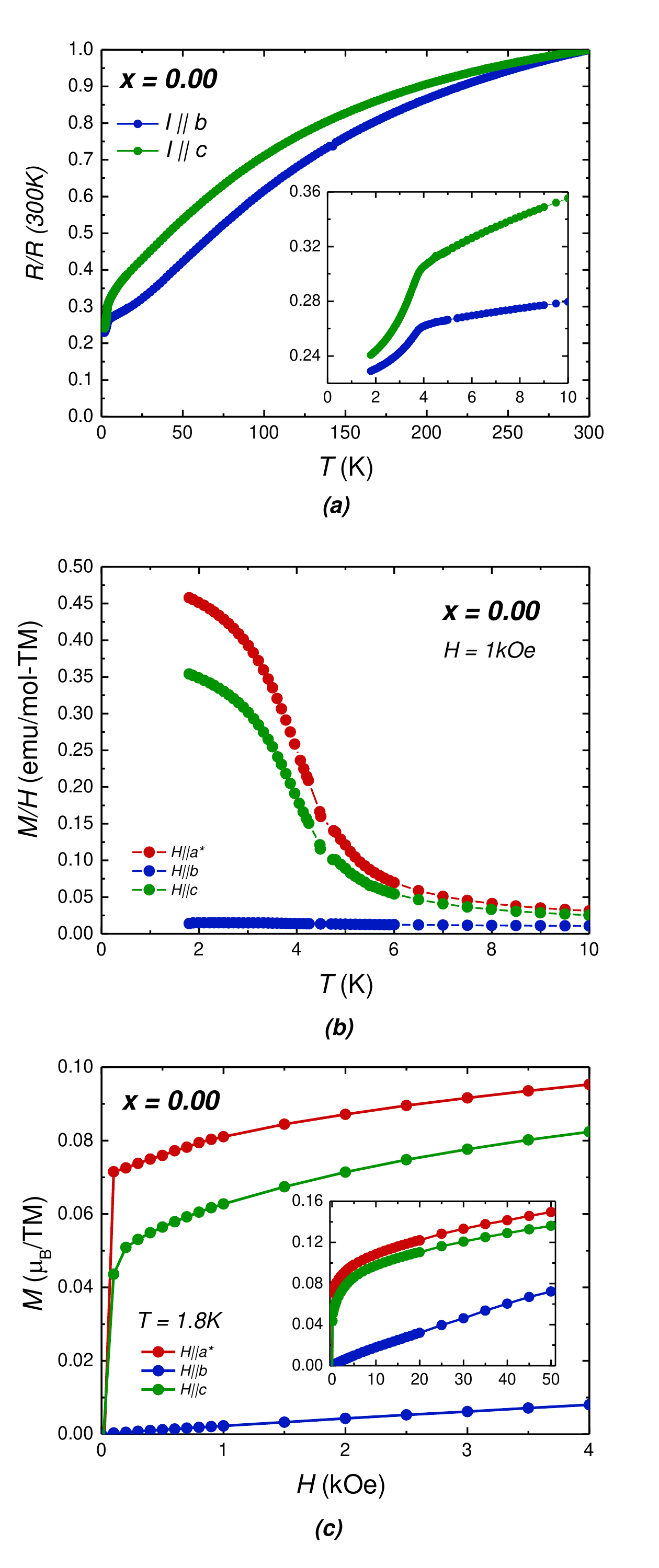}
    \caption{\footnotesize {(Color online) (a): Zero field, temperature dependent normalized resistance (Inset: The low temperature normalized $R(T)$ behavior shown for 1.8 K $\leq$ T $\leq$ 10 K), (b): Temperature dependent magnetization at H = 1 kOe for 1.8 K $\leq$ T $\leq$ 10 K, (c): Field dependent magnetization at a base temperature of 1.8 K shown for H $\leq$ 4 kOe and in the inset for H $\leq$ 4 kOe for $x$ = 0.00 in $\text{La}_{5}\text{(Co}_{1-x}\text {Ni}_{x})_2\text {Ge}_{3}$. }}
    \label{fig:All_0Ni}
\end{figure}

\begin{figure}[H]
    \centering
    \includegraphics[width=\linewidth]{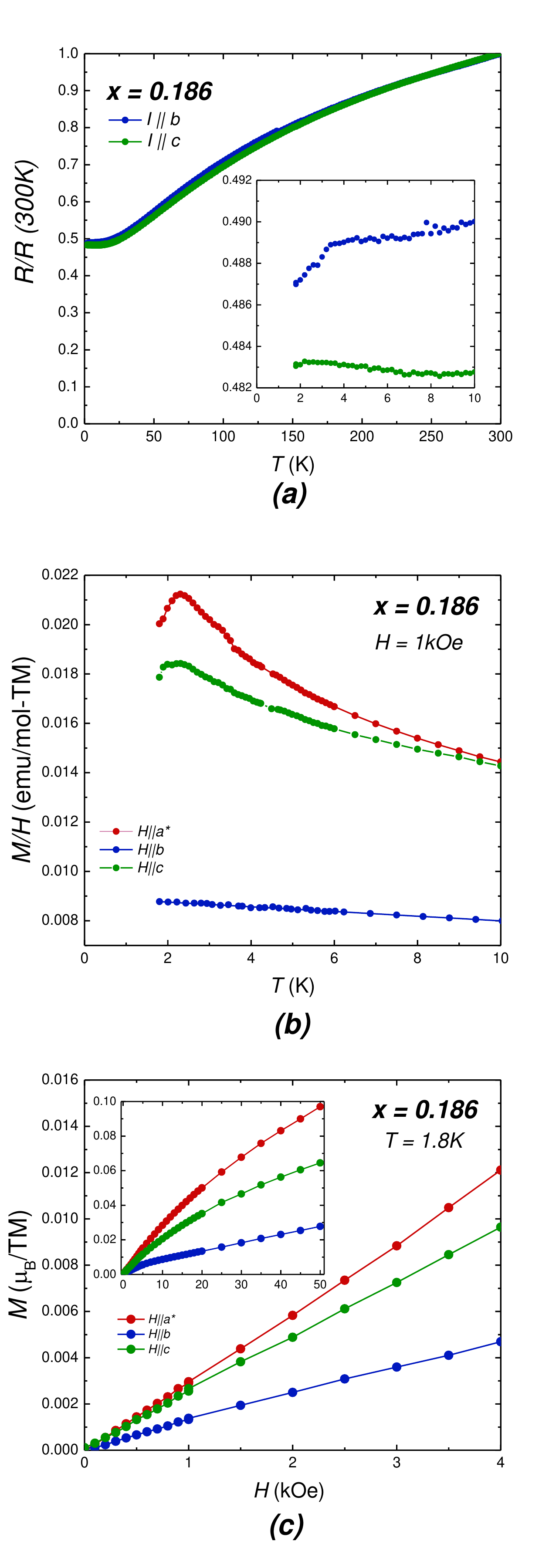}
    \caption{\footnotesize {(Color online) (a): Zero field, temperature dependent normalized resistance (Inset: The low temperature normalized $R(T)$ behavior shown for 1.8 K $\leq$ T $\leq$ 10 K), (b): Temperature dependent magnetization at H = 1 kOe for 1.8 K $\leq$ T $\leq$ 10 K, (c): Field dependent magnetization at a base temperature of 1.8 K shown for H $\leq$ 4 kOe and in the inset for H $\leq$ 4 kOe for $x$ = 0.186 in $\text{La}_{5}\text{(Co}_{1-x}\text {Ni}_{x})_2\text {Ge}_{3}$. }}
    \label{fig:All_15Ni}
\end{figure}

Temperature dependent normalized resistance, magnetization at H = 1 kOe as well as field dependent magnetization at T = 1.8 K data are plotted separately for each x in $\text{La}_{5}\text{(Co}_{1-x}\text {Ni}_{x})_2\text {Ge}_{3}$ as shown in Fig \ref{fig:All_0Ni} - Fig \ref{fig:All_15Ni}.

\begin{figure}[H]
    \centering
    \includegraphics[width=\linewidth]{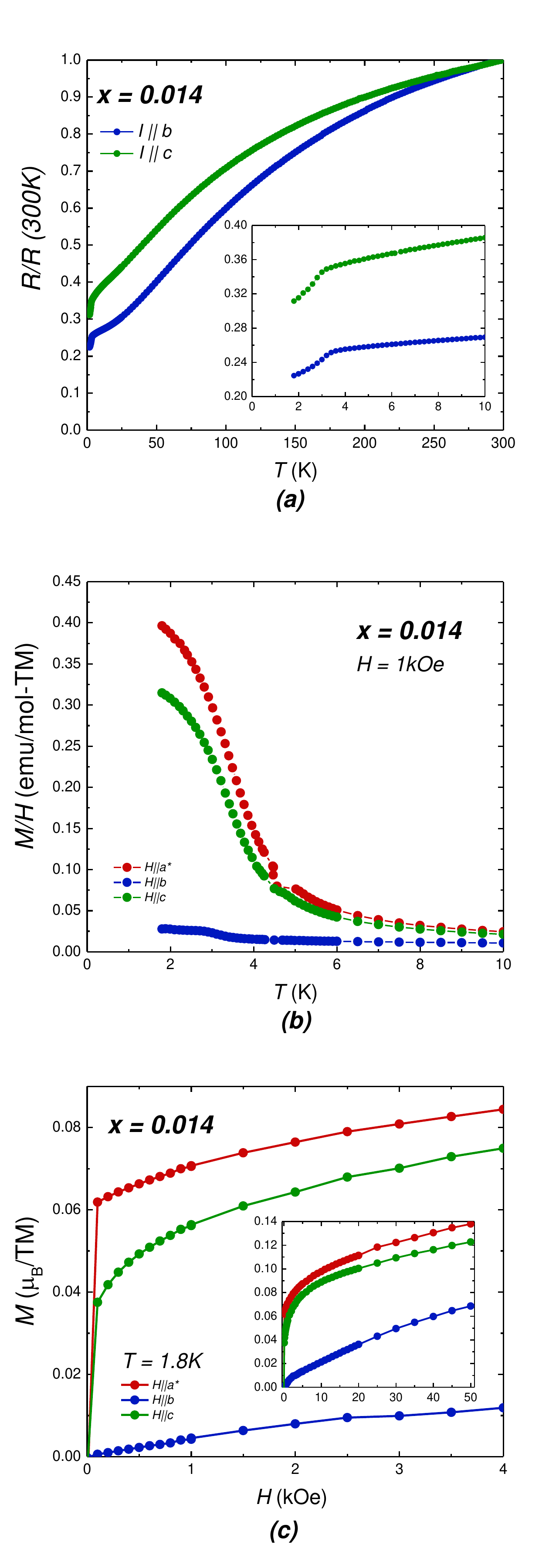}
    \caption{\footnotesize {(Color online) (a): Zero field, temperature dependent normalized resistance (Inset: The low temperature normalized $R(T)$ behavior shown for 1.8 K $\leq$ T $\leq$ 10 K), (b): Temperature dependent magnetization at H = 1 kOe for 1.8 K $\leq$ T $\leq$ 10 K, (c): Field dependent magnetization at a base temperature of 1.8 K shown for H $\leq$ 4 kOe and in the inset for H $\leq$ 4 kOe for $x$ = 0.014 in $\text{La}_{5}\text{(Co}_{1-x}\text {Ni}_{x})_2\text {Ge}_{3}$.}}
    \label{fig:All_1Ni}
\end{figure}

\begin{figure}[H]
    \centering
    \includegraphics[width=\linewidth]{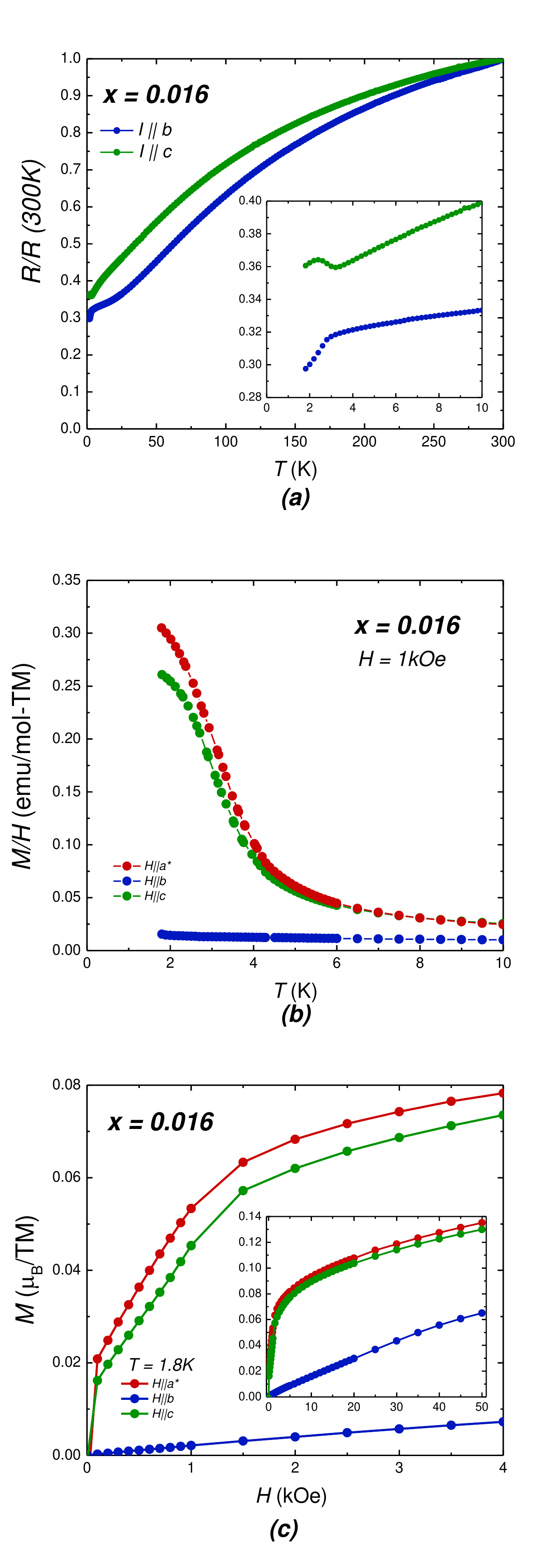}
    \caption{\footnotesize {(Color online) (a): Zero field, temperature dependent normalized resistance (Inset: The low temperature normalized $R(T)$ behavior shown for 1.8 K $\leq$ T $\leq$ 10 K), (b): Temperature dependent magnetization at H = 1 kOe for 1.8 K $\leq$ T $\leq$ 10 K, (c): Field dependent magnetization at a base temperature of 1.8 K shown for H $\leq$ 4 kOe and in the inset for H $\leq$ 4 kOe for $x$ = 0.016 in $\text{La}_{5}\text{(Co}_{1-x}\text {Ni}_{x})_2\text {Ge}_{3}$. }}
    \label{fig:All_1.75Ni}
\end{figure}

\begin{figure}[H]
    \centering
    \includegraphics[width=\linewidth]{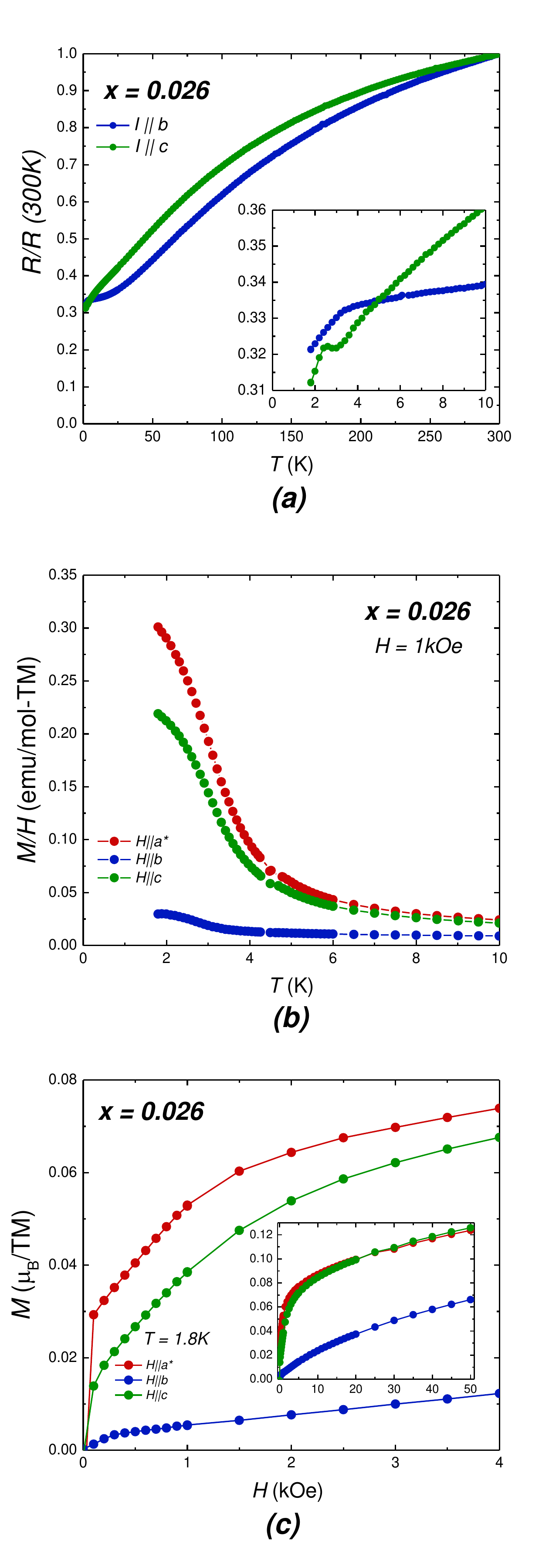}
    \caption{\footnotesize {(Color online) (a): Zero field, temperature dependent normalized resistance (Inset: The low temperature normalized $R(T)$ behavior shown for 1.8 K $\leq$ T $\leq$ 10 K), (b): Temperature dependent magnetization at H = 1 kOe for 1.8 K $\leq$ T $\leq$ 10 K, (c): Field dependent magnetization at a base temperature of 1.8 K shown for H $\leq$ 4 kOe and in the inset for H $\leq$ 4 kOe for $x$ = 0.026 in $\text{La}_{5}\text{(Co}_{1-x}\text {Ni}_{x})_2\text {Ge}_{3}$. }}
    \label{fig:All_2Ni}
\end{figure}

\begin{figure}[H]
    \centering
    \includegraphics[width=\linewidth]{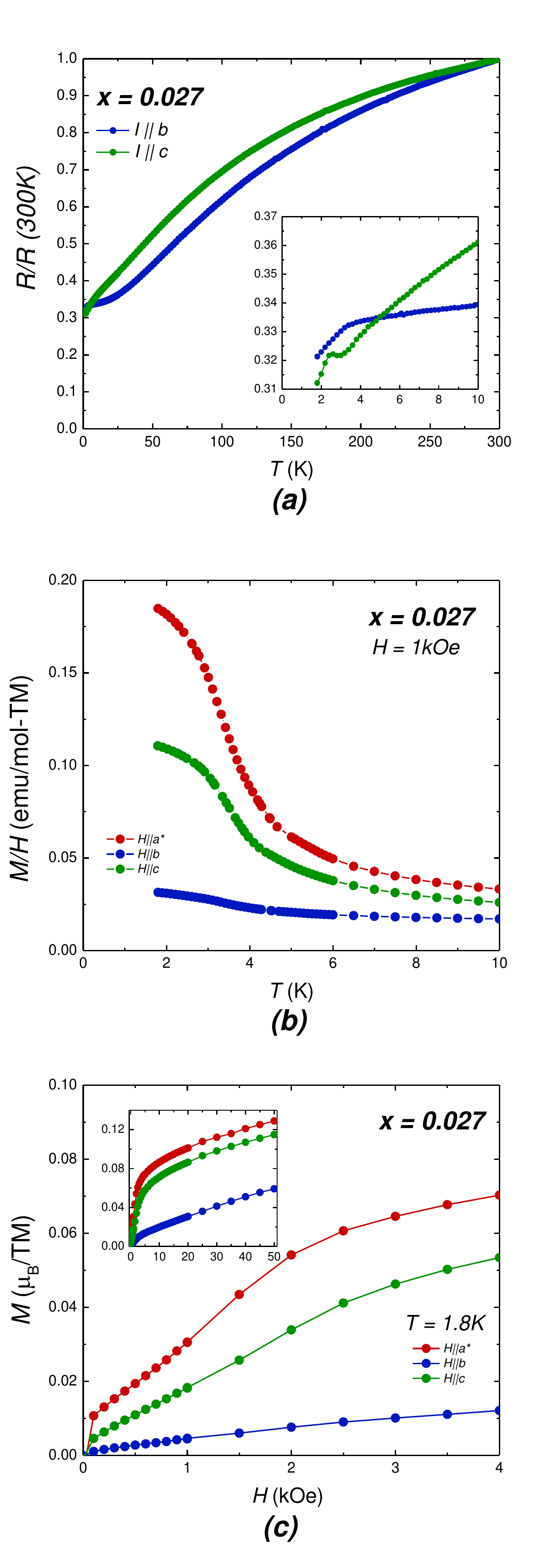}
    \caption{\footnotesize {(Color online) (a): Zero field, temperature dependent normalized resistance (Inset: The low temperature normalized $R(T)$ behavior shown for 1.8 K $\leq$ T $\leq$ 10 K), (b): Temperature dependent magnetization at H = 1 kOe for 1.8 K $\leq$ T $\leq$ 10 K, (c): Field dependent magnetization at a base temperature of 1.8 K shown for H $\leq$ 4 kOe and in the inset for H $\leq$ 4 kOe for $x$ = 0.027 in $\text{La}_{5}\text{(Co}_{1-x}\text {Ni}_{x})_2\text {Ge}_{3}$.}}
    \label{fig:All_2.25Ni}
\end{figure}

\begin{figure}[H]
    \centering
    \includegraphics[width=\linewidth]{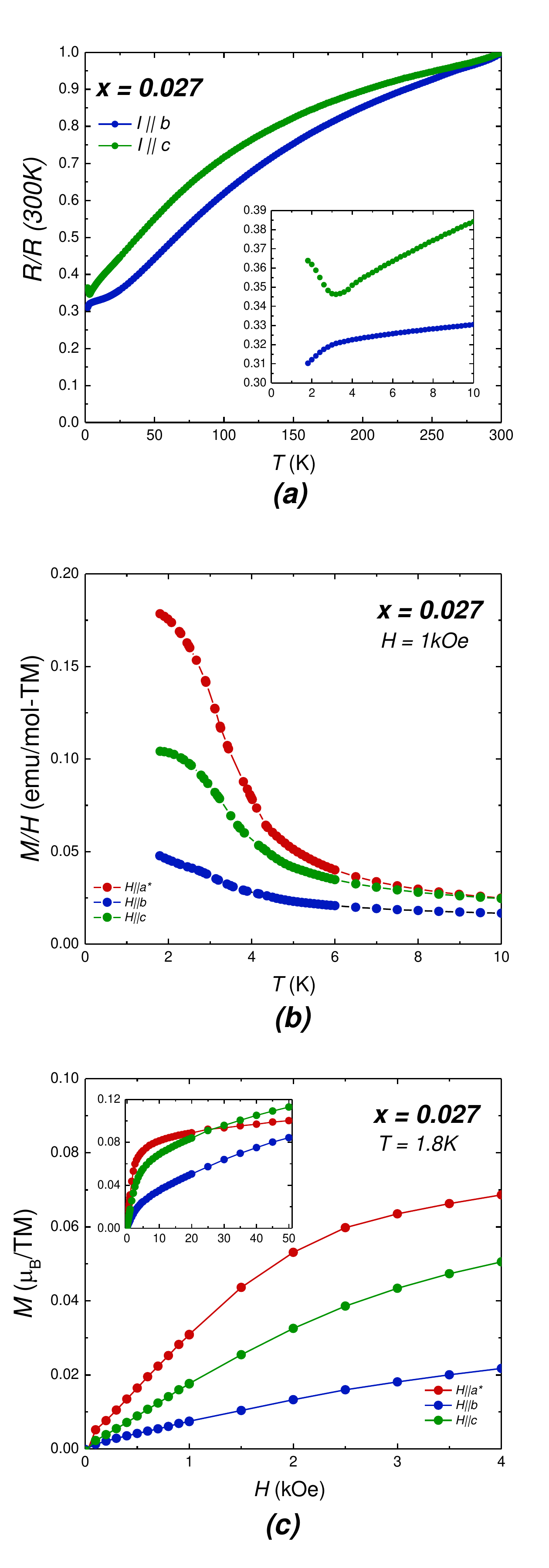}
    \caption{\footnotesize {(Color online) (a): Zero field, temperature dependent normalized resistance (Inset: The low temperature normalized $R(T)$ behavior shown for 1.8 K $\leq$ T $\leq$ 10 K), (b): Temperature dependent magnetization at H = 1 kOe for 1.8 K $\leq$ T $\leq$ 10 K, (c): Field dependent magnetization at a base temperature of 1.8 K shown for H $\leq$ 4 kOe and in the inset for H $\leq$ 4 kOe for $x$ = 0.027 in $\text{La}_{5}\text{(Co}_{1-x}\text {Ni}_{x})_2\text {Ge}_{3}$. }}
    \label{fig:All_2.5Ni}
\end{figure}

\begin{figure}[H]
    \centering
    \includegraphics[width=\linewidth]{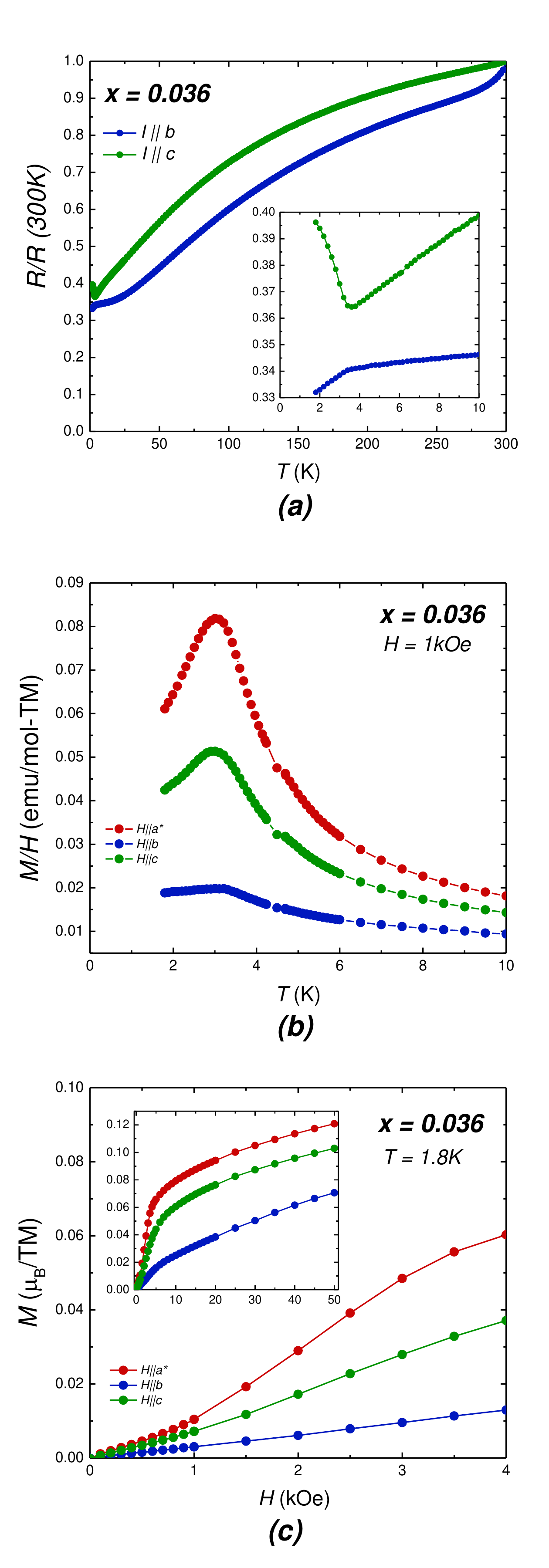}
    \caption{\footnotesize {(Color online) (a): Zero field, temperature dependent normalized resistance (Inset: The low temperature normalized $R(T)$ behavior shown for 1.8 K $\leq$ T $\leq$ 10 K), (b): Temperature dependent magnetization at H = 1 kOe for 1.8 K $\leq$ T $\leq$ 10 K, (c): Field dependent magnetization at a base temperature of 1.8 K shown for H $\leq$ 4 kOe and in the inset for H $\leq$ 4 kOe for $x$ = 0.036 in $\text{La}_{5}\text{(Co}_{1-x}\text {Ni}_{x})_2\text {Ge}_{3}$.}}
    \label{fig:All_3Ni}
\end{figure}

\begin{figure}[H]
    \centering
    \includegraphics[width=\linewidth]{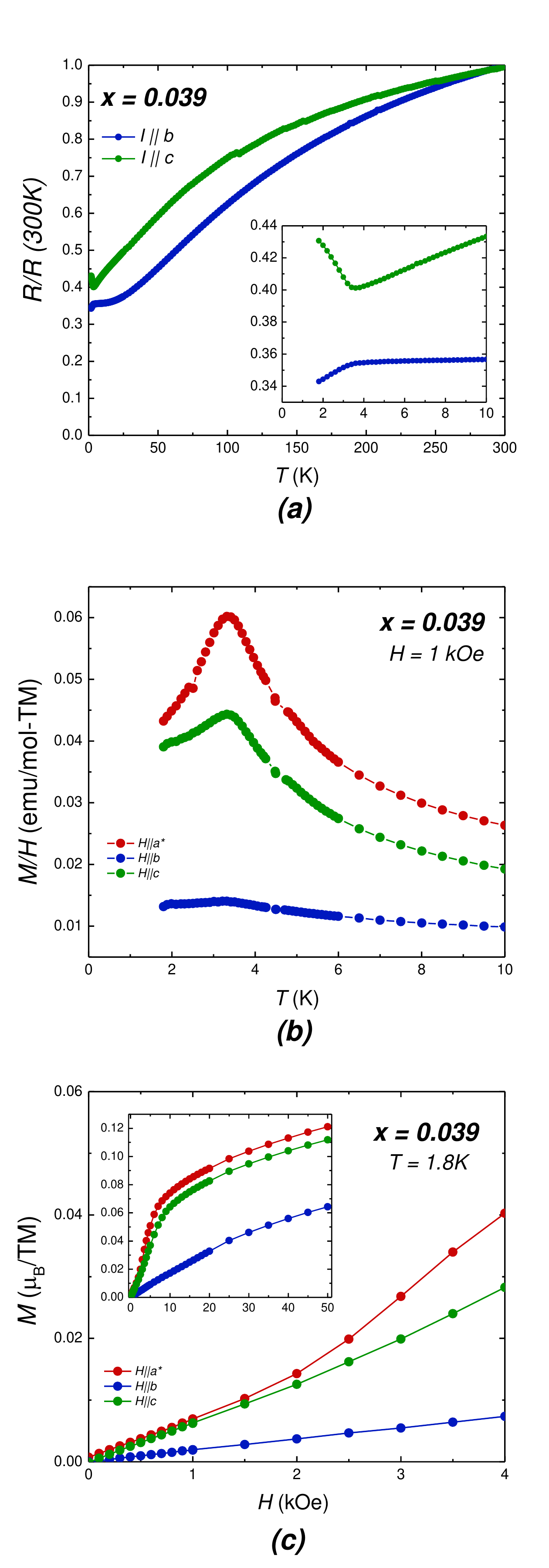}
    \caption{\footnotesize {(Color online) (a): Zero field, temperature dependent normalized resistance (Inset: The low temperature normalized $R(T)$ behavior shown for 1.8 K $\leq$ T $\leq$ 10 K), (b): Temperature dependent magnetization at H = 1 kOe for 1.8 K $\leq$ T $\leq$ 10 K, (c): Field dependent magnetization at a base temperature of 1.8 K shown for H $\leq$ 4 kOe and in the inset for H $\leq$ 4 kOe for $x$ = 0.039 in $\text{La}_{5}\text{(Co}_{1-x}\text {Ni}_{x})_2\text {Ge}_{3}$. }}
    \label{fig:All_4Ni}
\end{figure}

\begin{figure}[H]
    \centering
    \includegraphics[width=\linewidth]{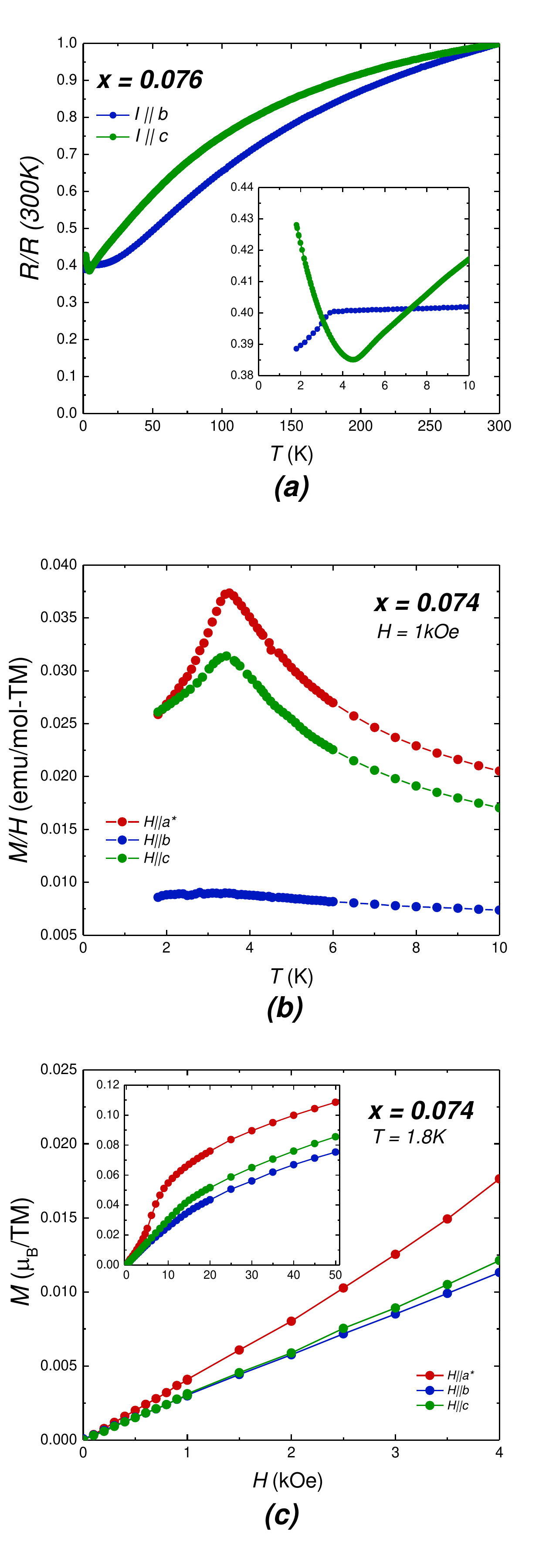}
    \caption{\footnotesize {(Color online) (a): Zero field, temperature dependent normalized resistance (Inset: The low temperature normalized $R(T)$ behavior shown for 1.8 K $\leq$ T $\leq$ 10 K) for x = 0.076, (b): Temperature dependent magnetization at H = 1 kOe for 1.8 K $\leq$ T $\leq$ 10 K, (c): Field dependent magnetization at a base temperature of 1.8 K shown for H $\leq$ 4 kOe and in the inset for H $\leq$ 4 kOe for $x$ = 0.074 in $\text{La}_{5}\text{(Co}_{1-x}\text {Ni}_{x})_2\text {Ge}_{3}$. }}
    \label{fig:All_6Ni}
\end{figure}

\begin{figure}[H]
    \centering
    \includegraphics[width=\linewidth]{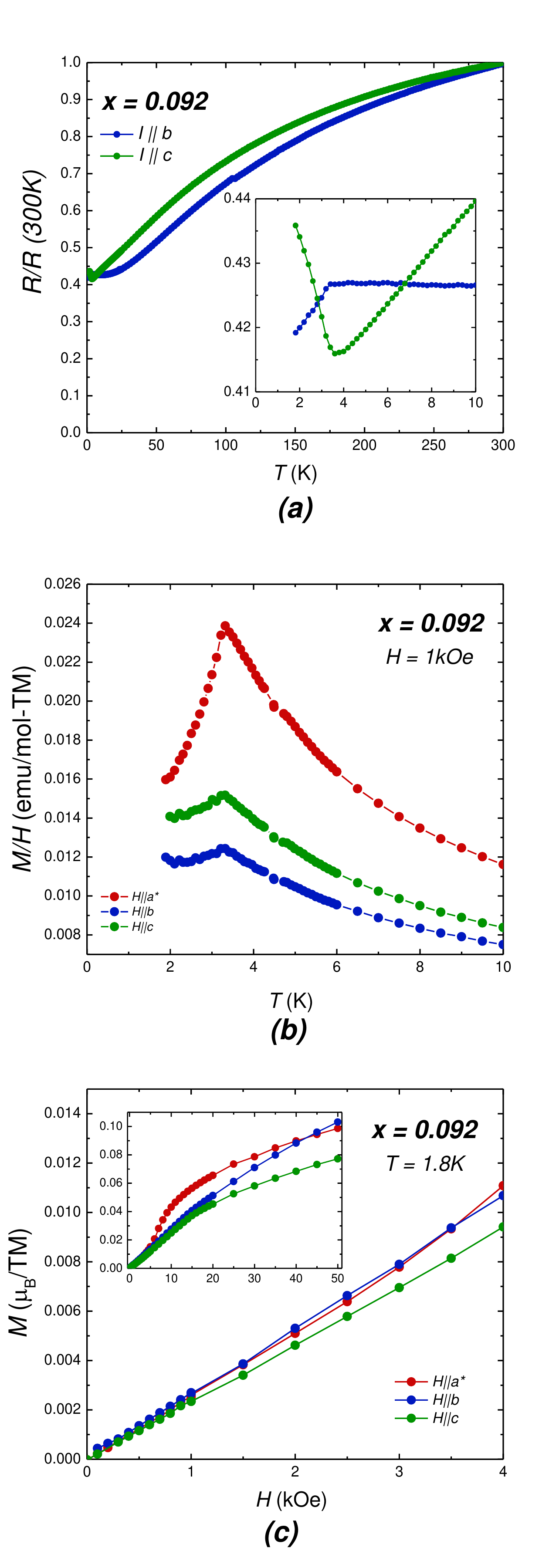}
    \caption{\footnotesize {(Color online) (a): Zero field, temperature dependent normalized resistance (Inset: The low temperature normalized $R(T)$ behavior shown for 1.8 K $\leq$ T $\leq$ 10 K), (b): Temperature dependent magnetization at H = 1 kOe for 1.8 K $\leq$ T $\leq$ 10 K, (c): Field dependent magnetization at a base temperature of 1.8 K shown for H $\leq$ 4 kOe and in the inset for H $\leq$ 4 kOe for $x$ = 0.092 in $\text{La}_{5}\text{(Co}_{1-x}\text {Ni}_{x})_2\text {Ge}_{3}$.}}
    \label{fig:All_8Ni}
\end{figure}

\begin{figure}[H]
    \centering
    \includegraphics[width=\linewidth]{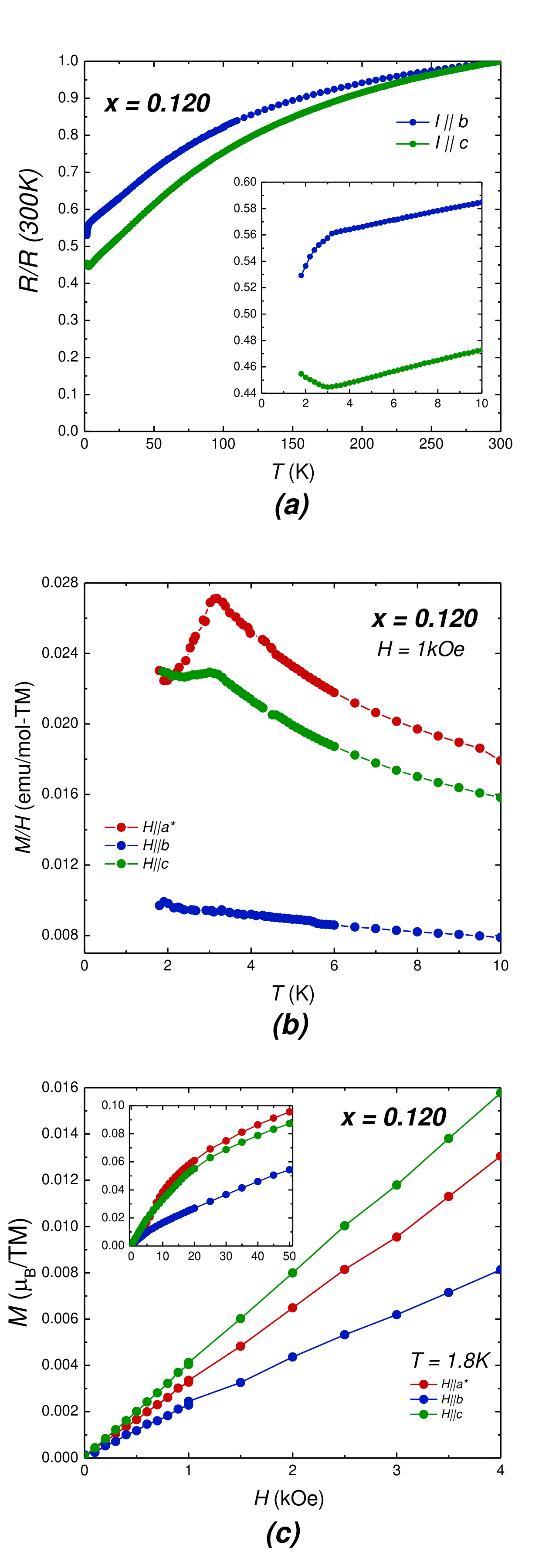}
    \caption{\footnotesize {(Color online) (a): Zero field, temperature dependent normalized resistance (Inset: The low temperature normalized $R(T)$ behavior shown for 1.8 K $\leq$ T $\leq$ 10 K), (b): Temperature dependent magnetization at H = 1 kOe for 1.8 K $\leq$ T $\leq$ 10 K, (c): Field dependent magnetization at a base temperature of 1.8 K shown for H $\leq$ 4 kOe and in the inset for H $\leq$ 4 kOe for $x$ = 0.120 in $\text{La}_{5}\text{(Co}_{1-x}\text {Ni}_{x})_2\text {Ge}_{3}$. }}
    \label{fig:All_10Ni}
\end{figure}

\begin{figure}[H]
    \centering
    \includegraphics[width=\linewidth]{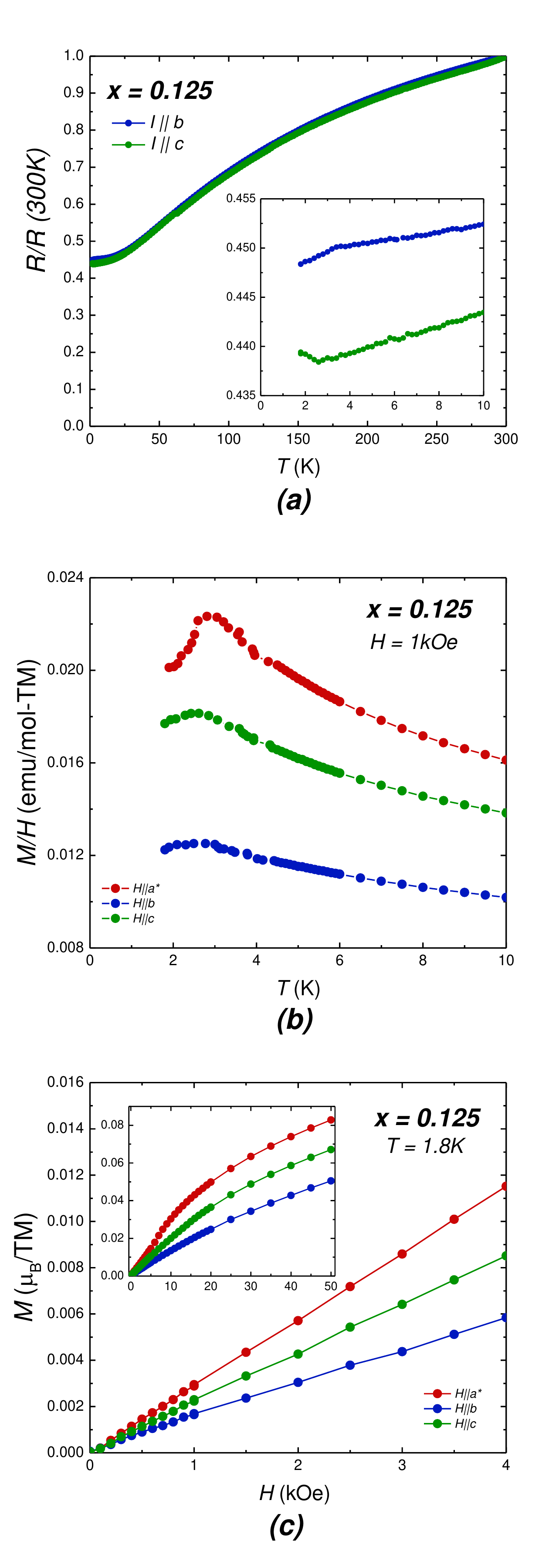}
    \caption{\footnotesize {(Color online) (a): Zero field, temperature dependent normalized resistance (Inset: The low temperature normalized $R(T)$ behavior shown for 1.8 K $\leq$ T $\leq$ 10 K), (b): Temperature dependent magnetization at H = 1 kOe for 1.8 K $\leq$ T $\leq$ 10 K, (c): Field dependent magnetization at a base temperature of 1.8 K shown for H $\leq$ 4 kOe and in the inset for H $\leq$ 4 kOe for $x$ = 0.125 in $\text{La}_{5}\text{(Co}_{1-x}\text {Ni}_{x})_2\text {Ge}_{3}$.}}
    \label{fig:All_12Ni}
\end{figure}

\begin{figure}[h]
     \centering
     \includegraphics[width=\linewidth]{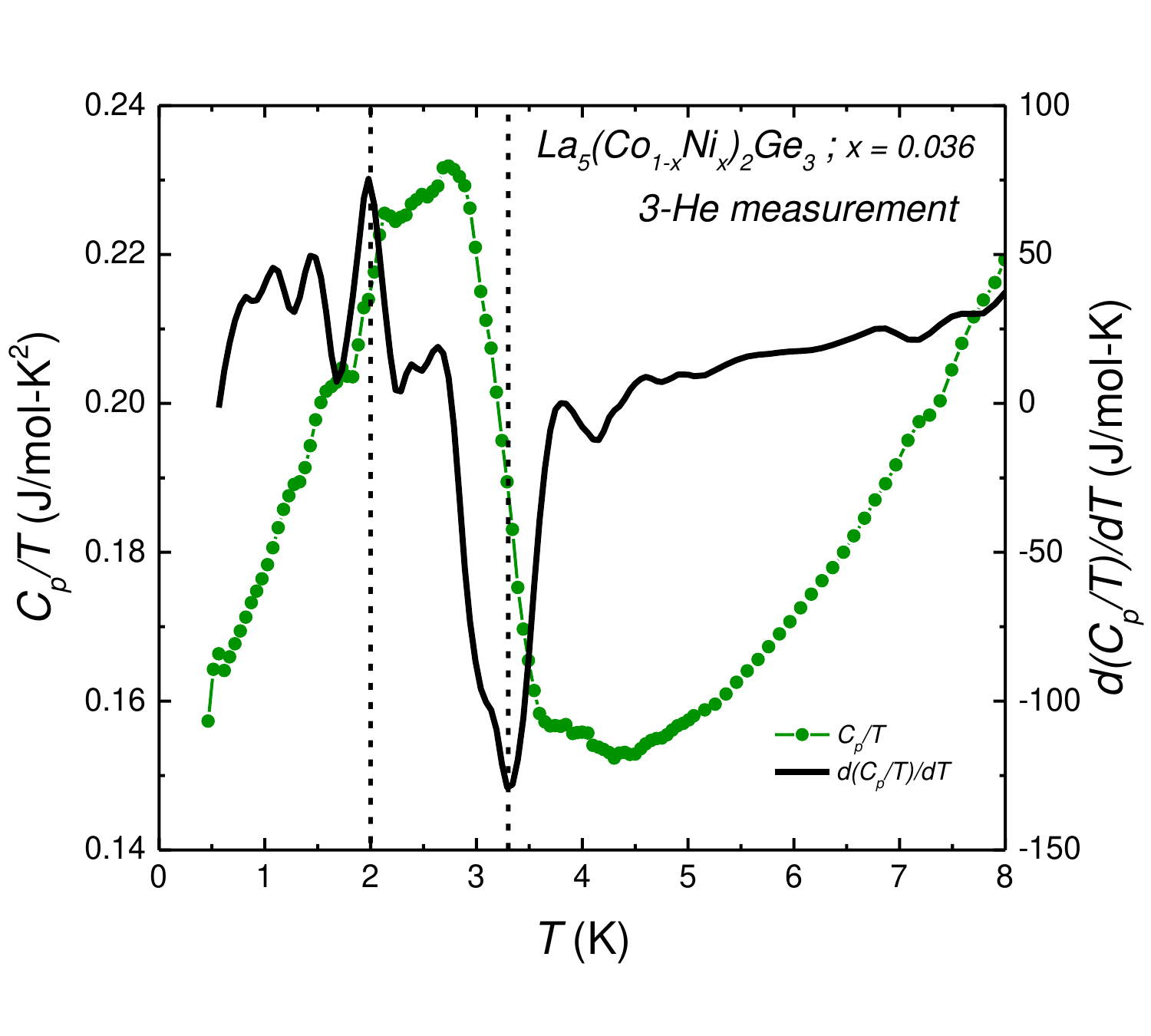}
     \caption{\footnotesize {The Specific heat data for x = 0.036 measured using a 3-He insert to PPMS along with its derivative plotted in the right hand axis. A sharp transition is observed at around 3.3 K, and another subtle one $\sim$ 2 K which is not observes in any other measurement.} }
     \label{fig:He3-cp}
 \end{figure}

As $x$ = 0.036 lies at the border where we start seeing only a single AFM transition in the $M(T)$ and opening of a super-zone gap in the $R(T)$ data, it was probed to low temperatures by measuring specific heat down to 0.5 K using a 3-He inset. Fig \ref{fig:He3-cp} shows transitions around $\sim$ 3.3 K, also observed in specific heat data down to base temperature of 1.8 K and another at $\sim$ 2 K, which is not observed in the data of previously discussed magnetization and resistance measurements. 

\begin{figure}[H]
    \centering
    \includegraphics[width=\linewidth]{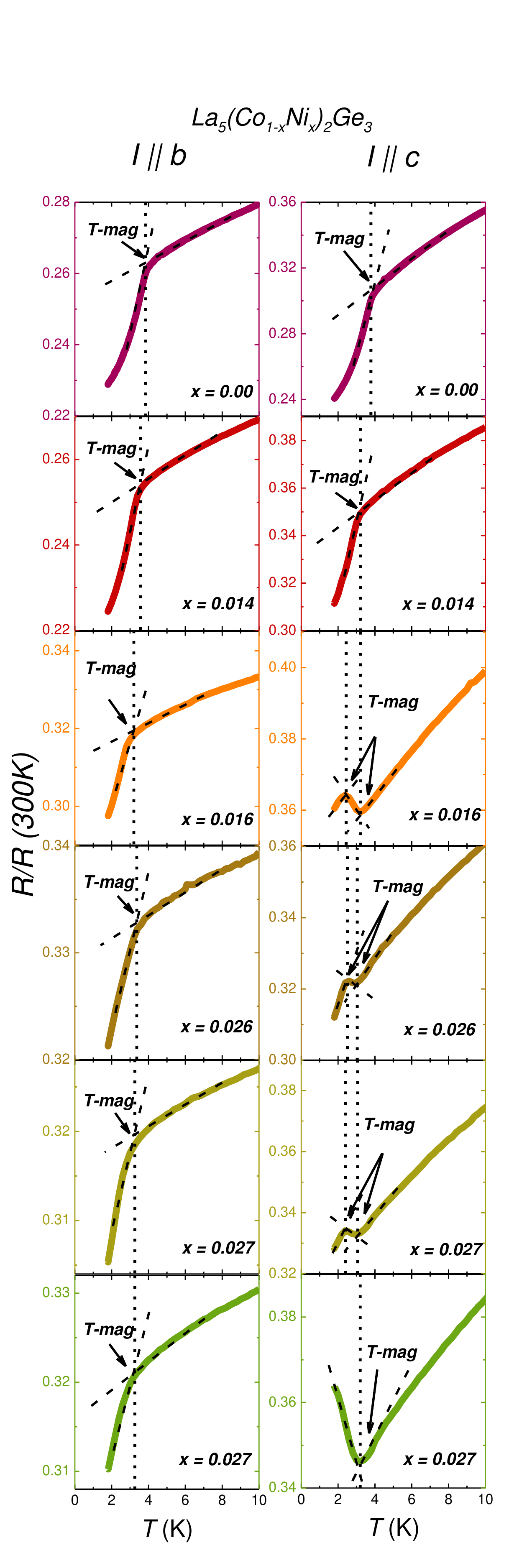}
    \caption{\footnotesize{ (Color online) The temperature dependent normalized resistance along both b- and c- axis of the low Ni substitutions which show ferromagnetic behavior in $M(H)$ measurements at T = 1.8 K. The transition temperature ($T_{mag}$) is marked using a vertical line as per the criteria used (See text for details).}}
    \label{fig:R(T)_low x_criteria}
\end{figure}

Turning to normalized resistance data, the transition temperature ($T_{mag}$) of the Ni substituted samples manifesting a FM ground state is determined from the intersection of the two dashed lines for measurements along I$||$b and I$||$c. $T_{mag}$ is marked using a vertical dashed line for each of the substitutions in Fig \ref{fig:R(T)_low x_criteria}. 

\par

\begin{figure}[H]
    \centering
    \includegraphics[width=\linewidth]{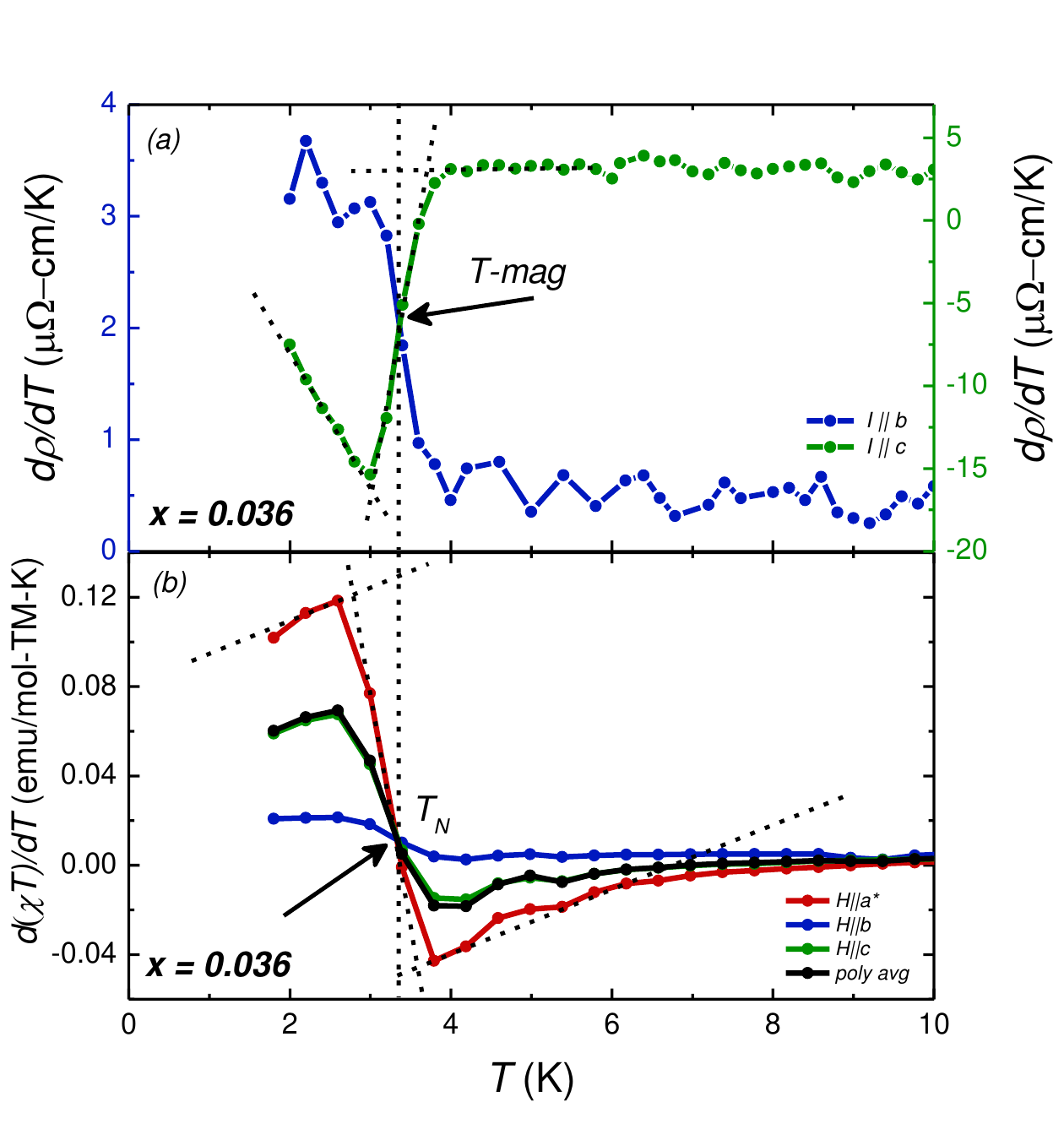}
    \caption{\footnotesize{ (Color online) (a) $\frac{d \rho}{dT}$ and (b) $\frac{d (\chi T)}{dT}$ for x = 0.036 which is used as a criteria to obtain the transition temperature for the AFM states by taking the point of maximum slope. In panel (a), the values for current along the b-direction is shown in the left axis and values for current along the c-axis is in the right axis. The criteria for obtaining the transition temperature is marked for I$||$c in (a) and for H$||$a* in (b). The vertical line in each of the panel are the ($T_{mag}$) and ($T_{N}$) are the average transition temperatures obtained from $\frac{d \rho}{dT}$ and $\frac{d (\chi T)}{dT}$ plots respectively.}}
    \label{fig:criteria_3Ni new}
\end{figure}

\begin{figure}[H]
    \centering
    \includegraphics[width=\linewidth]{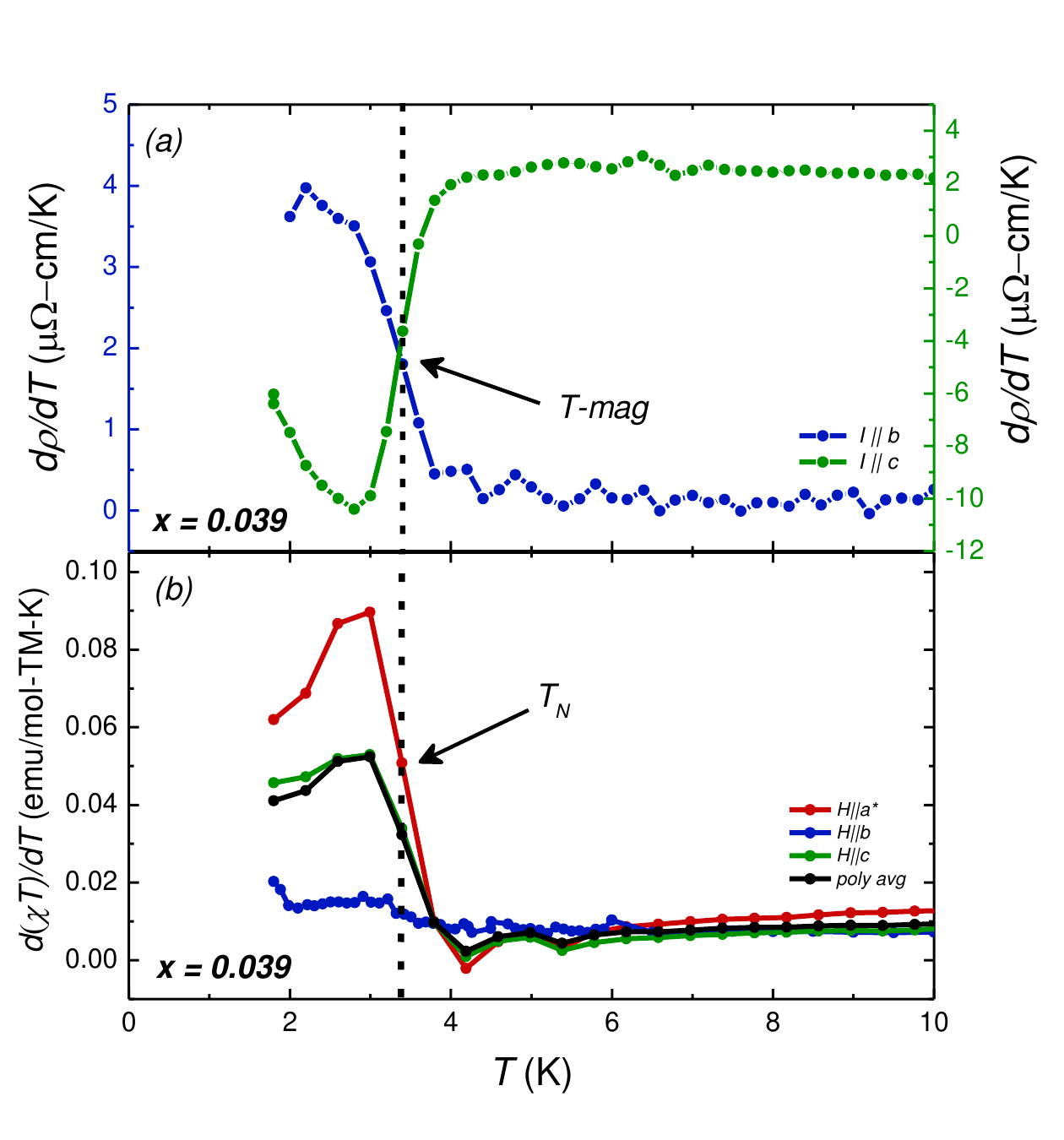}
    \caption{\footnotesize{ (Color online) (a) $\frac{d \rho}{dT}$ and (b) $\frac{d (\chi T)}{dT}$ for x = 0.039 which is used as a criteria to obtain the transition temperature for the AFM states by taking the point of maximum slope. In panel (a), the values for current along the b-direction is shown in the left axis and values for current along the c-axis is in the right axis. The vertical line in each of the panel are the ($T_{mag}$) and ($T_{N}$) are the average transition temperatures obtained from $\frac{d \rho}{dT}$ and $\frac{d (\chi T)}{dT}$ plots respectively.}}
    \label{fig:criteria_4Ni}
\end{figure}

\begin{figure}[H]
    \centering
    \includegraphics[width=\linewidth]{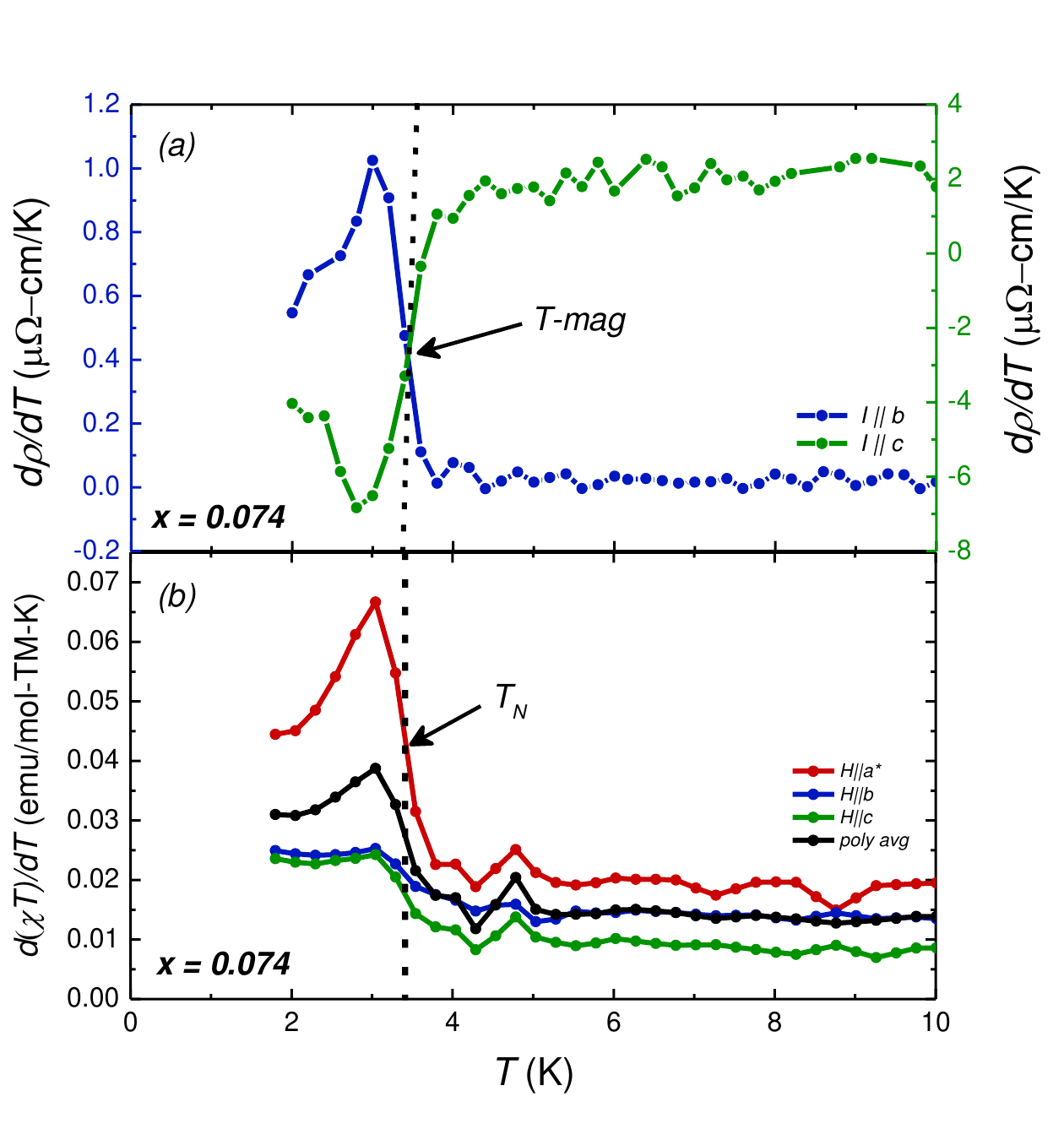}
    \caption{\footnotesize{ (Color online) (a) $\frac{d \rho}{dT}$ and (b) $\frac{d (\chi T)}{dT}$ for x = 0.074 which is used as a criteria to obtain the transition temperature for the AFM states by taking the point of maximum slope. In panel (a), the values for current along the b-direction is shown in the left axis and values for current along the c-axis is in the right axis. The vertical line in each of the panel are the ($T_{mag}$) and ($T_{N}$) are the average transition temperatures obtained from $\frac{d \rho}{dT}$ and $\frac{d (\chi T)}{dT}$ plots respectively.}}
    \label{fig:criteria_6Ni}
\end{figure}

\begin{figure}[H]
    \centering
    \includegraphics[width=\linewidth]{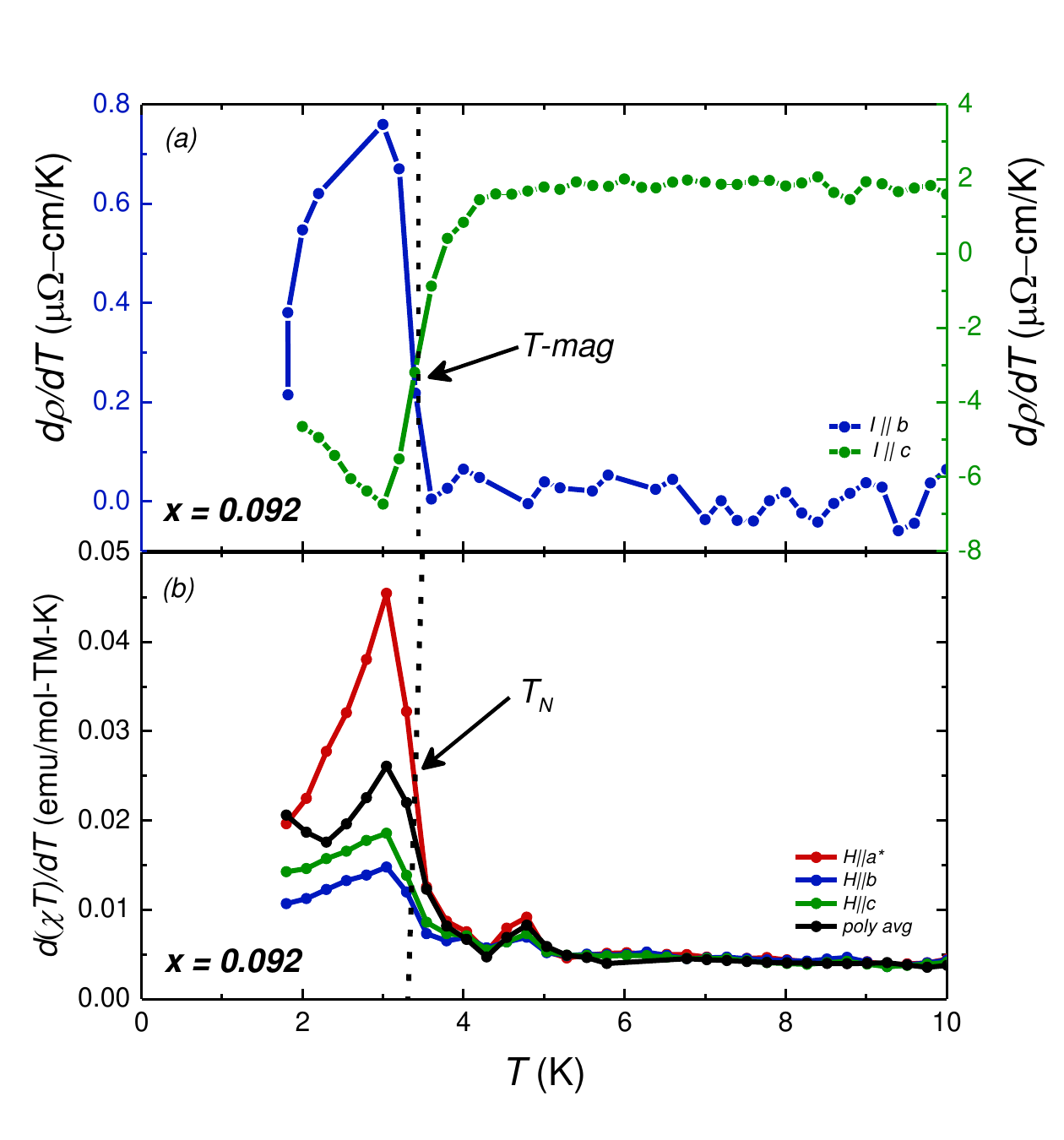}
    \caption{\footnotesize{  (Color online) (a) $\frac{d \rho}{dT}$ and (b) $\frac{d (\chi T)}{dT}$ for x = 0.092 which is used as a criteria to obtain the transition temperature for the AFM states by taking the point of maximum slope. In panel (a), the values for current along the b-direction is shown in the left axis and values for current along the c-axis is in the right axis. The vertical line in each of the panel are the ($T_{mag}$) and ($T_{N}$) are the average transition temperatures obtained from $\frac{d \rho}{dT}$ and $\frac{d (\chi T)}{dT}$ plots respectively.}}
    \label{fig:criteria_ 8Ni}
\end{figure}

\begin{figure}[H]
    \centering
    \includegraphics[width=\linewidth]{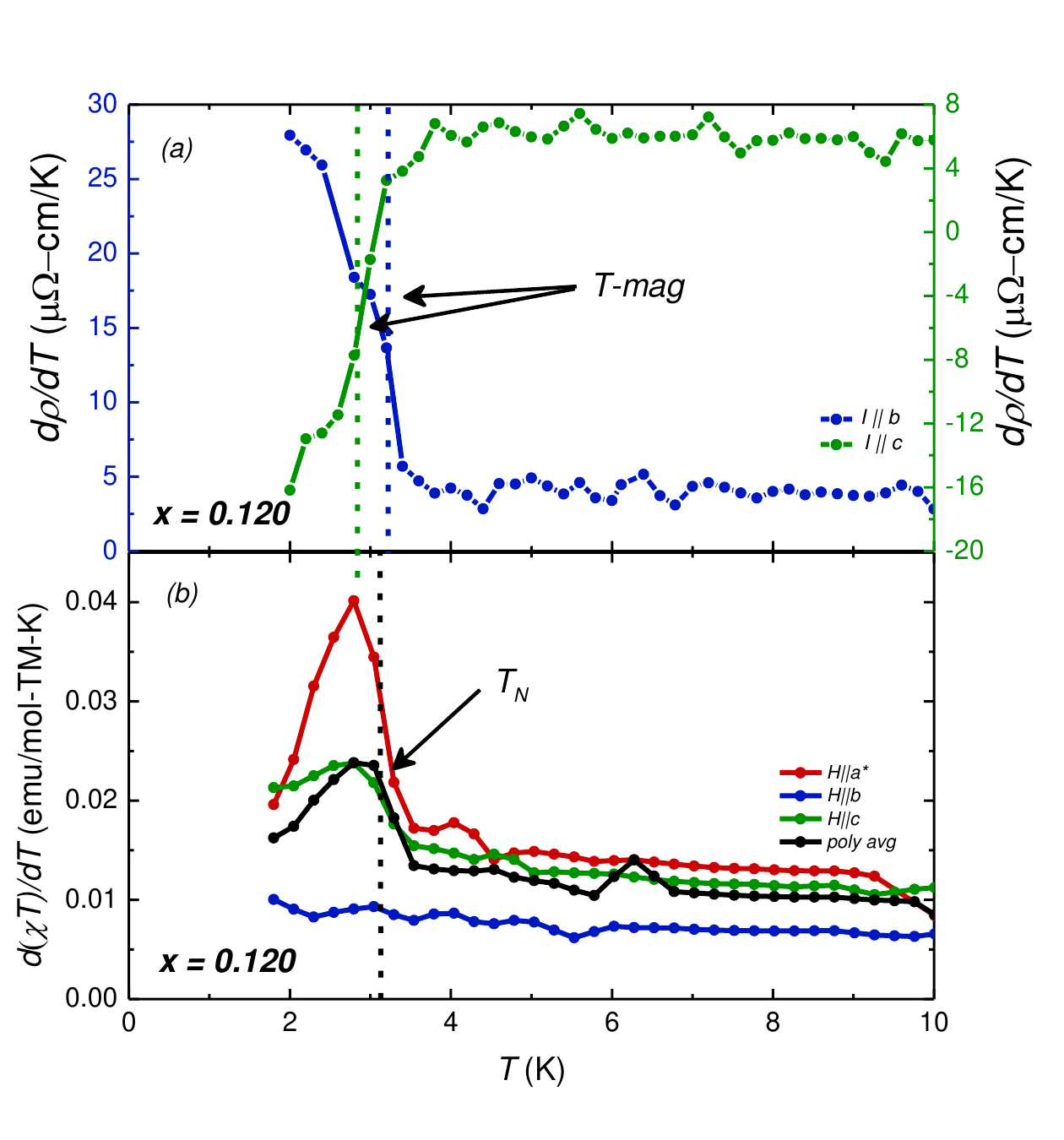}
    \caption{\footnotesize{ (Color online) (a) $\frac{d \rho}{dT}$ and (b) $\frac{d (\chi T)}{dT}$ for x = 0.120 which is used as a criteria to obtain the transition temperature for the AFM states by taking the point of maximum slope. In panel (a), the values for current along the b-direction is shown in the left axis and values for current along the c-axis is in the right axis. The vertical lines in each of the panel are the ($T_{mag}$) and ($T_{N}$) are the average transition temperatures obtained from $\frac{d \rho}{dT}$ and $\frac{d (\chi T)}{dT}$ plots for each direction respectively.}}
    \label{fig:criteria_10Ni}
\end{figure}

\begin{figure}[H]
    \centering
    \includegraphics[width=\linewidth]{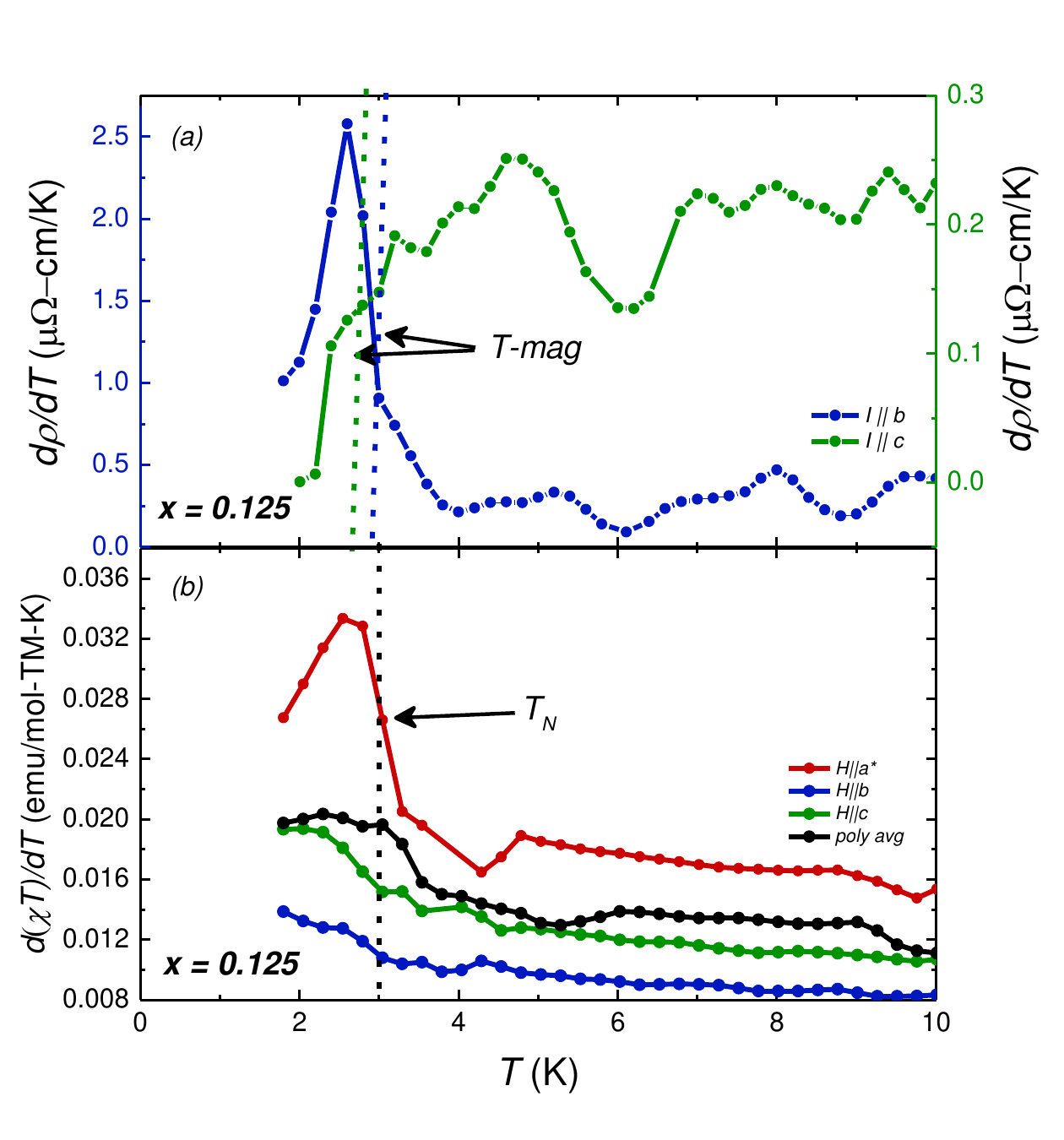}
    \caption{\footnotesize{ (Color online) (a) $\frac{d \rho}{dT}$ and (b) $\frac{d (\chi T)}{dT}$ for x = 0.125 which is used as a criteria to obtain the transition temperature for the AFM states by taking the point of maximum slope. In panel (a), the values for current along the b-direction is shown in the left axis and values for current along the c-axis is in the right axis. The vertical lines in each of the panel are the ($T_{mag}$) and ($T_{N}$) are the average transition temperatures obtained from $\frac{d \rho}{dT}$ and $\frac{d (\chi T)}{dT}$ plots for each direction respectively.}}
    \label{fig:criteria_12Ni}
\end{figure}

\begin{figure}[H]
    \centering
    \includegraphics[width=\linewidth]{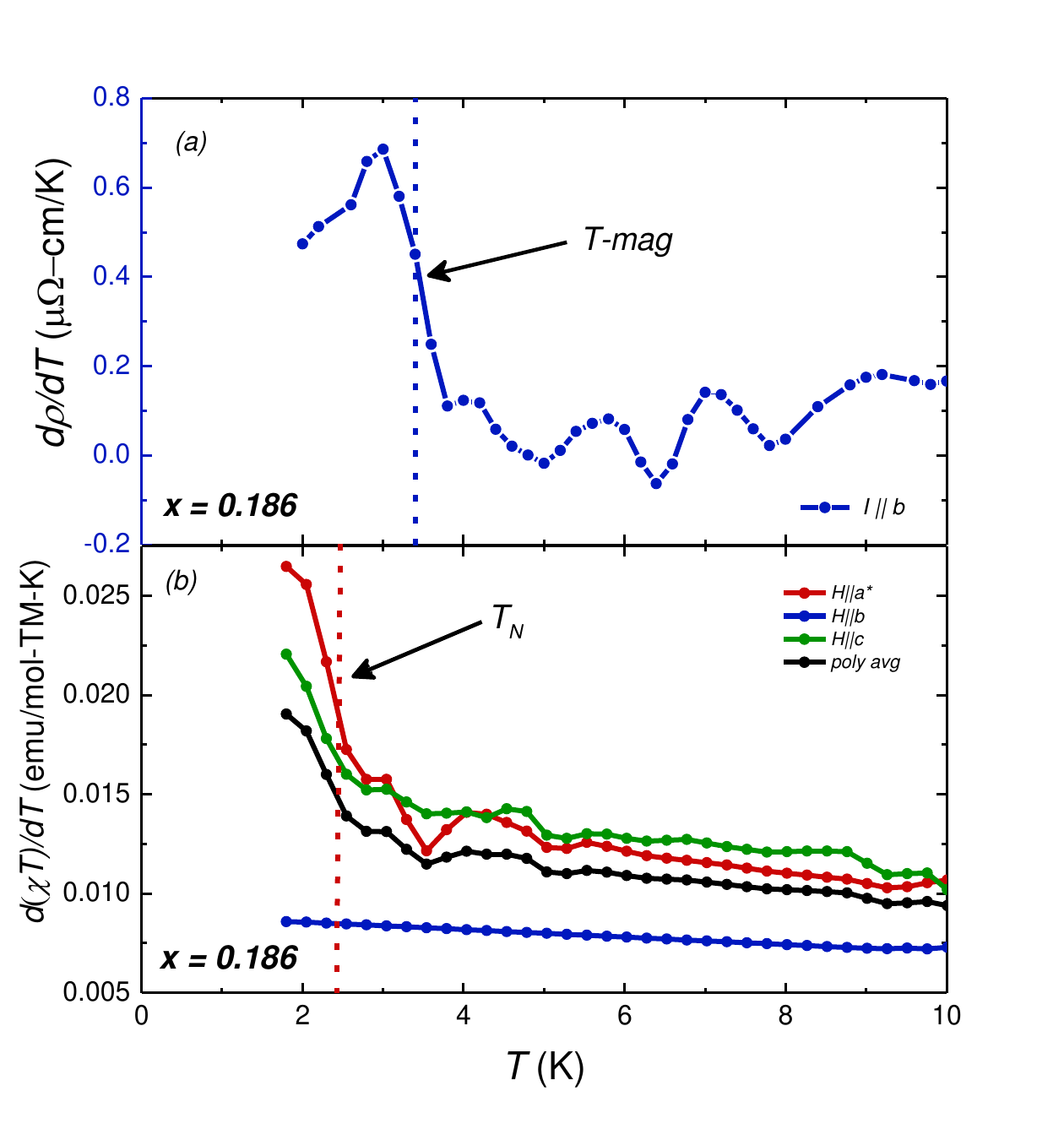}
    \caption{\footnotesize{(Color online) (a) $\frac{d \rho}{dT}$ and (b) $\frac{d (\chi T)}{dT}$ for x = 0.186 which is used as a criteria to obtain the transition temperature for the AFM states by taking the point of maximum slope. For the $\frac{d (\chi T)}{dT}$ along H$||$b, we do not get any visible feature. In panel (a), the values for current along the b-direction is only shown. The vertical line in each of the panel are the ($T_{mag}$) and ($T_{N}$) are the average transition temperatures obtained from $\frac{d \rho}{dT}$ and $\frac{d (\chi T)}{dT}$ plots respectively.}}
    \label{fig:criteria_15Ni}
\end{figure}

$T_{mag}$ for the AFM Ni substitutions is inferred from the $\frac{d \rho}{dT}$ plots and is determined by construction of three dashed lines in the low-, intermediate-, and high-temperature regimes. The intermediate temperature line goes through the point of maximum slope of the $\rho$(T). $\text{T}_{mag}$ is determined as the midpoint of the intersection points of the three dashed lines, and the uncertainty in $\text{T}_{mag}$ is obtained from the temperature difference of the two intersection points \cite{Xiang2021AvoidedLa5Co2Ge3}. $\text{T}_{mag}$ for the AFM Ni substitutions is marked using a vertical line in \ref{fig:criteria_3Ni new}(a) - \ref{fig:criteria_15Ni}(a).

\begin{figure}[H]
    \centering
    \includegraphics[width=\linewidth]{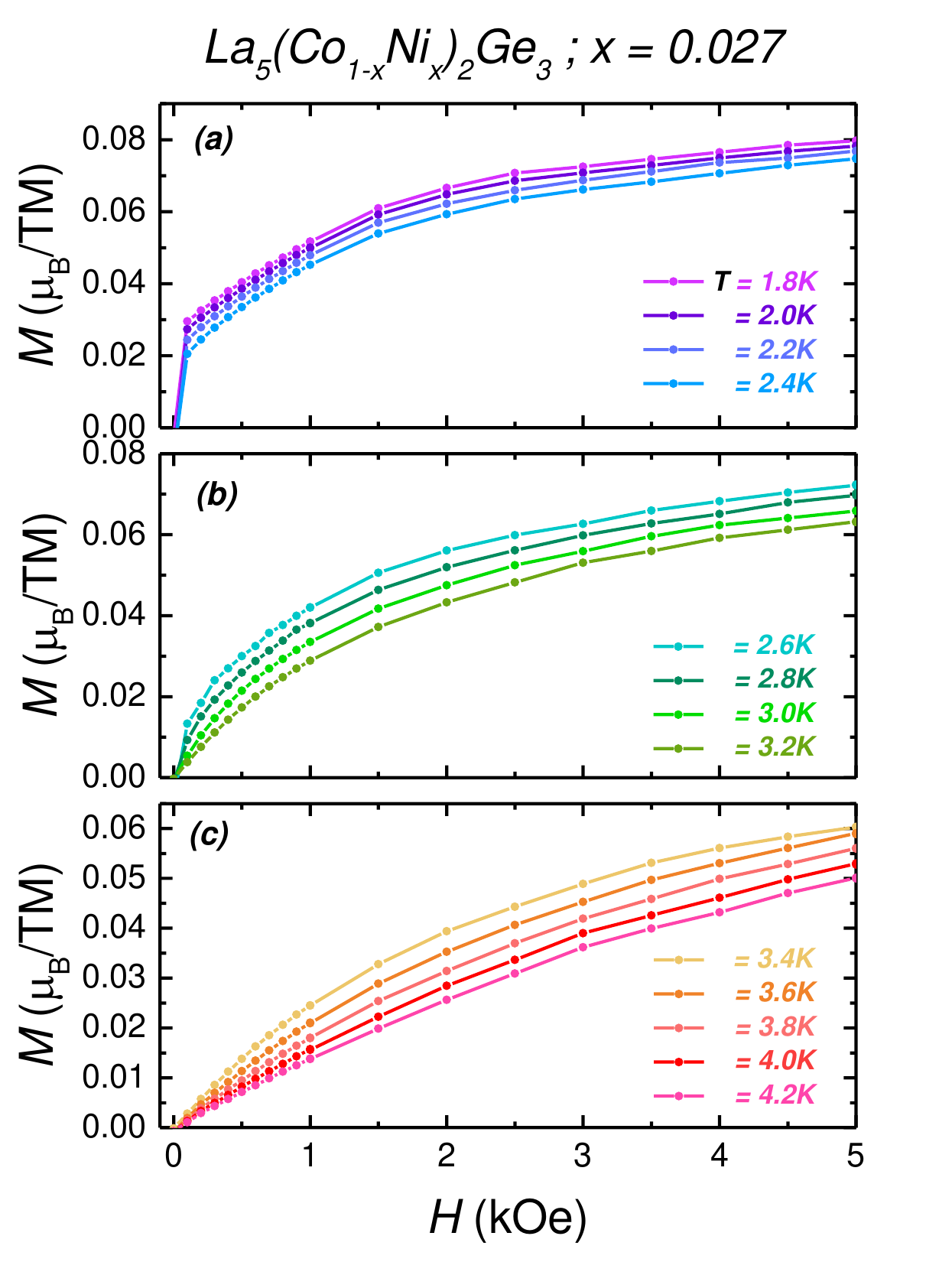}
    \caption{\footnotesize{ (Color online) $M(H)$ taken at different temperatures (1.8 K $\leq$ T $\leq$ 4.2 K) for x = 0.027 in $\text{La}_{5}\text{(Co}_{1-x}\text {Ni}_{x})_2\text {Ge}_{3}$ to show the evolution of ground state from a ferromagnetic (T $\leq$ 2.6 K) to a non ferromagnetic state.}}
    \label{fig:MH_2Ni}
\end{figure}

The $M(H)$ data on the $x$ = 0.027 sample measured for $T \leq$ 4.2 K are shown in Fig \ref{fig:MH_2Ni}. The Arrott plots ($M^{2}~ \text{vs}~ H/M$), shown in Fig \ref{fig:arrott}, are constructed based on Fig \ref{fig:MH_2Ni} data. A finite spontaneous moment clearly indicates FM behavior. Data taken for T $\geq$ 2.6 K has no finite intercept suggesting non-FM behavior (negative intercept in Fig \ref{fig:arrott}).

\par

The $\mu$SR experiments (both ZF and wTF) were performed in a spin-rotated mode, where the initial muon spin polarization ($P_{\mu}$) is turned by a certain angle relative to muon beam momentum \cite{Khasanov2020MagnetismStudies}. Such geometry is used to measure single crystalline samples.

\begin{figure}[H]
    \centering
    \includegraphics[width=\linewidth]{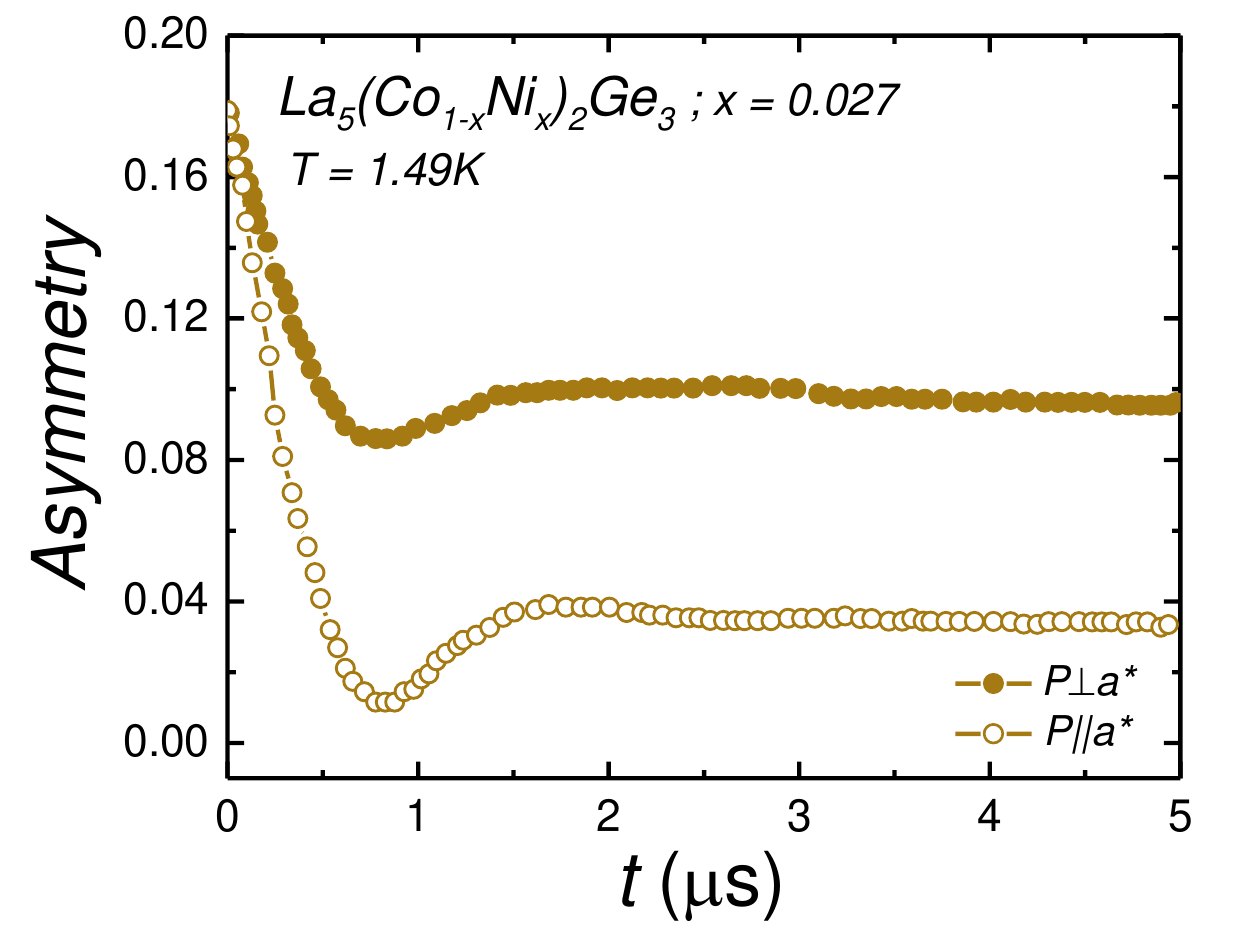}
    \caption{\footnotesize {Muon asymmetry for x = 0.027 collected at T $\sim$ 1.5 K with the component of the initial muon polarization perpendicular and parallel to a*. }}
    \label{fig:wTF zf 2Ni}
\end{figure}

\begin{figure}[h]
    \centering
    \includegraphics[width=\linewidth]{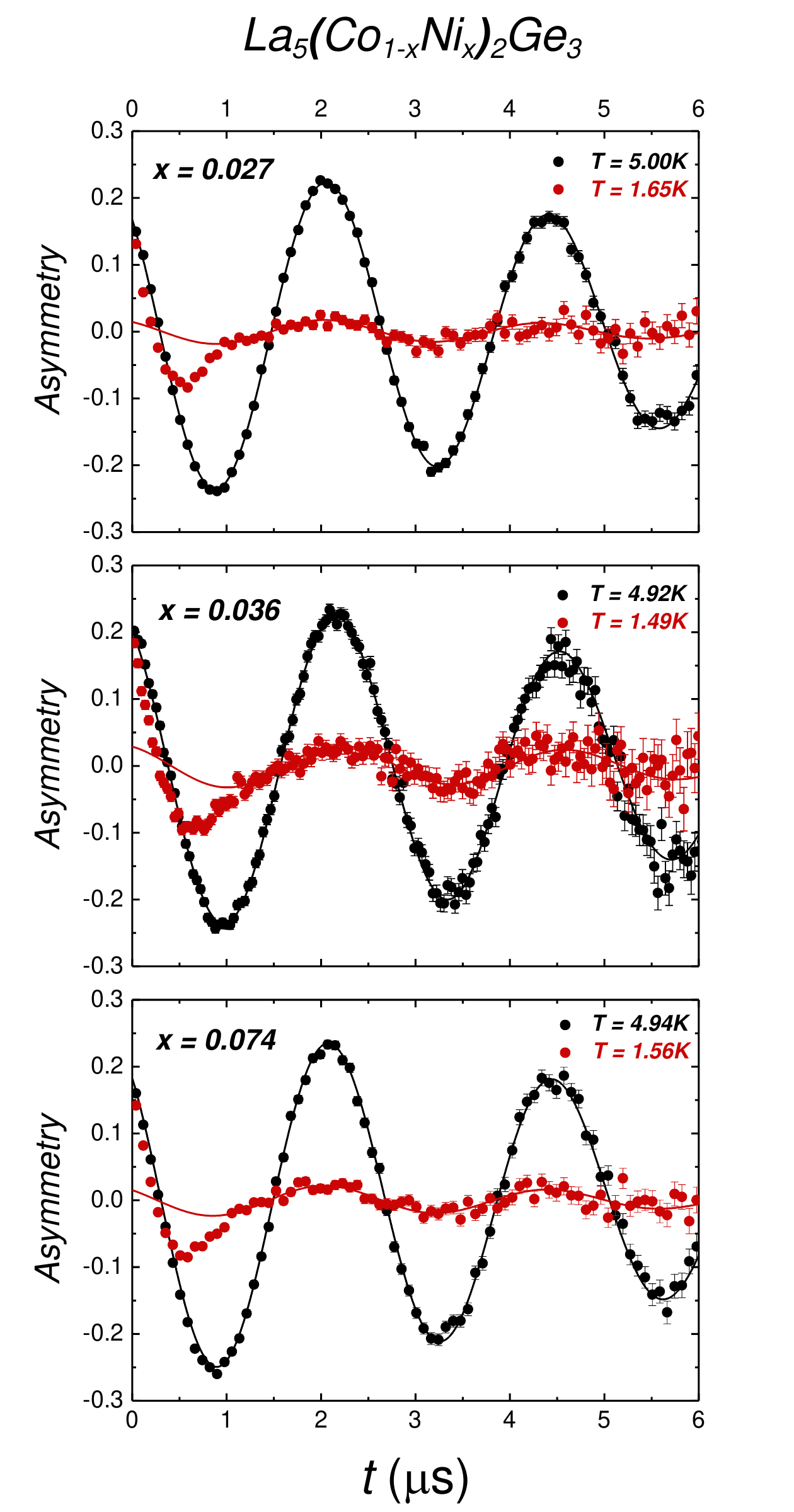}
    \caption{\footnotesize{wTF-$\mu$SR time dependent asymmetry for x = 0.027, 0.036 and 0.074 in $\text{La}_{5}\text{(Co}_{1-x}\text {Ni}_{x})_2\text {Ge}_{3}$. For each substitution we have data for temperatures above and below the magnetic transition as shown in black and red symbols separately. The solid lines are fit to Eq. \ref{eqn:wtf} (see text for details). }}
    \label{fig:wtf all}
\end{figure}

Weak transverse field (wTF) $\mu$SR experiments were done by applying an external field of $H_{ex}$ = 30 Oe perpendicular to the muon spin polarization to determine the onset temperature of the magnetic transitions in the samples. The polarized muons were implanted along the a* direction as discussed earlier. In the paramagnetic state, muons experience the spatially invariant external field and produce long-lived oscillations, reflecting the coherent muon precession around the external field. In the ordered state, the muon asymmetry has a complex precession due to vector combination of the internal field of the sample and the externally applied field. 

\par

As the wTF muon spectra was measured both above and below the magnetic transition ($\approx$ 3 K), the time evolution of the muon asymmetry can be assumed to have non-magnetic (nm) and magnetic (m) contributions to the data and can be written as a combination of the two contributions; 

\begin{equation}
    \begin{split}
        A(t)& = A(0)P(t)\\
            & = A_{nm}(0)P_{nm}(t) + A_{m}(0)P_{m}(t)
    \end{split}
    \label{eqn:nm and m}
\end{equation}

where, $A_{nm}(0)[A_{m}(0)]$ are the initial asymmetry and $P_{nm}(t)[P_{m}(t)]$, the time evolution of the muon spin polarization in the non-magnetic and [magnetic] states respectively.

\begin{figure}[h]
    \centering
    \includegraphics[width=\linewidth]{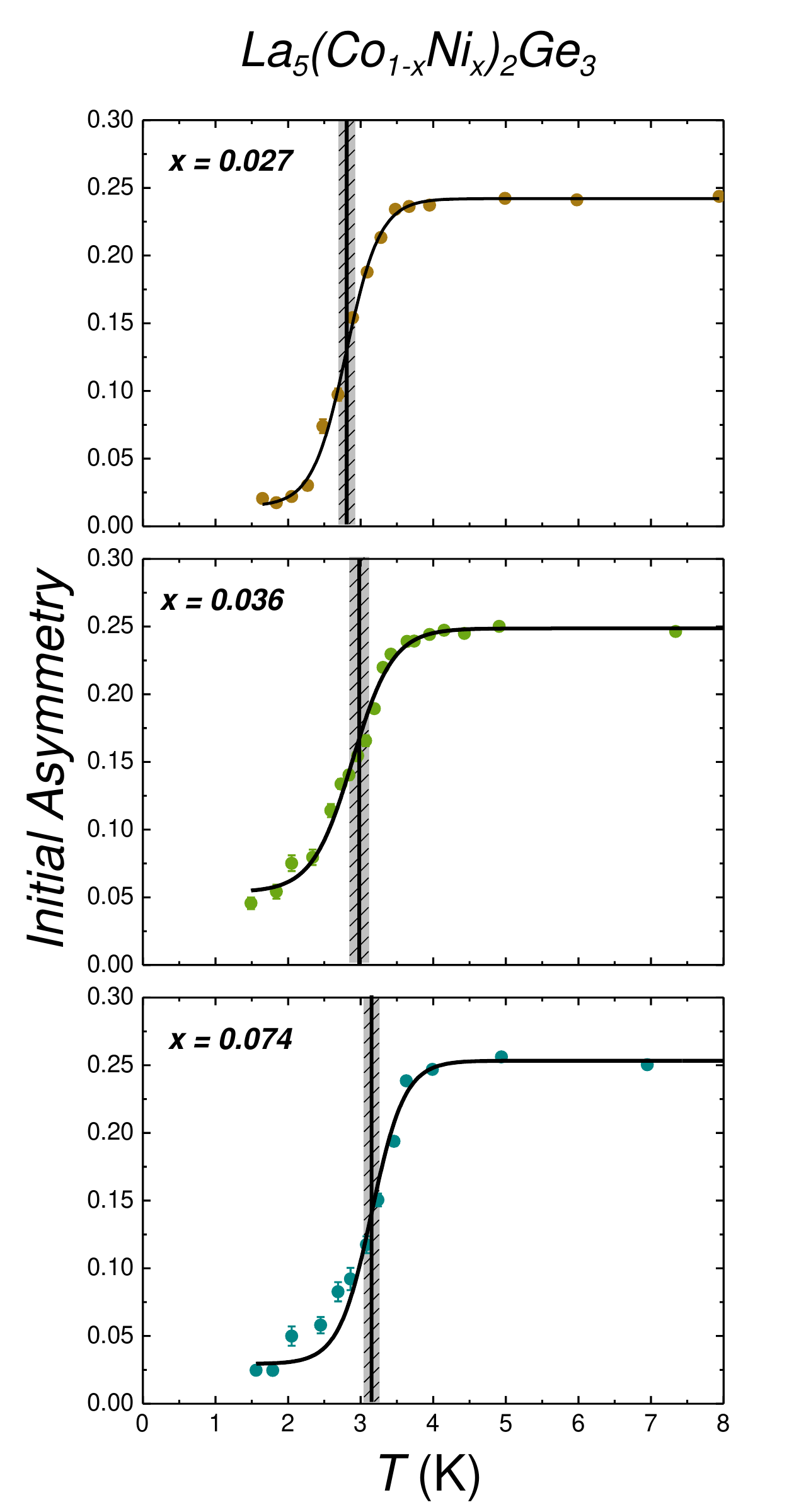}
    \caption{\footnotesize {Temperature dependence of the initial asymmetry A(0) obtained by fitting Eq.\ref{eqn:wtf} to the wTF muon spectra for x = 0.027, 0.036 and 0.074 in $\text{La}_{5}\text{(Co}_{1-x}\text {Ni}_{x})_2\text {Ge}_{3}$. The black solid line is the fit of Eq.\ref{eqn:T_A(0)} to A(0). The vertical black line in each panel is the $T_{mag}$ and the grey shaded area is the $\Delta T_{mag}$ obtained from the same fit. }}
    \label{fig:A0_T}
\end{figure}

The wTF data in the paramagnetic state can be analyzed using only the non-magnetic sample contribution to the $\mu$SR asymmetry. The magnetic contribution [$A_{m}(0)P_{m}(t)$] in the ordered state vanishes by $\sim 1 \mu s$ and thus we will also analyze the wTF ordered state data using the non-magnetic contribution for simplicity after avoiding the early time data. The non-magnetic wTF data is fit using; 
\begin{equation}
    \begin{split}
    A_{wTF}(t)& = A(0)P(t) \\
              & = A(0)cos(\gamma_{\mu} B_{ext} t~+~\phi)e^{-\sigma^{2}t^{2}/2}
    \end{split}
    \label{eqn:wtf}
\end{equation}

where, A(0) is the initial asymmetry, $\gamma_{\mu}/2 \pi$ = 135.5 MHz/T is the gyromagnetic ratio of the muons, $\phi$ the initial phase of the muon spin ensemble and $\sigma$ is the Gaussian relaxation rate caused by nuclear moments. The solid lines in Fig \ref{fig:wtf all} are fits to this equation. The black points for all the three substitutions represent data above the magnetic transition showing coherent muon precession whereas the red points are data taken with the sample in a magnetically ordered state. The fit lines for data taken in the magnetically ordered state (red line) were not fit to and do not pass through the early time data points (t $\sim$ 1 $\mu s$) as already discussed. \par

\begin{table*}[htbp]
  \centering
  \setlength{\tabcolsep}{10pt}
  \renewcommand{\arraystretch}{1.5}
    \begin{tabular}{r|rrrrrrr}
    \hline
    \hline
    \multicolumn{1}{c|}{$x$} & \multicolumn{1}{c}{$a (\AA)$} & \multicolumn{1}{c}{$b (\AA)$} & \multicolumn{1}{c}{$c (\AA)$} & \multicolumn{1}{c}{$\beta (\degree)$} & \multicolumn{1}{c}{$V (\AA)^{3}$} & \multicolumn{1}{c}{$wR$} & \multicolumn{1}{c}{$GOF$} \\
    \hline
    0.00 & 18.332(4) & 4.3421(2) & 13.256(3) & 103.99(1) & 1023.7(1) & 8.79  & 1.50 \\
    0.014(6) & 18.330(7) & 4.3409(4) & 13.254(5) & 103.96(2) & 1023.6(3) & 12.16 & 1.84 \\
    0.016(6) & 18.328(3) & 4.3421(2) & 13.260(2) & 103.99(1) & 1023.9(1) & 10.73 & 1.57 \\
    0.026(4) & 18.333(5) & 4.3409(3) & 13.253(3) & 104.27(1) & 1023.8(2) & 12.54 & 1.66 \\
    0.027(6) & 18.341(3) & 4.3417(2) & 13.258(2) & 103.96(1) & 1024.6(1) & 10.50 & 1.52 \\
    0.027(8) & 18.351(4) & 4.3415(3) & 13.253(3) & 103.89(2) & 1025.1(2) & 10.32 & 1.50 \\
    0.036(7) & 18.342(6) & 4.3411(4) & 13.249(4) & 103.95(2) & 1024.1(2) & 10.22 & 1.52 \\
    0.039(5) & 18.343(5) & 4.3417(3) & 13.253(3) & 103.94(2) & 1024.3(3) & 9.52  & 1.52 \\
    0.076(7) & 18.360(6) & 4.3386(4) & 13.259(4) &  103.93(3) & 1025.9(2) & 9.37  & 1.53 \\
    0.092(7) & 18.342(4) & 4.3385(3) & 13.258(3) & 103.99(2) & 1023.7(1) & 11.05 & 1.69 \\
    0.120(7) & 18.349(3) & 4.3389(2) & 13.265(2) & 104.21(1) & 1024.4(2) & 15.00 & 2.17 \\
    0.125(8) & 18.356(5) & 4.3378(3) & 13.255(3) &  103.92(2) & 1024.6(2) & 9.86  & 1.63 \\
    0.186(10) & 18.355(3) & 4.3364(2) & 13.262(2) &  104.05(1) & 1024.1(2) & 12.31 & 1.71 \\
    \hline
    \hline
    \end{tabular}%
    \caption[\footnotesize]{The values of the lattice parameters a,b,c, angle $\beta$ between a and c, and the volume of the unit cell V of the different x in $\text{La}_{5}\text{(Co}_{1-x}\text {Ni}_{x})_2\text {Ge}_{3}$ with the value of x indicated in the leftmost column. The lattice parameters are obtained from the refinement of the powder diffraction pattern. The wR and the GOF are the goodness of the fit obtained.}
  \label{tab:lattice}%
\end{table*}%

To study the change of the wTF muon data with Ni substitution, we track the temperature dependence of the initial time asymmetry A(0), inferred from Eq. \ref{eqn:wtf}, as a function of temperature for each x. Eq.\ref{eqn:T_A(0)} determines the magnetic ordering temperature, $T_{mag}$, and $\Delta T_{mag}$ is the width of the magnetic transition. 

\begin{equation}
    A(0,T) = A(0)\frac{1}{1+e^{([T_{mag}-T])/\Delta T_{mag})}}~+~A_{bg}(0)
    \label{eqn:T_A(0)}
\end{equation}

where, A(0) is the total initial asymmetry, $T_{mag}$ is the magnetic ordering temperature and $\Delta T_{mag}$ is the width of the magnetic transition. $A_{bg}(0)$ is the background contribution which represents the muons missing the sample and stopping in the holder, cryostat walls, etc.\par

Figure \ref{fig:A0_T} shows the temperature dependence of the initial asymmetry for the measured Ni substituted samples. The transition temperatures obtained from fitting the wTF data is similar to values obtained from the previously discussed measurements. However, no sign of two transitions for x = 0.027 is obtained from wTF $\mu$SR studies. The width of the transition for each of the substitutions is small suggesting magnetic order sets in uniformly in the samples. The value of $\Delta T_{mag}$ obtained from fitting Eq. \ref{eqn:T_A(0)} and is comparable to the widths of transitions we obtain from our resistance, magnetization and specific heat measurements.

\par

Table \ref{tab:lattice} shows the exact values of the Ni substitution levels obtained from EDS measurements on the single crystals of $\text{La}_{5}\text{(Co}_{1-x}\text {Ni}_{x})_2\text {Ge}_{3}$ along with lattice parameters a, b, c, the angle $\beta$, and volume of the unit cell obtained from the refinement of the powder X-Ray diffraction patterns of representative samples. The uncertainty associated with each parameter is also shown in parenthesises. 
\par

\bibliographystyle{apsrev4-1}
\bibliography{references.bib}

\begin{thebibliography}{61}%
\makeatletter
\providecommand \@ifxundefined [1]{%
 \@ifx{#1\undefined}
}%
\providecommand \@ifnum [1]{%
 \ifnum #1\expandafter \@firstoftwo
 \else \expandafter \@secondoftwo
 \fi
}%
\providecommand \@ifx [1]{%
 \ifx #1\expandafter \@firstoftwo
 \else \expandafter \@secondoftwo
 \fi
}%
\providecommand \natexlab [1]{#1}%
\providecommand \enquote  [1]{``#1''}%
\providecommand \bibnamefont  [1]{#1}%
\providecommand \bibfnamefont [1]{#1}%
\providecommand \citenamefont [1]{#1}%
\providecommand \href@noop [0]{\@secondoftwo}%
\providecommand \href [0]{\begingroup \@sanitize@url \@href}%
\providecommand \@href[1]{\@@startlink{#1}\@@href}%
\providecommand \@@href[1]{\endgroup#1\@@endlink}%
\providecommand \@sanitize@url [0]{\catcode `\\12\catcode `\$12\catcode `\&12\catcode `\#12\catcode `\^12\catcode `\_12\catcode `\%12\relax}%
\providecommand \@@startlink[1]{}%
\providecommand \@@endlink[0]{}%
\providecommand \url  [0]{\begingroup\@sanitize@url \@url }%
\providecommand \@url [1]{\endgroup\@href {#1}{\urlprefix }}%
\providecommand \urlprefix  [0]{URL }%
\providecommand \Eprint [0]{\href }%
\providecommand \doibase [0]{http://dx.doi.org/}%
\providecommand \selectlanguage [0]{\@gobble}%
\providecommand \bibinfo  [0]{\@secondoftwo}%
\providecommand \bibfield  [0]{\@secondoftwo}%
\providecommand \translation [1]{[#1]}%
\providecommand \BibitemOpen [0]{}%
\providecommand \bibitemStop [0]{}%
\providecommand \bibitemNoStop [0]{.\EOS\space}%
\providecommand \EOS [0]{\spacefactor3000\relax}%
\providecommand \BibitemShut  [1]{\csname bibitem#1\endcsname}%
\let\auto@bib@innerbib\@empty
\bibitem [{\citenamefont {Canfield}\ and\ \citenamefont {Bud'ko}(2016)}]{Canfield2016PreservedMagnetism}%
  \BibitemOpen
  \bibfield  {author} {\bibinfo {author} {\bibfnamefont {P.~C.}\ \bibnamefont {Canfield}}\ and\ \bibinfo {author} {\bibfnamefont {S.~L.}\ \bibnamefont {Bud'ko}},\ }\href {\doibase 10.1088/0034-4885/79/8/084506} {\bibfield  {journal} {\bibinfo  {journal} {Reports on Progress in Physics}\ }\textbf {\bibinfo {volume} {79}},\ \bibinfo {pages} {084506} (\bibinfo {year} {2016})}\BibitemShut {NoStop}%
\bibitem [{\citenamefont {Brando}\ \emph {et~al.}(2016)\citenamefont {Brando}, \citenamefont {Belitz}, \citenamefont {Grosche},\ and\ \citenamefont {Kirkpatrick}}]{Brando2016MetallicFerromagnets}%
  \BibitemOpen
  \bibfield  {author} {\bibinfo {author} {\bibfnamefont {M.}~\bibnamefont {Brando}}, \bibinfo {author} {\bibfnamefont {D.}~\bibnamefont {Belitz}}, \bibinfo {author} {\bibfnamefont {F.~M.}\ \bibnamefont {Grosche}}, \ and\ \bibinfo {author} {\bibfnamefont {T.~R.}\ \bibnamefont {Kirkpatrick}},\ }\href {\doibase 10.1103/RevModPhys.88.025006} {\bibfield  {journal} {\bibinfo  {journal} {Reviews of Modern Physics}\ }\textbf {\bibinfo {volume} {88}},\ \bibinfo {pages} {025006} (\bibinfo {year} {2016})}\BibitemShut {NoStop}%
\bibitem [{\citenamefont {Pfleiderer}\ \emph {et~al.}(2001{\natexlab{a}})\citenamefont {Pfleiderer}, \citenamefont {Julian},\ and\ \citenamefont {Lonzarich}}]{Pfleiderer2001Non-Fermi-liquidFerromagnets}%
  \BibitemOpen
  \bibfield  {author} {\bibinfo {author} {\bibfnamefont {C.}~\bibnamefont {Pfleiderer}}, \bibinfo {author} {\bibfnamefont {S.~R.}\ \bibnamefont {Julian}}, \ and\ \bibinfo {author} {\bibfnamefont {G.~G.}\ \bibnamefont {Lonzarich}},\ }\href {\doibase 10.1038/35106527} {\bibfield  {journal} {\bibinfo  {journal} {Nature}\ }\textbf {\bibinfo {volume} {414}},\ \bibinfo {pages} {427} (\bibinfo {year} {2001}{\natexlab{a}})}\BibitemShut {NoStop}%
\bibitem [{\citenamefont {Huy}\ \emph {et~al.}(2007)\citenamefont {Huy}, \citenamefont {Gasparini}, \citenamefont {de~Nijs}, \citenamefont {Huang}, \citenamefont {Klaasse}, \citenamefont {Gortenmulder}, \citenamefont {de~Visser}, \citenamefont {Hamann}, \citenamefont {G{\"{o}}rlach},\ and\ \citenamefont {L{\"{o}}hneysen}}]{Huy2007SuperconductivityUCoGe}%
  \BibitemOpen
  \bibfield  {author} {\bibinfo {author} {\bibfnamefont {N.~T.}\ \bibnamefont {Huy}}, \bibinfo {author} {\bibfnamefont {A.}~\bibnamefont {Gasparini}}, \bibinfo {author} {\bibfnamefont {D.~E.}\ \bibnamefont {de~Nijs}}, \bibinfo {author} {\bibfnamefont {Y.}~\bibnamefont {Huang}}, \bibinfo {author} {\bibfnamefont {J.~C.~P.}\ \bibnamefont {Klaasse}}, \bibinfo {author} {\bibfnamefont {T.}~\bibnamefont {Gortenmulder}}, \bibinfo {author} {\bibfnamefont {A.}~\bibnamefont {de~Visser}}, \bibinfo {author} {\bibfnamefont {A.}~\bibnamefont {Hamann}}, \bibinfo {author} {\bibfnamefont {T.}~\bibnamefont {G{\"{o}}rlach}}, \ and\ \bibinfo {author} {\bibfnamefont {H.~v.}\ \bibnamefont {L{\"{o}}hneysen}},\ }\href {\doibase 10.1103/PhysRevLett.99.067006} {\bibfield  {journal} {\bibinfo  {journal} {Physical Review Letters}\ }\textbf {\bibinfo {volume} {99}},\ \bibinfo {pages} {067006} (\bibinfo {year} {2007})}\BibitemShut {NoStop}%
\bibitem [{\citenamefont {Dagotto}(1994)}]{Dagotto1994}%
  \BibitemOpen
  \bibfield  {author} {\bibinfo {author} {\bibfnamefont {E.}~\bibnamefont {Dagotto}},\ }\href {\doibase 10.1103/RevModPhys.66.763} {\bibfield  {journal} {\bibinfo  {journal} {Reviews of Modern Physics}\ }\textbf {\bibinfo {volume} {66}},\ \bibinfo {pages} {763} (\bibinfo {year} {1994})}\BibitemShut {NoStop}%
\bibitem [{\citenamefont {Paglione}\ and\ \citenamefont {Greene}(2010)}]{Paglione2010High-temperatureMaterials}%
  \BibitemOpen
  \bibfield  {author} {\bibinfo {author} {\bibfnamefont {J.}~\bibnamefont {Paglione}}\ and\ \bibinfo {author} {\bibfnamefont {R.~L.}\ \bibnamefont {Greene}},\ }\href {\doibase 10.1038/nphys1759} {\bibfield  {journal} {\bibinfo  {journal} {Nature Physics}\ }\textbf {\bibinfo {volume} {6}},\ \bibinfo {pages} {645} (\bibinfo {year} {2010})}\BibitemShut {NoStop}%
\bibitem [{\citenamefont {Pfleiderer}\ \emph {et~al.}(2001{\natexlab{b}})\citenamefont {Pfleiderer}, \citenamefont {Uhlarz}, \citenamefont {Hayden}, \citenamefont {Vollmer}, \citenamefont {L{\"{o}}hneysen}, \citenamefont {Bernhoeft},\ and\ \citenamefont {Lonzarich}}]{Pfleiderer2001}%
  \BibitemOpen
  \bibfield  {author} {\bibinfo {author} {\bibfnamefont {C.}~\bibnamefont {Pfleiderer}}, \bibinfo {author} {\bibfnamefont {M.}~\bibnamefont {Uhlarz}}, \bibinfo {author} {\bibfnamefont {S.~M.}\ \bibnamefont {Hayden}}, \bibinfo {author} {\bibfnamefont {R.}~\bibnamefont {Vollmer}}, \bibinfo {author} {\bibfnamefont {H.~v.}\ \bibnamefont {L{\"{o}}hneysen}}, \bibinfo {author} {\bibfnamefont {N.~R.}\ \bibnamefont {Bernhoeft}}, \ and\ \bibinfo {author} {\bibfnamefont {G.~G.}\ \bibnamefont {Lonzarich}},\ }\href {\doibase 10.1038/35083531} {\bibfield  {journal} {\bibinfo  {journal} {Nature}\ }\textbf {\bibinfo {volume} {412}},\ \bibinfo {pages} {58} (\bibinfo {year} {2001}{\natexlab{b}})}\BibitemShut {NoStop}%
\bibitem [{\citenamefont {Gegenwart}\ \emph {et~al.}(2008)\citenamefont {Gegenwart}, \citenamefont {Si},\ and\ \citenamefont {Steglich}}]{Gegenwart2008QuantumMetals}%
  \BibitemOpen
  \bibfield  {author} {\bibinfo {author} {\bibfnamefont {P.}~\bibnamefont {Gegenwart}}, \bibinfo {author} {\bibfnamefont {Q.}~\bibnamefont {Si}}, \ and\ \bibinfo {author} {\bibfnamefont {F.}~\bibnamefont {Steglich}},\ }\href@noop {} {\bibfield  {journal} {\bibinfo  {journal} {Nature Physics}\ }\textbf {\bibinfo {volume} {4}},\ \bibinfo {pages} {186} (\bibinfo {year} {2008})}\BibitemShut {NoStop}%
\bibitem [{\citenamefont {L{\'{e}}vy}\ \emph {et~al.}(2007)\citenamefont {L{\'{e}}vy}, \citenamefont {Sheikin},\ and\ \citenamefont {Huxley}}]{Levy2007AcuteURhGe}%
  \BibitemOpen
  \bibfield  {author} {\bibinfo {author} {\bibfnamefont {F.}~\bibnamefont {L{\'{e}}vy}}, \bibinfo {author} {\bibfnamefont {I.}~\bibnamefont {Sheikin}}, \ and\ \bibinfo {author} {\bibfnamefont {A.}~\bibnamefont {Huxley}},\ }\href {\doibase 10.1038/nphys608} {\bibfield  {journal} {\bibinfo  {journal} {Nature Physics}\ }\textbf {\bibinfo {volume} {3}},\ \bibinfo {pages} {460} (\bibinfo {year} {2007})}\BibitemShut {NoStop}%
\bibitem [{\citenamefont {Uemura}\ \emph {et~al.}(2007)\citenamefont {Uemura}, \citenamefont {Goko}, \citenamefont {Gat-Malureanu}, \citenamefont {Carlo}, \citenamefont {Russo}, \citenamefont {Savici}, \citenamefont {Aczel}, \citenamefont {MacDougall}, \citenamefont {Rodriguez}, \citenamefont {Luke}, \citenamefont {Dunsiger}, \citenamefont {McCollam}, \citenamefont {Arai}, \citenamefont {Pfleiderer}, \citenamefont {B{\"{o}}ni}, \citenamefont {Yoshimura}, \citenamefont {Baggio-Saitovitch}, \citenamefont {Fontes}, \citenamefont {Larrea}, \citenamefont {Sushko},\ and\ \citenamefont {Sereni}}]{Uemura2007PhaseSr1xCaxRuO3}%
  \BibitemOpen
  \bibfield  {author} {\bibinfo {author} {\bibfnamefont {Y.~J.}\ \bibnamefont {Uemura}}, \bibinfo {author} {\bibfnamefont {T.}~\bibnamefont {Goko}}, \bibinfo {author} {\bibfnamefont {I.~M.}\ \bibnamefont {Gat-Malureanu}}, \bibinfo {author} {\bibfnamefont {J.~P.}\ \bibnamefont {Carlo}}, \bibinfo {author} {\bibfnamefont {P.~L.}\ \bibnamefont {Russo}}, \bibinfo {author} {\bibfnamefont {A.~T.}\ \bibnamefont {Savici}}, \bibinfo {author} {\bibfnamefont {A.}~\bibnamefont {Aczel}}, \bibinfo {author} {\bibfnamefont {G.~J.}\ \bibnamefont {MacDougall}}, \bibinfo {author} {\bibfnamefont {J.~A.}\ \bibnamefont {Rodriguez}}, \bibinfo {author} {\bibfnamefont {G.~M.}\ \bibnamefont {Luke}}, \bibinfo {author} {\bibfnamefont {S.~R.}\ \bibnamefont {Dunsiger}}, \bibinfo {author} {\bibfnamefont {A.}~\bibnamefont {McCollam}}, \bibinfo {author} {\bibfnamefont {J.}~\bibnamefont {Arai}}, \bibinfo {author} {\bibfnamefont {C.}~\bibnamefont {Pfleiderer}}, \bibinfo {author} {\bibfnamefont {P.}~\bibnamefont {B{\"{o}}ni}}, \bibinfo {author}
  {\bibfnamefont {K.}~\bibnamefont {Yoshimura}}, \bibinfo {author} {\bibfnamefont {E.}~\bibnamefont {Baggio-Saitovitch}}, \bibinfo {author} {\bibfnamefont {M.~B.}\ \bibnamefont {Fontes}}, \bibinfo {author} {\bibfnamefont {J.}~\bibnamefont {Larrea}}, \bibinfo {author} {\bibfnamefont {Y.~V.}\ \bibnamefont {Sushko}}, \ and\ \bibinfo {author} {\bibfnamefont {J.}~\bibnamefont {Sereni}},\ }\href {\doibase 10.1038/nphys488} {\bibfield  {journal} {\bibinfo  {journal} {Nature Physics}\ }\textbf {\bibinfo {volume} {3}},\ \bibinfo {pages} {29} (\bibinfo {year} {2007})}\BibitemShut {NoStop}%
\bibitem [{\citenamefont {Pfleiderer}\ \emph {et~al.}(2004)\citenamefont {Pfleiderer}, \citenamefont {Reznik}, \citenamefont {Pintschovius}, \citenamefont {L{\"{o}}hneysen}, \citenamefont {Garst},\ and\ \citenamefont {Rosch}}]{Pfleiderer2004PartialMnSi}%
  \BibitemOpen
  \bibfield  {author} {\bibinfo {author} {\bibfnamefont {C.}~\bibnamefont {Pfleiderer}}, \bibinfo {author} {\bibfnamefont {D.}~\bibnamefont {Reznik}}, \bibinfo {author} {\bibfnamefont {L.}~\bibnamefont {Pintschovius}}, \bibinfo {author} {\bibfnamefont {H.~v.}\ \bibnamefont {L{\"{o}}hneysen}}, \bibinfo {author} {\bibfnamefont {M.}~\bibnamefont {Garst}}, \ and\ \bibinfo {author} {\bibfnamefont {A.}~\bibnamefont {Rosch}},\ }\href {\doibase 10.1038/nature02232} {\bibfield  {journal} {\bibinfo  {journal} {Nature}\ }\textbf {\bibinfo {volume} {427}},\ \bibinfo {pages} {227} (\bibinfo {year} {2004})}\BibitemShut {NoStop}%
\bibitem [{\citenamefont {Ubaid-Kassis}\ \emph {et~al.}(2010)\citenamefont {Ubaid-Kassis}, \citenamefont {Vojta},\ and\ \citenamefont {Schroeder}}]{Ubaid-Kassis2010Quantum/math}%
  \BibitemOpen
  \bibfield  {author} {\bibinfo {author} {\bibfnamefont {S.}~\bibnamefont {Ubaid-Kassis}}, \bibinfo {author} {\bibfnamefont {T.}~\bibnamefont {Vojta}}, \ and\ \bibinfo {author} {\bibfnamefont {A.}~\bibnamefont {Schroeder}},\ }\href {\doibase 10.1103/PhysRevLett.104.066402} {\bibfield  {journal} {\bibinfo  {journal} {Physical Review Letters}\ }\textbf {\bibinfo {volume} {104}},\ \bibinfo {pages} {066402} (\bibinfo {year} {2010})}\BibitemShut {NoStop}%
\bibitem [{\citenamefont {Pfleiderer}(2009)}]{Pfleiderer2009SuperconductingCompounds}%
  \BibitemOpen
  \bibfield  {author} {\bibinfo {author} {\bibfnamefont {C.}~\bibnamefont {Pfleiderer}},\ }\href {\doibase 10.1103/RevModPhys.81.1551} {\bibfield  {journal} {\bibinfo  {journal} {Reviews of Modern Physics}\ }\textbf {\bibinfo {volume} {81}},\ \bibinfo {pages} {1551} (\bibinfo {year} {2009})}\BibitemShut {NoStop}%
\bibitem [{\citenamefont {Saxena}\ \emph {et~al.}(2000)\citenamefont {Saxena}, \citenamefont {Agarwal}, \citenamefont {Ahilan}, \citenamefont {Grosche}, \citenamefont {Haselwimmer}, \citenamefont {Steiner}, \citenamefont {Pugh}, \citenamefont {Walker}, \citenamefont {Julian}, \citenamefont {Monthoux}, \citenamefont {Lonzarich}, \citenamefont {Huxley}, \citenamefont {Sheikin}, \citenamefont {Braithwaite},\ and\ \citenamefont {Flouquet}}]{Saxena2000SuperconductivityUGe2}%
  \BibitemOpen
  \bibfield  {author} {\bibinfo {author} {\bibfnamefont {S.~S.}\ \bibnamefont {Saxena}}, \bibinfo {author} {\bibfnamefont {P.}~\bibnamefont {Agarwal}}, \bibinfo {author} {\bibfnamefont {K.}~\bibnamefont {Ahilan}}, \bibinfo {author} {\bibfnamefont {F.~M.}\ \bibnamefont {Grosche}}, \bibinfo {author} {\bibfnamefont {R.~K.~W.}\ \bibnamefont {Haselwimmer}}, \bibinfo {author} {\bibfnamefont {M.~J.}\ \bibnamefont {Steiner}}, \bibinfo {author} {\bibfnamefont {E.}~\bibnamefont {Pugh}}, \bibinfo {author} {\bibfnamefont {I.~R.}\ \bibnamefont {Walker}}, \bibinfo {author} {\bibfnamefont {S.~R.}\ \bibnamefont {Julian}}, \bibinfo {author} {\bibfnamefont {P.}~\bibnamefont {Monthoux}}, \bibinfo {author} {\bibfnamefont {G.~G.}\ \bibnamefont {Lonzarich}}, \bibinfo {author} {\bibfnamefont {A.}~\bibnamefont {Huxley}}, \bibinfo {author} {\bibfnamefont {I.}~\bibnamefont {Sheikin}}, \bibinfo {author} {\bibfnamefont {D.}~\bibnamefont {Braithwaite}}, \ and\ \bibinfo {author} {\bibfnamefont {J.}~\bibnamefont {Flouquet}},\ }\href
  {\doibase 10.1038/35020500} {\bibfield  {journal} {\bibinfo  {journal} {Nature}\ }\textbf {\bibinfo {volume} {406}},\ \bibinfo {pages} {587} (\bibinfo {year} {2000})}\BibitemShut {NoStop}%
\bibitem [{\citenamefont {Aoki}\ \emph {et~al.}(2001)\citenamefont {Aoki}, \citenamefont {Huxley}, \citenamefont {Ressouche}, \citenamefont {Braithwaite}, \citenamefont {Flouquet}, \citenamefont {Brison}, \citenamefont {Lhotel},\ and\ \citenamefont {Paulsen}}]{Aoki2001CoexistenceURhGe}%
  \BibitemOpen
  \bibfield  {author} {\bibinfo {author} {\bibfnamefont {D.}~\bibnamefont {Aoki}}, \bibinfo {author} {\bibfnamefont {A.}~\bibnamefont {Huxley}}, \bibinfo {author} {\bibfnamefont {E.}~\bibnamefont {Ressouche}}, \bibinfo {author} {\bibfnamefont {D.}~\bibnamefont {Braithwaite}}, \bibinfo {author} {\bibfnamefont {J.}~\bibnamefont {Flouquet}}, \bibinfo {author} {\bibfnamefont {J.-P.}\ \bibnamefont {Brison}}, \bibinfo {author} {\bibfnamefont {E.}~\bibnamefont {Lhotel}}, \ and\ \bibinfo {author} {\bibfnamefont {C.}~\bibnamefont {Paulsen}},\ }\href {\doibase 10.1038/35098048} {\bibfield  {journal} {\bibinfo  {journal} {Nature}\ }\textbf {\bibinfo {volume} {413}},\ \bibinfo {pages} {613} (\bibinfo {year} {2001})}\BibitemShut {NoStop}%
\bibitem [{\citenamefont {Cheng}\ \emph {et~al.}(2015)\citenamefont {Cheng}, \citenamefont {Matsubayashi}, \citenamefont {Wu}, \citenamefont {Sun}, \citenamefont {Lin}, \citenamefont {Luo},\ and\ \citenamefont {Uwatoko}}]{Cheng2015PressureMnP}%
  \BibitemOpen
  \bibfield  {author} {\bibinfo {author} {\bibfnamefont {J.-G.}\ \bibnamefont {Cheng}}, \bibinfo {author} {\bibfnamefont {K.}~\bibnamefont {Matsubayashi}}, \bibinfo {author} {\bibfnamefont {W.}~\bibnamefont {Wu}}, \bibinfo {author} {\bibfnamefont {J.}~\bibnamefont {Sun}}, \bibinfo {author} {\bibfnamefont {F.}~\bibnamefont {Lin}}, \bibinfo {author} {\bibfnamefont {J.}~\bibnamefont {Luo}}, \ and\ \bibinfo {author} {\bibfnamefont {Y.}~\bibnamefont {Uwatoko}},\ }\href {\doibase 10.1103/PhysRevLett.114.117001} {\bibfield  {journal} {\bibinfo  {journal} {Physical Review Letters}\ }\textbf {\bibinfo {volume} {114}},\ \bibinfo {pages} {117001} (\bibinfo {year} {2015})}\BibitemShut {NoStop}%
\bibitem [{\citenamefont {Ran}\ \emph {et~al.}(2019)\citenamefont {Ran}, \citenamefont {Eckberg}, \citenamefont {Ding}, \citenamefont {Furukawa}, \citenamefont {Metz}, \citenamefont {Saha}, \citenamefont {Liu}, \citenamefont {Zic}, \citenamefont {Kim}, \citenamefont {Paglione},\ and\ \citenamefont {Butch}}]{Ran2019NearlySuperconductivity}%
  \BibitemOpen
  \bibfield  {author} {\bibinfo {author} {\bibfnamefont {S.}~\bibnamefont {Ran}}, \bibinfo {author} {\bibfnamefont {C.}~\bibnamefont {Eckberg}}, \bibinfo {author} {\bibfnamefont {Q.-P.}\ \bibnamefont {Ding}}, \bibinfo {author} {\bibfnamefont {Y.}~\bibnamefont {Furukawa}}, \bibinfo {author} {\bibfnamefont {T.}~\bibnamefont {Metz}}, \bibinfo {author} {\bibfnamefont {S.~R.}\ \bibnamefont {Saha}}, \bibinfo {author} {\bibfnamefont {I.-L.}\ \bibnamefont {Liu}}, \bibinfo {author} {\bibfnamefont {M.}~\bibnamefont {Zic}}, \bibinfo {author} {\bibfnamefont {H.}~\bibnamefont {Kim}}, \bibinfo {author} {\bibfnamefont {J.}~\bibnamefont {Paglione}}, \ and\ \bibinfo {author} {\bibfnamefont {N.~P.}\ \bibnamefont {Butch}},\ }\href {\doibase 10.1126/science.aav8645} {\bibfield  {journal} {\bibinfo  {journal} {Science}\ }\textbf {\bibinfo {volume} {365}},\ \bibinfo {pages} {684} (\bibinfo {year} {2019})}\BibitemShut {NoStop}%
\bibitem [{\citenamefont {Shibauchi}\ \emph {et~al.}(2014)\citenamefont {Shibauchi}, \citenamefont {Carrington},\ and\ \citenamefont {Matsuda}}]{Shibauchi2014APnictides}%
  \BibitemOpen
  \bibfield  {author} {\bibinfo {author} {\bibfnamefont {T.}~\bibnamefont {Shibauchi}}, \bibinfo {author} {\bibfnamefont {A.}~\bibnamefont {Carrington}}, \ and\ \bibinfo {author} {\bibfnamefont {Y.}~\bibnamefont {Matsuda}},\ }\href {\doibase 10.1146/annurev-conmatphys-031113-133921} {\bibfield  {journal} {\bibinfo  {journal} {Annual Review of Condensed Matter Physics}\ }\textbf {\bibinfo {volume} {5}},\ \bibinfo {pages} {113} (\bibinfo {year} {2014})}\BibitemShut {NoStop}%
\bibitem [{\citenamefont {L{\"{o}}hneysen}\ \emph {et~al.}(1994)\citenamefont {L{\"{o}}hneysen}, \citenamefont {Pietrus}, \citenamefont {Portisch}, \citenamefont {Schlager}, \citenamefont {Schr{\"{o}}der}, \citenamefont {Sieck},\ and\ \citenamefont {Trappmann}}]{Lohneysen1994Non-Fermi-liquidInstabilityb}%
  \BibitemOpen
  \bibfield  {author} {\bibinfo {author} {\bibfnamefont {H.~v.}\ \bibnamefont {L{\"{o}}hneysen}}, \bibinfo {author} {\bibfnamefont {T.}~\bibnamefont {Pietrus}}, \bibinfo {author} {\bibfnamefont {G.}~\bibnamefont {Portisch}}, \bibinfo {author} {\bibfnamefont {H.~G.}\ \bibnamefont {Schlager}}, \bibinfo {author} {\bibfnamefont {A.}~\bibnamefont {Schr{\"{o}}der}}, \bibinfo {author} {\bibfnamefont {M.}~\bibnamefont {Sieck}}, \ and\ \bibinfo {author} {\bibfnamefont {T.}~\bibnamefont {Trappmann}},\ }\href {\doibase 10.1103/PhysRevLett.72.3262} {\bibfield  {journal} {\bibinfo  {journal} {Physical Review Letters}\ }\textbf {\bibinfo {volume} {72}},\ \bibinfo {pages} {3262} (\bibinfo {year} {1994})}\BibitemShut {NoStop}%
\bibitem [{\citenamefont {Friedemann}\ \emph {et~al.}(2009)\citenamefont {Friedemann}, \citenamefont {Westerkamp}, \citenamefont {Brando}, \citenamefont {Oeschler}, \citenamefont {Wirth}, \citenamefont {Gegenwart}, \citenamefont {Krellner}, \citenamefont {Geibel},\ and\ \citenamefont {Steglich}}]{Friedemann2009DetachingInYbRh2Si2}%
  \BibitemOpen
  \bibfield  {author} {\bibinfo {author} {\bibfnamefont {S.}~\bibnamefont {Friedemann}}, \bibinfo {author} {\bibfnamefont {T.}~\bibnamefont {Westerkamp}}, \bibinfo {author} {\bibfnamefont {M.}~\bibnamefont {Brando}}, \bibinfo {author} {\bibfnamefont {N.}~\bibnamefont {Oeschler}}, \bibinfo {author} {\bibfnamefont {S.}~\bibnamefont {Wirth}}, \bibinfo {author} {\bibfnamefont {P.}~\bibnamefont {Gegenwart}}, \bibinfo {author} {\bibfnamefont {C.}~\bibnamefont {Krellner}}, \bibinfo {author} {\bibfnamefont {C.}~\bibnamefont {Geibel}}, \ and\ \bibinfo {author} {\bibfnamefont {F.}~\bibnamefont {Steglich}},\ }\href {\doibase 10.1038/nphys1299} {\bibfield  {journal} {\bibinfo  {journal} {Nature Physics}\ }\textbf {\bibinfo {volume} {5}},\ \bibinfo {pages} {465} (\bibinfo {year} {2009})}\BibitemShut {NoStop}%
\bibitem [{\citenamefont {Schmiedeshoff}\ \emph {et~al.}(2011)\citenamefont {Schmiedeshoff}, \citenamefont {Mun}, \citenamefont {Lounsbury}, \citenamefont {Tracy}, \citenamefont {Palm}, \citenamefont {Hannahs}, \citenamefont {Park}, \citenamefont {Murphy}, \citenamefont {Bud’ko},\ and\ \citenamefont {Canfield}}]{Schmiedeshoff2011MultipleYbAgGe}%
  \BibitemOpen
  \bibfield  {author} {\bibinfo {author} {\bibfnamefont {G.~M.}\ \bibnamefont {Schmiedeshoff}}, \bibinfo {author} {\bibfnamefont {E.~D.}\ \bibnamefont {Mun}}, \bibinfo {author} {\bibfnamefont {A.~W.}\ \bibnamefont {Lounsbury}}, \bibinfo {author} {\bibfnamefont {S.~J.}\ \bibnamefont {Tracy}}, \bibinfo {author} {\bibfnamefont {E.~C.}\ \bibnamefont {Palm}}, \bibinfo {author} {\bibfnamefont {S.~T.}\ \bibnamefont {Hannahs}}, \bibinfo {author} {\bibfnamefont {J.-H.}\ \bibnamefont {Park}}, \bibinfo {author} {\bibfnamefont {T.~P.}\ \bibnamefont {Murphy}}, \bibinfo {author} {\bibfnamefont {S.~L.}\ \bibnamefont {Bud’ko}}, \ and\ \bibinfo {author} {\bibfnamefont {P.~C.}\ \bibnamefont {Canfield}},\ }\href {\doibase 10.1103/PhysRevB.83.180408} {\bibfield  {journal} {\bibinfo  {journal} {Physical Review B}\ }\textbf {\bibinfo {volume} {83}},\ \bibinfo {pages} {180408} (\bibinfo {year} {2011})}\BibitemShut {NoStop}%
\bibitem [{\citenamefont {Tokiwa}\ \emph {et~al.}(2013)\citenamefont {Tokiwa}, \citenamefont {Garst}, \citenamefont {Gegenwart}, \citenamefont {Bud’ko},\ and\ \citenamefont {Canfield}}]{Tokiwa2013QuantumYbAgGe}%
  \BibitemOpen
  \bibfield  {author} {\bibinfo {author} {\bibfnamefont {Y.}~\bibnamefont {Tokiwa}}, \bibinfo {author} {\bibfnamefont {M.}~\bibnamefont {Garst}}, \bibinfo {author} {\bibfnamefont {P.}~\bibnamefont {Gegenwart}}, \bibinfo {author} {\bibfnamefont {S.~L.}\ \bibnamefont {Bud’ko}}, \ and\ \bibinfo {author} {\bibfnamefont {P.~C.}\ \bibnamefont {Canfield}},\ }\href {\doibase 10.1103/PhysRevLett.111.116401} {\bibfield  {journal} {\bibinfo  {journal} {Physical Review Letters}\ }\textbf {\bibinfo {volume} {111}},\ \bibinfo {pages} {116401} (\bibinfo {year} {2013})}\BibitemShut {NoStop}%
\bibitem [{\citenamefont {Mun}\ \emph {et~al.}(2013)\citenamefont {Mun}, \citenamefont {Bud'ko}, \citenamefont {Martin}, \citenamefont {Kim}, \citenamefont {Tanatar}, \citenamefont {Park}, \citenamefont {Murphy}, \citenamefont {Schmiedeshoff}, \citenamefont {Dilley}, \citenamefont {Prozorov},\ and\ \citenamefont {Canfield}}]{Mun2013Magnetic-field-tunedYbPtBi}%
  \BibitemOpen
  \bibfield  {author} {\bibinfo {author} {\bibfnamefont {E.~D.}\ \bibnamefont {Mun}}, \bibinfo {author} {\bibfnamefont {S.~L.}\ \bibnamefont {Bud'ko}}, \bibinfo {author} {\bibfnamefont {C.}~\bibnamefont {Martin}}, \bibinfo {author} {\bibfnamefont {H.}~\bibnamefont {Kim}}, \bibinfo {author} {\bibfnamefont {M.~A.}\ \bibnamefont {Tanatar}}, \bibinfo {author} {\bibfnamefont {J.-H.}\ \bibnamefont {Park}}, \bibinfo {author} {\bibfnamefont {T.}~\bibnamefont {Murphy}}, \bibinfo {author} {\bibfnamefont {G.~M.}\ \bibnamefont {Schmiedeshoff}}, \bibinfo {author} {\bibfnamefont {N.}~\bibnamefont {Dilley}}, \bibinfo {author} {\bibfnamefont {R.}~\bibnamefont {Prozorov}}, \ and\ \bibinfo {author} {\bibfnamefont {P.~C.}\ \bibnamefont {Canfield}},\ }\href {\doibase 10.1103/PhysRevB.87.075120} {\bibfield  {journal} {\bibinfo  {journal} {Physical Review B}\ }\textbf {\bibinfo {volume} {87}},\ \bibinfo {pages} {075120} (\bibinfo {year} {2013})}\BibitemShut {NoStop}%
\bibitem [{\citenamefont {Nandi}\ \emph {et~al.}(2010)\citenamefont {Nandi}, \citenamefont {Kim}, \citenamefont {Kreyssig}, \citenamefont {Fernandes}, \citenamefont {Pratt}, \citenamefont {Thaler}, \citenamefont {Ni}, \citenamefont {Bud’ko}, \citenamefont {Canfield}, \citenamefont {Schmalian}, \citenamefont {McQueeney},\ and\ \citenamefont {Goldman}}]{Nandi2010AnomalousCrystals}%
  \BibitemOpen
  \bibfield  {author} {\bibinfo {author} {\bibfnamefont {S.}~\bibnamefont {Nandi}}, \bibinfo {author} {\bibfnamefont {M.~G.}\ \bibnamefont {Kim}}, \bibinfo {author} {\bibfnamefont {A.}~\bibnamefont {Kreyssig}}, \bibinfo {author} {\bibfnamefont {R.~M.}\ \bibnamefont {Fernandes}}, \bibinfo {author} {\bibfnamefont {D.~K.}\ \bibnamefont {Pratt}}, \bibinfo {author} {\bibfnamefont {A.}~\bibnamefont {Thaler}}, \bibinfo {author} {\bibfnamefont {N.}~\bibnamefont {Ni}}, \bibinfo {author} {\bibfnamefont {S.~L.}\ \bibnamefont {Bud’ko}}, \bibinfo {author} {\bibfnamefont {P.~C.}\ \bibnamefont {Canfield}}, \bibinfo {author} {\bibfnamefont {J.}~\bibnamefont {Schmalian}}, \bibinfo {author} {\bibfnamefont {R.~J.}\ \bibnamefont {McQueeney}}, \ and\ \bibinfo {author} {\bibfnamefont {A.~I.}\ \bibnamefont {Goldman}},\ }\href {\doibase 10.1103/PhysRevLett.104.057006} {\bibfield  {journal} {\bibinfo  {journal} {Physical Review Letters}\ }\textbf {\bibinfo {volume} {104}},\ \bibinfo {pages} {057006} (\bibinfo {year}
  {2010})}\BibitemShut {NoStop}%
\bibitem [{\citenamefont {Belitz}\ \emph {et~al.}(1999)\citenamefont {Belitz}, \citenamefont {Kirkpatrick},\ and\ \citenamefont {Vojta}}]{Belitz1999FirstFerromagnets}%
  \BibitemOpen
  \bibfield  {author} {\bibinfo {author} {\bibfnamefont {D.}~\bibnamefont {Belitz}}, \bibinfo {author} {\bibfnamefont {T.~R.}\ \bibnamefont {Kirkpatrick}}, \ and\ \bibinfo {author} {\bibfnamefont {T.}~\bibnamefont {Vojta}},\ }\href {\doibase 10.1103/PhysRevLett.82.4707} {\bibfield  {journal} {\bibinfo  {journal} {Physical Review Letters}\ }\textbf {\bibinfo {volume} {82}},\ \bibinfo {pages} {4707} (\bibinfo {year} {1999})}\BibitemShut {NoStop}%
\bibitem [{\citenamefont {Chubukov}\ \emph {et~al.}(2004)\citenamefont {Chubukov}, \citenamefont {P{\'{e}}pin},\ and\ \citenamefont {Rech}}]{Chubukov2004InstabilityFerromagnets}%
  \BibitemOpen
  \bibfield  {author} {\bibinfo {author} {\bibfnamefont {A.~V.}\ \bibnamefont {Chubukov}}, \bibinfo {author} {\bibfnamefont {C.}~\bibnamefont {P{\'{e}}pin}}, \ and\ \bibinfo {author} {\bibfnamefont {J.}~\bibnamefont {Rech}},\ }\href {\doibase 10.1103/PhysRevLett.92.147003} {\bibfield  {journal} {\bibinfo  {journal} {Physical Review Letters}\ }\textbf {\bibinfo {volume} {92}},\ \bibinfo {pages} {147003} (\bibinfo {year} {2004})}\BibitemShut {NoStop}%
\bibitem [{\citenamefont {Conduit}\ \emph {et~al.}(2009)\citenamefont {Conduit}, \citenamefont {Green},\ and\ \citenamefont {Simons}}]{Conduit2009InhomogeneousFerromagnetism}%
  \BibitemOpen
  \bibfield  {author} {\bibinfo {author} {\bibfnamefont {G.~J.}\ \bibnamefont {Conduit}}, \bibinfo {author} {\bibfnamefont {A.~G.}\ \bibnamefont {Green}}, \ and\ \bibinfo {author} {\bibfnamefont {B.~D.}\ \bibnamefont {Simons}},\ }\href {\doibase 10.1103/PhysRevLett.103.207201} {\bibfield  {journal} {\bibinfo  {journal} {Physical Review Letters}\ }\textbf {\bibinfo {volume} {103}},\ \bibinfo {pages} {207201} (\bibinfo {year} {2009})}\BibitemShut {NoStop}%
\bibitem [{\citenamefont {Karahasanovic}\ \emph {et~al.}(2012)\citenamefont {Karahasanovic}, \citenamefont {Kr{\"{u}}ger},\ and\ \citenamefont {Green}}]{Karahasanovic2012QuantumPoints}%
  \BibitemOpen
  \bibfield  {author} {\bibinfo {author} {\bibfnamefont {U.}~\bibnamefont {Karahasanovic}}, \bibinfo {author} {\bibfnamefont {F.}~\bibnamefont {Kr{\"{u}}ger}}, \ and\ \bibinfo {author} {\bibfnamefont {A.~G.}\ \bibnamefont {Green}},\ }\href {\doibase 10.1103/PhysRevB.85.165111} {\bibfield  {journal} {\bibinfo  {journal} {Physical Review B - Condensed Matter and Materials Physics}\ }\textbf {\bibinfo {volume} {85}},\ \bibinfo {pages} {165111} (\bibinfo {year} {2012})}\BibitemShut {NoStop}%
\bibitem [{\citenamefont {Pedder}\ \emph {et~al.}(2013)\citenamefont {Pedder}, \citenamefont {Kr{\"{u}}ger},\ and\ \citenamefont {Green}}]{Pedder2013ResummationPoints}%
  \BibitemOpen
  \bibfield  {author} {\bibinfo {author} {\bibfnamefont {C.~J.}\ \bibnamefont {Pedder}}, \bibinfo {author} {\bibfnamefont {F.}~\bibnamefont {Kr{\"{u}}ger}}, \ and\ \bibinfo {author} {\bibfnamefont {A.~G.}\ \bibnamefont {Green}},\ }\href {\doibase 10.1103/PhysRevB.88.165109} {\bibfield  {journal} {\bibinfo  {journal} {Physical Review B}\ }\textbf {\bibinfo {volume} {88}},\ \bibinfo {pages} {165109} (\bibinfo {year} {2013})}\BibitemShut {NoStop}%
\bibitem [{\citenamefont {Belitz}\ and\ \citenamefont {Kirkpatrick}(2017)}]{Belitz2017QuantumMagnets}%
  \BibitemOpen
  \bibfield  {author} {\bibinfo {author} {\bibfnamefont {D.}~\bibnamefont {Belitz}}\ and\ \bibinfo {author} {\bibfnamefont {T.~R.}\ \bibnamefont {Kirkpatrick}},\ }\href {\doibase 10.1103/PhysRevLett.119.267202} {\bibfield  {journal} {\bibinfo  {journal} {Physical Review Letters}\ }\textbf {\bibinfo {volume} {119}},\ \bibinfo {pages} {267202} (\bibinfo {year} {2017})}\BibitemShut {NoStop}%
\bibitem [{\citenamefont {Rech}\ \emph {et~al.}(2006)\citenamefont {Rech}, \citenamefont {P{\'{e}}pin},\ and\ \citenamefont {Chubukov}}]{Rech2006QuantumPoint}%
  \BibitemOpen
  \bibfield  {author} {\bibinfo {author} {\bibfnamefont {J.}~\bibnamefont {Rech}}, \bibinfo {author} {\bibfnamefont {C.}~\bibnamefont {P{\'{e}}pin}}, \ and\ \bibinfo {author} {\bibfnamefont {A.~V.}\ \bibnamefont {Chubukov}},\ }\href {\doibase 10.1103/PhysRevB.74.195126} {\bibfield  {journal} {\bibinfo  {journal} {Physical Review B - Condensed Matter and Materials Physics}\ }\textbf {\bibinfo {volume} {74}},\ \bibinfo {pages} {195126} (\bibinfo {year} {2006})}\BibitemShut {NoStop}%
\bibitem [{\citenamefont {Wysoki{\'{n}}ski}(2019)}]{Wysokinski2019MechanismMagnets}%
  \BibitemOpen
  \bibfield  {author} {\bibinfo {author} {\bibfnamefont {M.~M.}\ \bibnamefont {Wysoki{\'{n}}ski}},\ }\href {\doibase 10.1038/s41598-019-55658-x} {\bibfield  {journal} {\bibinfo  {journal} {Scientific Reports}\ }\textbf {\bibinfo {volume} {9}},\ \bibinfo {pages} {19461} (\bibinfo {year} {2019})}\BibitemShut {NoStop}%
\bibitem [{\citenamefont {Huxley}\ \emph {et~al.}(2000)\citenamefont {Huxley}, \citenamefont {Sheikin},\ and\ \citenamefont {Braithwaite}}]{Huxley2000MetamagneticUGe2}%
  \BibitemOpen
  \bibfield  {author} {\bibinfo {author} {\bibfnamefont {A.}~\bibnamefont {Huxley}}, \bibinfo {author} {\bibfnamefont {I.}~\bibnamefont {Sheikin}}, \ and\ \bibinfo {author} {\bibfnamefont {D.}~\bibnamefont {Braithwaite}},\ }\href {\doibase 10.1016/S0921-4526(99)02545-4} {\bibfield  {journal} {\bibinfo  {journal} {Physica B: Condensed Matter}\ }\textbf {\bibinfo {volume} {284-288}},\ \bibinfo {pages} {1277} (\bibinfo {year} {2000})}\BibitemShut {NoStop}%
\bibitem [{\citenamefont {Pfleiderer}\ and\ \citenamefont {Huxley}(2002)}]{Pfleiderer2002PressurePresented}%
  \BibitemOpen
  \bibfield  {author} {\bibinfo {author} {\bibfnamefont {C.}~\bibnamefont {Pfleiderer}}\ and\ \bibinfo {author} {\bibfnamefont {A.~D.}\ \bibnamefont {Huxley}},\ }\href {\doibase 10.1103/PhysRevLett.89.147005} {\bibfield  {journal} {\bibinfo  {journal} {Physical Review Letters}\ }\textbf {\bibinfo {volume} {89}},\ \bibinfo {pages} {147005} (\bibinfo {year} {2002})}\BibitemShut {NoStop}%
\bibitem [{\citenamefont {Uhlarz}\ \emph {et~al.}(2004)\citenamefont {Uhlarz}, \citenamefont {Pfleiderer},\ and\ \citenamefont {Hayden}}]{Uhlarz2004QuantumZrZn2}%
  \BibitemOpen
  \bibfield  {author} {\bibinfo {author} {\bibfnamefont {M.}~\bibnamefont {Uhlarz}}, \bibinfo {author} {\bibfnamefont {C.}~\bibnamefont {Pfleiderer}}, \ and\ \bibinfo {author} {\bibfnamefont {S.~M.}\ \bibnamefont {Hayden}},\ }\href {\doibase 10.1103/PhysRevLett.93.256404} {\bibfield  {journal} {\bibinfo  {journal} {Physical Review Letters}\ }\textbf {\bibinfo {volume} {93}},\ \bibinfo {pages} {256404} (\bibinfo {year} {2004})}\BibitemShut {NoStop}%
\bibitem [{\citenamefont {Niklowitz}\ \emph {et~al.}(2005)\citenamefont {Niklowitz}, \citenamefont {Beckers}, \citenamefont {Lonzarich}, \citenamefont {Knebel}, \citenamefont {Salce}, \citenamefont {Thomasson}, \citenamefont {Bernhoeft}, \citenamefont {Braithwaite},\ and\ \citenamefont {Flouquet}}]{Niklowitz2005Spin-fluctuation-dominatedPressure}%
  \BibitemOpen
  \bibfield  {author} {\bibinfo {author} {\bibfnamefont {P.~G.}\ \bibnamefont {Niklowitz}}, \bibinfo {author} {\bibfnamefont {F.}~\bibnamefont {Beckers}}, \bibinfo {author} {\bibfnamefont {G.~G.}\ \bibnamefont {Lonzarich}}, \bibinfo {author} {\bibfnamefont {G.}~\bibnamefont {Knebel}}, \bibinfo {author} {\bibfnamefont {B.}~\bibnamefont {Salce}}, \bibinfo {author} {\bibfnamefont {J.}~\bibnamefont {Thomasson}}, \bibinfo {author} {\bibfnamefont {N.}~\bibnamefont {Bernhoeft}}, \bibinfo {author} {\bibfnamefont {D.}~\bibnamefont {Braithwaite}}, \ and\ \bibinfo {author} {\bibfnamefont {J.}~\bibnamefont {Flouquet}},\ }\href {\doibase 10.1103/PhysRevB.72.024424} {\bibfield  {journal} {\bibinfo  {journal} {Physical Review B - Condensed Matter and Materials Physics}\ }\textbf {\bibinfo {volume} {72}},\ \bibinfo {pages} {024424} (\bibinfo {year} {2005})}\BibitemShut {NoStop}%
\bibitem [{\citenamefont {Kaluarachchi}\ \emph {et~al.}(2017)\citenamefont {Kaluarachchi}, \citenamefont {Bud'ko}, \citenamefont {Canfield},\ and\ \citenamefont {Taufour}}]{Kaluarachchi2017TricriticalPressure}%
  \BibitemOpen
  \bibfield  {author} {\bibinfo {author} {\bibfnamefont {U.~S.}\ \bibnamefont {Kaluarachchi}}, \bibinfo {author} {\bibfnamefont {S.~L.}\ \bibnamefont {Bud'ko}}, \bibinfo {author} {\bibfnamefont {P.~C.}\ \bibnamefont {Canfield}}, \ and\ \bibinfo {author} {\bibfnamefont {V.}~\bibnamefont {Taufour}},\ }\href {\doibase 10.1038/s41467-017-00699-x} {\bibfield  {journal} {\bibinfo  {journal} {Nature Communications}\ }\textbf {\bibinfo {volume} {8}},\ \bibinfo {pages} {546} (\bibinfo {year} {2017})}\BibitemShut {NoStop}%
\bibitem [{\citenamefont {Gati}\ \emph {et~al.}(2021)\citenamefont {Gati}, \citenamefont {Wilde}, \citenamefont {Khasanov}, \citenamefont {Xiang}, \citenamefont {Dissanayake}, \citenamefont {Gupta}, \citenamefont {Matsuda}, \citenamefont {Ye}, \citenamefont {Haberl}, \citenamefont {Kaluarachchi}, \citenamefont {McQueeney}, \citenamefont {Kreyssig}, \citenamefont {Bud'ko},\ and\ \citenamefont {Canfield}}]{Gati2021FormationLaCrGe3}%
  \BibitemOpen
  \bibfield  {author} {\bibinfo {author} {\bibfnamefont {E.}~\bibnamefont {Gati}}, \bibinfo {author} {\bibfnamefont {J.~M.}\ \bibnamefont {Wilde}}, \bibinfo {author} {\bibfnamefont {R.}~\bibnamefont {Khasanov}}, \bibinfo {author} {\bibfnamefont {L.}~\bibnamefont {Xiang}}, \bibinfo {author} {\bibfnamefont {S.}~\bibnamefont {Dissanayake}}, \bibinfo {author} {\bibfnamefont {R.}~\bibnamefont {Gupta}}, \bibinfo {author} {\bibfnamefont {M.}~\bibnamefont {Matsuda}}, \bibinfo {author} {\bibfnamefont {F.}~\bibnamefont {Ye}}, \bibinfo {author} {\bibfnamefont {B.}~\bibnamefont {Haberl}}, \bibinfo {author} {\bibfnamefont {U.}~\bibnamefont {Kaluarachchi}}, \bibinfo {author} {\bibfnamefont {R.~J.}\ \bibnamefont {McQueeney}}, \bibinfo {author} {\bibfnamefont {A.}~\bibnamefont {Kreyssig}}, \bibinfo {author} {\bibfnamefont {S.~L.}\ \bibnamefont {Bud'ko}}, \ and\ \bibinfo {author} {\bibfnamefont {P.~C.}\ \bibnamefont {Canfield}},\ }\href {\doibase 10.1103/PhysRevB.103.075111} {\bibfield  {journal} {\bibinfo  {journal}
  {Physical Review B}\ }\textbf {\bibinfo {volume} {103}},\ \bibinfo {pages} {075111} (\bibinfo {year} {2021})}\BibitemShut {NoStop}%
\bibitem [{\citenamefont {Kotegawa}\ \emph {et~al.}(2019)\citenamefont {Kotegawa}, \citenamefont {Matsuoka}, \citenamefont {Uga}, \citenamefont {Takemura}, \citenamefont {Manago}, \citenamefont {Chikuchi}, \citenamefont {Sugawara}, \citenamefont {Tou},\ and\ \citenamefont {Harima}}]{Kotegawa2019Indicationsub4/sub}%
  \BibitemOpen
  \bibfield  {author} {\bibinfo {author} {\bibfnamefont {H.}~\bibnamefont {Kotegawa}}, \bibinfo {author} {\bibfnamefont {E.}~\bibnamefont {Matsuoka}}, \bibinfo {author} {\bibfnamefont {T.}~\bibnamefont {Uga}}, \bibinfo {author} {\bibfnamefont {M.}~\bibnamefont {Takemura}}, \bibinfo {author} {\bibfnamefont {M.}~\bibnamefont {Manago}}, \bibinfo {author} {\bibfnamefont {N.}~\bibnamefont {Chikuchi}}, \bibinfo {author} {\bibfnamefont {H.}~\bibnamefont {Sugawara}}, \bibinfo {author} {\bibfnamefont {H.}~\bibnamefont {Tou}}, \ and\ \bibinfo {author} {\bibfnamefont {H.}~\bibnamefont {Harima}},\ }\href {\doibase 10.7566/JPSJ.88.093702} {\bibfield  {journal} {\bibinfo  {journal} {Journal of the Physical Society of Japan}\ }\textbf {\bibinfo {volume} {88}},\ \bibinfo {pages} {093702} (\bibinfo {year} {2019})}\BibitemShut {NoStop}%
\bibitem [{\citenamefont {Niklowitz}\ \emph {et~al.}(2019)\citenamefont {Niklowitz}, \citenamefont {Hirschberger}, \citenamefont {Lucas}, \citenamefont {Cermak}, \citenamefont {Schneidewind}, \citenamefont {Faulhaber}, \citenamefont {Mignot}, \citenamefont {Duncan}, \citenamefont {Neubauer}, \citenamefont {Pfleiderer},\ and\ \citenamefont {Grosche}}]{Niklowitz2019Ultrasmall/math}%
  \BibitemOpen
  \bibfield  {author} {\bibinfo {author} {\bibfnamefont {P.}~\bibnamefont {Niklowitz}}, \bibinfo {author} {\bibfnamefont {M.}~\bibnamefont {Hirschberger}}, \bibinfo {author} {\bibfnamefont {M.}~\bibnamefont {Lucas}}, \bibinfo {author} {\bibfnamefont {P.}~\bibnamefont {Cermak}}, \bibinfo {author} {\bibfnamefont {A.}~\bibnamefont {Schneidewind}}, \bibinfo {author} {\bibfnamefont {E.}~\bibnamefont {Faulhaber}}, \bibinfo {author} {\bibfnamefont {J.-M.}\ \bibnamefont {Mignot}}, \bibinfo {author} {\bibfnamefont {W.}~\bibnamefont {Duncan}}, \bibinfo {author} {\bibfnamefont {A.}~\bibnamefont {Neubauer}}, \bibinfo {author} {\bibfnamefont {C.}~\bibnamefont {Pfleiderer}}, \ and\ \bibinfo {author} {\bibfnamefont {F.}~\bibnamefont {Grosche}},\ }\href {\doibase 10.1103/PhysRevLett.123.247203} {\bibfield  {journal} {\bibinfo  {journal} {Physical Review Letters}\ }\textbf {\bibinfo {volume} {123}},\ \bibinfo {pages} {247203} (\bibinfo {year} {2019})}\BibitemShut {NoStop}%
\bibitem [{\citenamefont {Jesche}\ \emph {et~al.}(2017)\citenamefont {Jesche}, \citenamefont {Ball{\'{e}}}, \citenamefont {Kliemt}, \citenamefont {Geibel}, \citenamefont {Brando},\ and\ \citenamefont {Krellner}}]{Jesche2017AvoidedCeFePO}%
  \BibitemOpen
  \bibfield  {author} {\bibinfo {author} {\bibfnamefont {A.}~\bibnamefont {Jesche}}, \bibinfo {author} {\bibfnamefont {T.}~\bibnamefont {Ball{\'{e}}}}, \bibinfo {author} {\bibfnamefont {K.}~\bibnamefont {Kliemt}}, \bibinfo {author} {\bibfnamefont {C.}~\bibnamefont {Geibel}}, \bibinfo {author} {\bibfnamefont {M.}~\bibnamefont {Brando}}, \ and\ \bibinfo {author} {\bibfnamefont {C.}~\bibnamefont {Krellner}},\ }\href {\doibase 10.1002/pssb.201600169} {\bibfield  {journal} {\bibinfo  {journal} {Physica Status Solidi (B) Basic Research}\ }\textbf {\bibinfo {volume} {254}},\ \bibinfo {pages} {1600169} (\bibinfo {year} {2017})}\BibitemShut {NoStop}%
\bibitem [{\citenamefont {Kirkpatrick}\ and\ \citenamefont {Belitz}(2020)}]{Kirkpatrick2020FerromagneticSystems}%
  \BibitemOpen
  \bibfield  {author} {\bibinfo {author} {\bibfnamefont {T.}~\bibnamefont {Kirkpatrick}}\ and\ \bibinfo {author} {\bibfnamefont {D.}~\bibnamefont {Belitz}},\ }\href {\doibase 10.1103/PhysRevLett.124.147201} {\bibfield  {journal} {\bibinfo  {journal} {Physical Review Letters}\ }\textbf {\bibinfo {volume} {124}},\ \bibinfo {pages} {147201} (\bibinfo {year} {2020})}\BibitemShut {NoStop}%
\bibitem [{\citenamefont {Saunders}\ \emph {et~al.}(2020)\citenamefont {Saunders}, \citenamefont {Xiang}, \citenamefont {Khasanov}, \citenamefont {Kong}, \citenamefont {Lin}, \citenamefont {Bud'ko},\ and\ \citenamefont {Canfield}}]{Saunders2020ExceedinglyLa5Co2Ge3}%
  \BibitemOpen
  \bibfield  {author} {\bibinfo {author} {\bibfnamefont {S.~M.}\ \bibnamefont {Saunders}}, \bibinfo {author} {\bibfnamefont {L.}~\bibnamefont {Xiang}}, \bibinfo {author} {\bibfnamefont {R.}~\bibnamefont {Khasanov}}, \bibinfo {author} {\bibfnamefont {T.}~\bibnamefont {Kong}}, \bibinfo {author} {\bibfnamefont {Q.}~\bibnamefont {Lin}}, \bibinfo {author} {\bibfnamefont {S.~L.}\ \bibnamefont {Bud'ko}}, \ and\ \bibinfo {author} {\bibfnamefont {P.~C.}\ \bibnamefont {Canfield}},\ }\href {\doibase 10.1103/PhysRevB.101.214405} {\bibfield  {journal} {\bibinfo  {journal} {Physical Review B}\ }\textbf {\bibinfo {volume} {101}},\ \bibinfo {pages} {214405} (\bibinfo {year} {2020})}\BibitemShut {NoStop}%
\bibitem [{\citenamefont {Lin}\ \emph {et~al.}(2017)\citenamefont {Lin}, \citenamefont {Aguirre}, \citenamefont {Saunders}, \citenamefont {Hackett}, \citenamefont {Liu}, \citenamefont {Taufour}, \citenamefont {Paudyal}, \citenamefont {Bud'ko}, \citenamefont {Canfield},\ and\ \citenamefont {Miller}}]{Lin2017PolarClusters}%
  \BibitemOpen
  \bibfield  {author} {\bibinfo {author} {\bibfnamefont {Q.}~\bibnamefont {Lin}}, \bibinfo {author} {\bibfnamefont {K.}~\bibnamefont {Aguirre}}, \bibinfo {author} {\bibfnamefont {S.~M.}\ \bibnamefont {Saunders}}, \bibinfo {author} {\bibfnamefont {T.~A.}\ \bibnamefont {Hackett}}, \bibinfo {author} {\bibfnamefont {Y.}~\bibnamefont {Liu}}, \bibinfo {author} {\bibfnamefont {V.}~\bibnamefont {Taufour}}, \bibinfo {author} {\bibfnamefont {D.}~\bibnamefont {Paudyal}}, \bibinfo {author} {\bibfnamefont {S.}~\bibnamefont {Bud'ko}}, \bibinfo {author} {\bibfnamefont {P.~C.}\ \bibnamefont {Canfield}}, \ and\ \bibinfo {author} {\bibfnamefont {G.~J.}\ \bibnamefont {Miller}},\ }\href {\doibase 10.1002/chem.201702798} {\bibfield  {journal} {\bibinfo  {journal} {Chemistry – A European Journal}\ }\textbf {\bibinfo {volume} {23}},\ \bibinfo {pages} {10516} (\bibinfo {year} {2017})}\BibitemShut {NoStop}%
\bibitem [{\citenamefont {Rhodes}\ and\ \citenamefont {Wohlfarth}(1963)}]{Rhodes1963TheAlloys}%
  \BibitemOpen
  \bibfield  {author} {\bibinfo {author} {\bibfnamefont {P.}~\bibnamefont {Rhodes}}\ and\ \bibinfo {author} {\bibfnamefont {E.~P.}\ \bibnamefont {Wohlfarth}},\ }\href {\doibase 10.1098/rspa.1963.0086} {\bibfield  {journal} {\bibinfo  {journal} {Proceedings of the Royal Society of London. Series A. Mathematical and Physical Sciences}\ }\textbf {\bibinfo {volume} {273}},\ \bibinfo {pages} {247} (\bibinfo {year} {1963})}\BibitemShut {NoStop}%
\bibitem [{\citenamefont {Santiago}\ \emph {et~al.}(2017)\citenamefont {Santiago}, \citenamefont {Huang},\ and\ \citenamefont {Morosan}}]{Santiago2017ItinerantMetals}%
  \BibitemOpen
  \bibfield  {author} {\bibinfo {author} {\bibfnamefont {J.~M.}\ \bibnamefont {Santiago}}, \bibinfo {author} {\bibfnamefont {C.-L.}\ \bibnamefont {Huang}}, \ and\ \bibinfo {author} {\bibfnamefont {E.}~\bibnamefont {Morosan}},\ }\href {\doibase 10.1088/1361-648X/aa7889} {\bibfield  {journal} {\bibinfo  {journal} {Journal of Physics: Condensed Matter}\ }\textbf {\bibinfo {volume} {29}},\ \bibinfo {pages} {373002} (\bibinfo {year} {2017})}\BibitemShut {NoStop}%
\bibitem [{\citenamefont {Xiang}\ \emph {et~al.}(2021)\citenamefont {Xiang}, \citenamefont {Gati}, \citenamefont {Bud'ko}, \citenamefont {Saunders},\ and\ \citenamefont {Canfield}}]{Xiang2021AvoidedLa5Co2Ge3}%
  \BibitemOpen
  \bibfield  {author} {\bibinfo {author} {\bibfnamefont {L.}~\bibnamefont {Xiang}}, \bibinfo {author} {\bibfnamefont {E.}~\bibnamefont {Gati}}, \bibinfo {author} {\bibfnamefont {S.~L.}\ \bibnamefont {Bud'ko}}, \bibinfo {author} {\bibfnamefont {S.~M.}\ \bibnamefont {Saunders}}, \ and\ \bibinfo {author} {\bibfnamefont {P.~C.}\ \bibnamefont {Canfield}},\ }\href {\doibase 10.1103/PhysRevB.103.054419} {\bibfield  {journal} {\bibinfo  {journal} {Physical Review B}\ }\textbf {\bibinfo {volume} {103}},\ \bibinfo {pages} {054419} (\bibinfo {year} {2021})}\BibitemShut {NoStop}%
\bibitem [{\citenamefont {Canfield}\ and\ \citenamefont {Fisher}(2001)}]{Canfield2001High-temperatureQuasicrystals}%
  \BibitemOpen
  \bibfield  {author} {\bibinfo {author} {\bibfnamefont {P.~C.}\ \bibnamefont {Canfield}}\ and\ \bibinfo {author} {\bibfnamefont {I.~R.}\ \bibnamefont {Fisher}},\ }\href {\doibase 10.1016/S0022-0248(01)00827-2} {\bibfield  {journal} {\bibinfo  {journal} {Journal of Crystal Growth}\ }\textbf {\bibinfo {volume} {225}},\ \bibinfo {pages} {155} (\bibinfo {year} {2001})}\BibitemShut {NoStop}%
\bibitem [{\citenamefont {Canfield}(2020)}]{Canfield2020NewPhysics}%
  \BibitemOpen
  \bibfield  {author} {\bibinfo {author} {\bibfnamefont {P.~C.}\ \bibnamefont {Canfield}},\ }\href {\doibase 10.1088/1361-6633/ab514b} {\bibfield  {journal} {\bibinfo  {journal} {Reports on Progress in Physics}\ }\textbf {\bibinfo {volume} {83}},\ \bibinfo {pages} {016501} (\bibinfo {year} {2020})}\BibitemShut {NoStop}%
\bibitem [{\citenamefont {Canfield}\ and\ \citenamefont {Fisk}(1992)}]{Canfield1992GrowthFluxes}%
  \BibitemOpen
  \bibfield  {author} {\bibinfo {author} {\bibfnamefont {P.~C.}\ \bibnamefont {Canfield}}\ and\ \bibinfo {author} {\bibfnamefont {Z.}~\bibnamefont {Fisk}},\ }\href {\doibase 10.1080/13642819208215073} {\bibfield  {journal} {\bibinfo  {journal} {Philosophical Magazine B}\ }\textbf {\bibinfo {volume} {65}},\ \bibinfo {pages} {1117} (\bibinfo {year} {1992})}\BibitemShut {NoStop}%
\bibitem [{\citenamefont {Newbury}\ and\ \citenamefont {Ritchie}(2014)}]{Newbury2014RigorousDTSA-II}%
  \BibitemOpen
  \bibfield  {author} {\bibinfo {author} {\bibfnamefont {D.~E.}\ \bibnamefont {Newbury}}\ and\ \bibinfo {author} {\bibfnamefont {N.~W.~M.}\ \bibnamefont {Ritchie}},\ }in\ \href {\doibase 10.1117/12.2065842} {\emph {\bibinfo {booktitle} {Scanning Microscopies 2014}}},\ Vol.\ \bibinfo {volume} {9236}\ (\bibinfo  {publisher} {SPIE},\ \bibinfo {year} {2014})\ p.\ \bibinfo {pages} {92360H}\BibitemShut {NoStop}%
\bibitem [{\citenamefont {Toby}\ and\ \citenamefont {Von~Dreele}(2013)}]{Toby2013Package}%
  \BibitemOpen
  \bibfield  {author} {\bibinfo {author} {\bibfnamefont {B.~H.}\ \bibnamefont {Toby}}\ and\ \bibinfo {author} {\bibfnamefont {R.~B.}\ \bibnamefont {Von~Dreele}},\ }\href {\doibase 10.1107/S0021889813003531} {\bibfield  {journal} {\bibinfo  {journal} {Journal of Applied Crystallography}\ }\textbf {\bibinfo {volume} {46}},\ \bibinfo {pages} {544} (\bibinfo {year} {2013})}\BibitemShut {NoStop}%
\bibitem [{\citenamefont {Amato}\ \emph {et~al.}(2017)\citenamefont {Amato}, \citenamefont {Luetkens}, \citenamefont {Sedlak}, \citenamefont {Stoykov}, \citenamefont {Scheuermann}, \citenamefont {Elender}, \citenamefont {Raselli},\ and\ \citenamefont {Graf}}]{Amato2017TheBeam}%
  \BibitemOpen
  \bibfield  {author} {\bibinfo {author} {\bibfnamefont {A.}~\bibnamefont {Amato}}, \bibinfo {author} {\bibfnamefont {H.}~\bibnamefont {Luetkens}}, \bibinfo {author} {\bibfnamefont {K.}~\bibnamefont {Sedlak}}, \bibinfo {author} {\bibfnamefont {A.}~\bibnamefont {Stoykov}}, \bibinfo {author} {\bibfnamefont {R.}~\bibnamefont {Scheuermann}}, \bibinfo {author} {\bibfnamefont {M.}~\bibnamefont {Elender}}, \bibinfo {author} {\bibfnamefont {A.}~\bibnamefont {Raselli}}, \ and\ \bibinfo {author} {\bibfnamefont {D.}~\bibnamefont {Graf}},\ }\href {\doibase 10.1063/1.4986045} {\bibfield  {journal} {\bibinfo  {journal} {Review of Scientific Instruments}\ }\textbf {\bibinfo {volume} {88}},\ \bibinfo {pages} {093301} (\bibinfo {year} {2017})}\BibitemShut {NoStop}%
\bibitem [{\citenamefont {Monceau}(2012)}]{Monceau2012ElectronicOverview}%
  \BibitemOpen
  \bibfield  {author} {\bibinfo {author} {\bibfnamefont {P.}~\bibnamefont {Monceau}},\ }\href {\doibase 10.1080/00018732.2012.719674} {\bibfield  {journal} {\bibinfo  {journal} {Advances in Physics}\ }\textbf {\bibinfo {volume} {61}},\ \bibinfo {pages} {325} (\bibinfo {year} {2012})}\BibitemShut {NoStop}%
\bibitem [{\citenamefont {Bud’ko}\ and\ \citenamefont {Canfield}(2000)}]{Budko2000RotationalState}%
  \BibitemOpen
  \bibfield  {author} {\bibinfo {author} {\bibfnamefont {S.~L.}\ \bibnamefont {Bud’ko}}\ and\ \bibinfo {author} {\bibfnamefont {P.~C.}\ \bibnamefont {Canfield}},\ }\href {\doibase 10.1103/PhysRevB.61.R14932} {\bibfield  {journal} {\bibinfo  {journal} {Physical Review B}\ }\textbf {\bibinfo {volume} {61}},\ \bibinfo {pages} {R14932} (\bibinfo {year} {2000})}\BibitemShut {NoStop}%
\bibitem [{\citenamefont {Elliott}\ and\ \citenamefont {Wedgwood}(1963)}]{Elliott1963TheoryMetals}%
  \BibitemOpen
  \bibfield  {author} {\bibinfo {author} {\bibfnamefont {R.~J.}\ \bibnamefont {Elliott}}\ and\ \bibinfo {author} {\bibfnamefont {F.~A.}\ \bibnamefont {Wedgwood}},\ }\href@noop {} {\bibfield  {journal} {\bibinfo  {journal} {Proceedings of the Physical Society}\ }\textbf {\bibinfo {volume} {81}},\ \bibinfo {pages} {846} (\bibinfo {year} {1963})}\BibitemShut {NoStop}%
\bibitem [{\citenamefont {Fisher}\ and\ \citenamefont {Langer}(1968)}]{Fisher1968ResistivePoints}%
  \BibitemOpen
  \bibfield  {author} {\bibinfo {author} {\bibfnamefont {M.~E.}\ \bibnamefont {Fisher}}\ and\ \bibinfo {author} {\bibfnamefont {J.~S.}\ \bibnamefont {Langer}},\ }\href {\doibase 10.1103/PhysRevLett.20.665} {\bibfield  {journal} {\bibinfo  {journal} {Physical Review Letters}\ }\textbf {\bibinfo {volume} {20}},\ \bibinfo {pages} {665} (\bibinfo {year} {1968})}\BibitemShut {NoStop}%
\bibitem [{\citenamefont {Arrott}(1957)}]{Arrott1957CriterionIsotherms}%
  \BibitemOpen
  \bibfield  {author} {\bibinfo {author} {\bibfnamefont {A.}~\bibnamefont {Arrott}},\ }\href {\doibase 10.1103/PhysRev.108.1394} {\bibfield  {journal} {\bibinfo  {journal} {Physical Review}\ }\textbf {\bibinfo {volume} {108}},\ \bibinfo {pages} {1394} (\bibinfo {year} {1957})}\BibitemShut {NoStop}%
\bibitem [{\citenamefont {Fisher}(1962)}]{Fisher1962RelationAntiferromagnet}%
  \BibitemOpen
  \bibfield  {author} {\bibinfo {author} {\bibfnamefont {M.~E.}\ \bibnamefont {Fisher}},\ }\href {\doibase 10.1080/14786436208213705} {\bibfield  {journal} {\bibinfo  {journal} {Philosophical Magazine}\ }\textbf {\bibinfo {volume} {7}},\ \bibinfo {pages} {1731} (\bibinfo {year} {1962})}\BibitemShut {NoStop}%
\bibitem [{\citenamefont {R{\'{e}}otier}\ and\ \citenamefont {Yaouanc}(1997)}]{Reotier1997MuonMaterials}%
  \BibitemOpen
  \bibfield  {author} {\bibinfo {author} {\bibfnamefont {P.~D.~d.}\ \bibnamefont {R{\'{e}}otier}}\ and\ \bibinfo {author} {\bibfnamefont {A.}~\bibnamefont {Yaouanc}},\ }\href {\doibase 10.1088/0953-8984/9/43/002} {\bibfield  {journal} {\bibinfo  {journal} {Journal of Physics: Condensed Matter}\ }\textbf {\bibinfo {volume} {9}},\ \bibinfo {pages} {9113} (\bibinfo {year} {1997})}\BibitemShut {NoStop}%
\bibitem [{\citenamefont {Khasanov}\ \emph {et~al.}(2020)\citenamefont {Khasanov}, \citenamefont {Simutis}, \citenamefont {Pashkevich}, \citenamefont {Shevtsova}, \citenamefont {Meier}, \citenamefont {Xu}, \citenamefont {Bud'ko}, \citenamefont {Kogan},\ and\ \citenamefont {Canfield}}]{Khasanov2020MagnetismStudies}%
  \BibitemOpen
  \bibfield  {author} {\bibinfo {author} {\bibfnamefont {R.}~\bibnamefont {Khasanov}}, \bibinfo {author} {\bibfnamefont {G.}~\bibnamefont {Simutis}}, \bibinfo {author} {\bibfnamefont {Y.~G.}\ \bibnamefont {Pashkevich}}, \bibinfo {author} {\bibfnamefont {T.}~\bibnamefont {Shevtsova}}, \bibinfo {author} {\bibfnamefont {W.~R.}\ \bibnamefont {Meier}}, \bibinfo {author} {\bibfnamefont {M.}~\bibnamefont {Xu}}, \bibinfo {author} {\bibfnamefont {S.~L.}\ \bibnamefont {Bud'ko}}, \bibinfo {author} {\bibfnamefont {V.~G.}\ \bibnamefont {Kogan}}, \ and\ \bibinfo {author} {\bibfnamefont {P.~C.}\ \bibnamefont {Canfield}},\ }\href {\doibase 10.1103/PhysRevB.102.094504} {\bibfield  {journal} {\bibinfo  {journal} {Physical Review B}\ }\textbf {\bibinfo {volume} {102}},\ \bibinfo {pages} {094504} (\bibinfo {year} {2020})}\BibitemShut {NoStop}%
\end{thebibliography}%

\end{document}